\documentclass[12pt]{article}
\usepackage{epsfig,amssymb}
\usepackage{latexsym}

\newcommand{\bq}{\begin{equation}}
\newcommand{\eq}{\end{equation}}
\newcommand{\bqa}{\begin{eqnarray}}
\newcommand{\eqa}{\end{eqnarray}}
\newcommand{\ben}{\begin{enumerate}}
\newcommand{\een}{\end{enumerate}}
\newcommand{\bc}{\begin{center}}
\newcommand{\ec}{\end{center}}
\newcommand{\bqb}{\begin{eqnarray*}}
\newcommand{\eqb}{\end{eqnarray*}}
\newcommand{\psl}{\rlap / p}

\newcommand{\eesl}{\rlap / \epsilon}

\def\pr#1#2#3{Phys. Rev. ${\bf{#1}}$, #2 (#3)}
\def\prl#1#2#3{Phys. Rev. Lett. ${\bf{#1}}$, #2 (#3)}
\def\pl#1#2#3{Phys. Lett. ${\bf{#1}}$, #2 (#3)}
\def\prep#1#2#3{Phys. Rept. ${\bf{#1}}$, #2 (#3)}

\def\np#1#2#3{Nucl. Phys. ${\bf{#1}}$, #2 (#3)}
\def\npps#1#2#3{Nucl. Phys. A, Proc. Suppl. ${\bf{#1}}$, #2 (#3)}
\def\zp#1#2#3{Z. f. Phys. ${\bf{#1}}$, #2 (#3)}

\def\ijmp#1#2#3{Int. J. Mod. Phys. ${\bf{#1}}$, #2 (#3)}

\def\aop#1#2#3{Annals of Phys. ${\bf{#1}}$, #2 (#3)}

\def\polon#1#2#3{Acta Phys. Polon. ${\bf{#1}}$, #2 (#3)}

\begin{document}
\pagenumbering{arabic}
\thispagestyle{empty}
\def\thefootnote{\fnsymbol{footnote}}
\setcounter{footnote}{1}

\begin{flushright}
April 21, 2015;\\  corrected version\\
 \end{flushright}

\vspace{2cm}

\begin{center}
{\Large {\bf Specific  supersimple properties of $e^-e^+\to  \gamma H$\\
 at high energy}}.\\
 \vspace{1cm}
{\large G.J. Gounaris$^a$ and F.M. Renard$^b$}\\
\vspace{0.2cm}
$^a$Department of Theoretical Physics, Aristotle
University of Thessaloniki,\\
Gr-54124, Thessaloniki, Greece.\\
\vspace{0.2cm}
$^b$Laboratoire Univers et Particules de Montpellier,
UMR 5299\\
Universit\'{e} Montpellier II, Place Eug\`{e}ne Bataillon CC072\\
 F-34095 Montpellier Cedex 5.\\
\end{center}

\vspace*{1.cm}
\begin{center}
{\bf Abstract}
\end{center}

We study the process $e^-e^+\to  \gamma H$, where $H$ represents $H_{SM}$, $h^0$
or $H^0$. This process occurs at the one loop level in the standard model (SM) or in
the minimal supersymmetric standard model (MSSM). We establish supersimple (sim)
 high energy expressions for all helicity amplitudes of this process,
 and we identify their level of accuracy  for    describing
the various polarized and unpolarized observables, and   for distinguishing
 SM from MSSM or another   beyond the standard model (BSM).
We pay a special attention to transverse $e^{\mp}$ polarization and azimuthal
dependencies induced by  the imaginary parts of the amplitudes, which are
relatively important in this process.

\vspace{0.5cm}

\def\thefootnote{\arabic{footnote}}
\setcounter{footnote}{0}
\clearpage

\section{Introduction}

After the discovery \cite{Higgsdiscov} of the Higgs boson \cite{Higgs},
detailed experimental and theoretical studies are necessary
for checking its properties and dynamics \cite{Higgsearch, ILCH}.
Within this aim, we have recently studied the process $e^-e^+\to  Z H$ \cite {simZH},
where  $H$ represents $(H_{SM}, h^0, H^0)$. This process occurs at the one loop level
in the standard model (SM) or the minimal supersymmetric standard model (MSSM), and is
observable at future linear colliders  \cite{ILC, CLIC}, and
also at future circular colliders; see \cite{FCC-ee}.
We have analyzed the contents of its amplitudes in SM and  MSSM,
their consequences for the various observables, and we have established simple expressions
which approximate them at high energy.\\

In the present paper we consider the process $e^-e^+\to \gamma H$ which,
contrarily to $e^-e^+\to ZH$, has no Born term and should be
relatively more affected by anomalous effects.
The basic amplitudes are arising at electroweak
one loop order \cite{Barroso,Abbasabadi}.
These contributions contain specific SM or MSSM parts that we  discuss.
Due to the absence of real Born terms, the imaginary parts of the one
loop amplitudes play here an important role and we  look for the
consequences of this feature.

We start from  the complete computation of these one loop amplitudes
in SM and  MSSM, in order to dispose of the exact expressions in terms
of Passarino-Veltman (PV) functions \cite{PV}.
From their expansions at high energy \cite{asPV}, we then establish simple approximate
expressions called supersimple (sim), as in the other similar recent studies on
$(gg\to \gamma \gamma, ~\gamma Z, ~ZZ, W^-W^+)$, $ug\to d W^+$ and $(e^-e^+\to t\bar t, ~W^-W^+,~ ZH) $,
\cite{super,ttbar,WW,simZH}. As before,
we expect that these simple expressions will be useful for making  quick estimates of
the amplitudes and  meaningful comparisons with experimental results.

In addition, during the analysis of
$e^-e^+\to ZH$  \cite{simZH}, we have found that some observables are especially sensitive
to the underlying dynamics; i.e. SM,  MSSM or another BSM.
Our aim is therefore to check whether this is also the case for  $e^-e^+\to \gamma H$.
Particularly because $e^-e^+\to \gamma H$  is determined by the helicity violating (HV)
amplitudes\footnote{In our case, HV are the amplitudes
that at high energies violate (\ref{heli-cons}); see below.}
at high energies, in contrast to the processes aforementioned in the preceding paragraph, which are
dominated by their helicity conserving (HC) amplitudes  \cite{heli1, heli2}.
As we show in the Appendix, one consequence of this is that
 the augmented Sudakov-type quadratic logarithms in  $e^-e^+\to \gamma H$ do not retain the universal
structure observed in  $(gg\to \gamma \gamma, ~\gamma Z, ~ZZ, W^-W^+)$, $ug\to d W^+$,
$(e^-e^+\to t\bar t, ~W^-W^+,~ ZH) $,  \cite{super,ttbar,WW,simZH}.

This way, we have  calculated various polarized and unpolarized cross sections and asymmetries,
and studied their sensitivity to the underlying dynamics.
In addition, we also consider the possibility of
transversally polarized $e^-e^+$ beams, where the presence of imaginary parts in the
$e^-e^+\to \gamma H$ amplitudes leads to a particular azimuthal $\sin2\phi$
dependence,  supplying further dynamical tests.\\

The contents of the paper are the following: In Sect.2 we give the notation for
the  kinematics and the helicity amplitudes. In Sect.3 we
present the electroweak (EW) one loop contributions and establish their supersimple
expressions, explicitly written in the Appendix. Sect.4 is devoted to the description
of other BSM effects in terms of effective couplings. The various observables
are defined in Sect.5. The numerical analysis is presented  in Sect.6 with
many illustrations, while the conclusions are given in Sect.7.\\

\section{Kinematics and helicity amplitudes}

The kinematics of the process
\bq
e^-_\lambda (l) ~ e^+_{\lambda'} (l') \to \gamma_\tau (p)~ H(p')~~, \label{process}
\eq
is defined in terms of the helicities $(\lambda, \lambda')$ of the incoming $(e^-, e^+)$ beams,
and  the helicity $\tau$ of the outgoing $\gamma$, whose polarization vector is $\epsilon$.
The momenta of the various incoming and outgoing particles are $(l,l',p,p')$, and we also use
\bqa
 && s=(l+l')^2=(p+p')^2 ~~,~~ t=(l-p)^2=(l'-p')^2 ,~~ u=(l-p')^2=(l'-p)^2~~,~~ \nonumber \\
 && p_{\gamma}=E_{\gamma}={s-m^2_H\over2\sqrt{s}}=\beta_{\gamma}{s\over2}~~, \label{kinematics}
\eqa
where $p_\gamma=E_\gamma = p$ denote the equal values of
the three-momentum and  energy of the outgoing photon,
and $\theta$ is the angle between the direction of the incoming $e^-$
 and the outgoing $\gamma$.\\

 The general invariant amplitude is written as

\bq
A=\sum_i N_{i=1,3,4}(s,t) I_i +N'_1(s,t)J_1
\eq
using the forms
\bqa
&&I_1=\bar v(e^+_{\lambda'})~\eesl~u(e^-_\lambda)
~~,~~
I_3=\bar v(e^+_{\lambda'})~\epsilon \cdot l'\psl~~u(e^-_\lambda)~~,~~
I_4=\bar v(e^+_{\lambda'})~\epsilon\cdot l\psl~u(e^-_\lambda) ~~,  \label{I1I3I4-forms} \\
&& J_1=-i\bar v(e^+_{\lambda'})
\epsilon^{\mu\nu\rho\sigma}\gamma_{\mu}\epsilon_{\nu}p'_{\rho}p_{\sigma} u(e^-_\lambda)
~~, \label{J1-forms}
\eqa
already used in \cite{simZH}. Note that the $J_1$ form  only appears in
the case of CP-violating couplings; see Sect.4.

The scalar functions $N_{i=1,3,4}(s,t),N'_1(s,t)$ are obtained by computation
of specific diagrams.
One then gets the corresponding helicity amplitudes $F_{\lambda, \tau}(s,\theta)$
by usual expansion of the Dirac spinors appearing in  the forms
(\ref{I1I3I4-forms}, \ref{J1-forms}),
using the standard Jacob-Wick conventions \cite{JW};  note that  $\tau=\pm1$
and that
\bq
\lambda =-\lambda'=\mp {1\over 2} ~~~, \label{e-massless}
\eq
when one neglects the electron mass.
The notation $(\lambda=L,R)$ for these two cases of $e^{\mp}$ helicity is also used.
This way one then gets
\bqa
&& I_1  \to  \delta_{\lambda,L}\sqrt{s\over2}(\tau \cos\theta-1)
-\delta_{\lambda,R}\sqrt{s\over2}(\tau \cos\theta+1)) ~,~ \nonumber \\
&& I_3 = -I_4  \to  (\delta_{\lambda,R}-\delta_{\lambda,L}){ps\over2\sqrt{2}}\tau\sin^2\theta
~,~ \nonumber \\
&& J_1\to  2\lambda {ps\over\sqrt{2}}(\cos\theta+2\lambda\tau) ~. \label{I1I3I4J1-helicities}
\eqa

Due to  (\ref{e-massless}), the  helicity amplitudes $F_{\lambda, \tau}(s,\theta)$
violate the  high energy Helicity Conservation (HCns)
rule \cite{heli1, heli2} which requires
\bq
\lambda +\lambda'=\tau ~~. \label{heli-cons}
\eq
They are thus called helicity violating (HV) amplitudes,
and they  are indeed vanishing at high energy in  MSSM or SM.\\

When CP is conserved, the additional constraint
\bq
F_{\lambda,\tau}(s,\theta)=F_{\lambda,-\tau}(s,\pi-\theta) ~~, \label{CP-conservation}
\eq
also holds,  reducing the independent amplitudes to only two.\\

\section{ The one loop EW corrections and their supersimple (sim) expressions
in SM and MSSM. }

The one-loop amplitudes of the process $e^-e^+\to  \gamma H$
consist of  a reduced set of the diagrams appearing
in the $ZH$ case (replacing of course the $Z$ couplings by the $\gamma$ ones).
Having no Born terms,
there is no self-energy corrections nor renormalization counter terms.
There are also no $s$-channel
initial triangles, because of the absence of any  $H\gamma\gamma, HZ\gamma$ couplings.

Thus, the one loop diagrams in the SM case consist  only of final triangles
in $s$-channel, up and down triangles in $t$ and $u$ channels, direct, crossed and twisted
boxes, and specific diagrams involving 4-leg bosonic couplings;
 see \cite{Barroso,Abbasabadi,Denner}.
 In the MSSM case  we also have  corresponding diagrams involving
supersymmetric partners like
sleptons, squarks, charginos, neutralinos and additional Higgses.
As in \cite{simZH},  we always restrict to  CP-conserving SM or MSSM couplings,
 so that  (\ref{CP-conservation}) is satisfied.

We have computed all these contributions in terms of PV functions \cite{PV}.
This gives  the exact basic contributions.
We then compute their high energy expansions using the forms  in  \cite{asPV}.
The results thus obtained, called sim results, are given in the Appendix.
As already mentioned in Sect.2, the $e^-e^+\to  \gamma H$ amplitudes
are of helicity violating (HV) type \cite{heli1, heli2}. Because of this, for SM or MSSM,
they are suppressed    at high energies like $M/\sqrt{s}$, tempered by terms  involving
single and double  logarithms. \\

\section{Other BSM effects}

Apart from MSSM, another possibility of BSM physics is by inserting
anomalous couplings to the SM gauge and Higgs bosons.
There are various types of such models.
One consequence of them  for the process  $e^-e^+\to  \gamma H$ is the
appearance of Born terms with intermediate $\gamma, Z$ exchanges and
final $\gamma\gamma H$ and $\gamma ZH$ couplings.
Our aim is just to see how the generated amplitudes and observables can differ
from the one loop SM or MSSM predictions, and at what level of accuracy the
sim expressions are adequate for that purpose.

As an example of such a BSM model we consider the description \cite{anom}
of the anomalous couplings of the gauge and SM Higgs field given by  the 2 CP-conserving operators
$O_{UW}$, $O_{UB}$  and the 2 CP-violating ones $\bar O_{UW}$, $\bar O_{UB}$,
\bqa
&& {\cal O}_{UW}=\frac{1}{v^2}\left (\Phi^\dag\Phi-\frac{v^2}{2} \right )\overrightarrow{W}^{\mu\nu}
\overrightarrow{W}_{\mu\nu} ~~,~~
{\cal O}_{UB} =\frac{4}{v^2}\left (\Phi^\dag\Phi-\frac{v^2}{2} \right )B^{\mu\nu}B_{\mu\nu}~~,
\nonumber \\
&& {\bar {\cal O}}_{UW}=\frac{1}{v^2}  (\Phi^\dag\Phi )\overrightarrow{W}^{\mu\nu}
\widetilde{\overrightarrow{W}}_{\mu\nu} ~~, ~~
{\bar{\cal O}}_{UB} =\frac{4}{v^2} (\Phi^\dag\Phi )B^{\mu\nu}\widetilde{B}_{\mu\nu}~~,
\label{d-operators}
\eqa
 where $\Phi$ is the SM Higgs doublet field, with the  vacuum expectation value of its
 neutral component satisfying   $<\phi^0>\sqrt{2}\equiv v=(G_F\sqrt 2)^{-1}$ at tree level.
 Inserting the operators (\ref{d-operators}) in the SM lagrangian
 induces a BSM term  given by
\bq
\delta {\cal L}_{BSM}= d_{UW} {\cal O}_{UW}+ d_{UB} {\cal O}_{UB}+
\bar d_{UW} {\bar {\cal O}}_{UW}+ \bar d_{UB}{\bar{\cal O}}_{UB} ~~, \label{d-BSM}
\eq
which in turn creates the anomalous $\gamma \gamma H_{SM}$ and $\gamma Z H_{SM}$ couplings
\bqa
d_{\gamma Z}=s_Wc_W(d_{UW}-d_{UB}) &,& d_{\gamma \gamma }=(d_{UB}s^2_W+d_{UW})c^2_B ~~, \nonumber \\
\bar d_{\gamma Z}=s_Wc_W(\bar d_{UW}-\bar d_{UB}) &,&
\bar d_{\gamma \gamma }=(\bar d_{UB}s^2_W+\bar d_{UW})c^2_B ~~. \label{dVV'H-couplings}
\eqa

Denoting   $V=\gamma,Z$ and using
\bq
 g^{\gamma}_{eL}=g^{\gamma}_{eR}=q_e=-1 ~~,~~
  g^Z_{eL}={-1+2s^2_W\over2s_Wc_W} ~~,~~  g^Z_{eR}={s_W\over c_W}, ~~ \label{eV-couplings}
\eq
the induced (Born type
with $V$ exchange) anomalous contribution to the invariant amplitude becomes
\bq
A=-~\sum_V{e^2\over(s-m^2_V)}~[g^V_1~I_1+g^{\prime V}_1~J_1]~
[g^V_{eL}P_L+g^V_{eR}P_R] ~~, \label{BSM-amplitudes}
\eq
where the CP-conserving part appears with the $I_1$ form defined in (\ref{I1I3I4-forms})
and the couplings
\bqa
g^{\gamma}_1={2E_{\gamma}\sqrt{s}\over m_Zs_Wc_W}d_{\gamma \gamma} &,&
g^{Z}_1=~{2E_{\gamma}\sqrt{s}\over m_Zs_Wc_W}d_{\gamma Z} ~~,\label{gVi-effc}
\eqa
while the CP violating part is given by $J_1$ of  (\ref{J1-forms}) and the couplings
\bqa
g^{\prime \gamma}_1={2\over m_Zs_Wc_W}\bar d_{\gamma \gamma}  &,&
g^{\prime Z}_1=~{2\over m_Zs_Wc_W}\bar d_{\gamma Z} ~~. \label{gVi-effv}
\eqa

In the illustrations presented in Sect. 6, we choose  the
CP conserving  and the CP-violating BSM couplings in (\ref{d-BSM}), so that the BSM amplitudes are
comparable  to the   1 loop SM  results,
in the high energy domain ($ \lesssim 5$ TeV)  considered here.
More explicitly, we then show separately  two cases with non-vanishing couplings,
 respectively called
\bqa
(d_{UW}=0.00017 ~,~  \bar d_{UW}=0.0001) & \Rightarrow & {\rm W ~ eff} ~~, \label{Weff} \\
(d_{UB}=0.00017 ~,~ \bar d_{UB}=0.0001) & \Rightarrow &  {\rm B ~ eff} ~~. \label{Beff}
\eqa

Note that the (HCns) rule \cite{heli1, heli2} does not apply to such anomalous
 non renormalizable contributions. Because of this, such BSM amplitudes
 are not suppressed at high energy.\\

\section{Observables and amplitude analysis}

Contrary to the $ZH$ case, where real amplitude contributions coming from the Born terms
dominate, in the $\gamma H$ case the four independent helicity amplitudes are complex
and one would need 8 independent observables in order to make a complete analysis.
This would require longitudinal and transverse initial $e^{\mp}$,
as well as final $\gamma$ polarizations; these last ones being probably
very difficult to measure.

When CP is conserved, only two independent complex helicity amplitudes occur,
and one would need only 4 observables (at a given energy and all angles) in order
to make the amplitude analysis.

In presenting these observables we use the same notation as in \cite{simZH}.
A lower index like $L$ or $R$  refers to the initial $e^-$ polarization  and corresponds to
$\lambda=-{1\over2}$ or $\lambda=+{1\over2}$ respectively. The final photon polarization
is  denoted by an  index $\gamma_\tau$ for $\tau=\pm 1$. Quantities like
$\sigma(\lambda, \tau)$ can then be also denoted as
$\sigma^{\gamma_\tau}_{L,R}$; see also immediately after (\ref{e-massless}).\\

The differential  unpolarized cross sections and the corresponding
integrated ones over all $\theta$ angles are respectively given by
\bq
{d\sigma\over d\cos\theta}={\beta_{\gamma}\over 128\pi s}
\sum_{\lambda \tau}|F_{\lambda \tau}(\theta)|^2 ~~, \label{dsigma-unpol}
\eq
and
\bq
\sigma=\int^{1}_{-1} d\cos\theta {d\sigma\over d\cos\theta}~~, \label{sigma-unpol}
\eq
where $\beta_{\gamma}$ is defined in (\ref{kinematics}) and the summations apply  over
$\lambda=\pm{1\over2}$ and $\tau=\pm1$.

Correspondingly, unpolarized (or polarized with the adequate index)
cross sections  integrated over the forward (with respect to the $e^-$-beam)
or the backward region, are respectively denoted as $\sigma_F$ and $\sigma_B$.\\

\noindent
\underline{Polarization asymmetries} \\
These contain initial $e^-_L$-$e^-_R$ asymmetries,
with the $\gamma$ helicity  being either not observed,
or   chosen to have specific value  $\tau=\pm 1$
\bqa
&& A_{LR}={\sigma_L-\sigma_R\over \sigma_L+\sigma_R} ~~, \label{ALR-unpolZ} \\
&& A_{LR}(\tau)={\sigma_L(\tau)-\sigma_R(\tau)\over \sigma_L(\tau)+\sigma_R(\tau)} ~~;
~~ \label{ALR-tau}
\eqa
or asymmetries defined with respect the final $\gamma$ polarization,
with the $e^{\mp}$ beams selected to be either unpolarized,  or the electrons being
either  purely $e^-_L$ or purely $e^-_R$
\bqa
&& A^{{\rm pol}~\gamma}={\sigma (\tau=-)-\sigma(\tau=+)\over \sigma (\tau=-)
+\sigma^(\tau =+)}\equiv {\sigma^{\gamma_-}-\sigma^{\gamma_+}\over \sigma^{\gamma_-}
+\sigma^{\gamma_+}}  ~~, \label{A-polgamma} \\
&& A^{{\rm pol}~\gamma}(\lambda)={\sigma(\lambda,\tau=-) -\sigma(\lambda, \tau=+)
\over \sigma(\lambda,\tau=-) +\sigma(\lambda, \tau=+)}\equiv
{\sigma^{\gamma_-}_{\lambda}-\sigma^{\gamma_+}_{\lambda}
\over \sigma^{\gamma_-}_{\lambda}+\sigma^{\gamma_+}_{\lambda}}   ~~.  \label{Alambda-polgamma}
\eqa

\noindent
\underline{Forward-backward asymmetries}\\
In the unpolarized beam case, when the final photon polarization is  not looked at,
these are defined as
\bq
A_{FB}={\sigma_F-\sigma_B\over \sigma_F+\sigma_B} ~~; \label{AFB-polZ}
\eq
while  for any definite $e^-$ and photon helicity they are defined as
\bq
A_{FB}(\lambda, \tau)={\sigma_F(\lambda, \tau)-\sigma_B(\lambda, \tau)
\over \sigma_F(\lambda, \tau)+\sigma_B(\lambda, \tau)} ~~. \label{AFB-lambdatau}
\eq

Combining (\ref{Alambda-polgamma}, \ref{AFB-lambdatau}), one
obtains a peculiar forward-backward asymmetry of the
above $\gamma$ transverse polarization asymmetry,
\bq
A^{{\rm pol}~\gamma}_{FB}=
{(\sigma^{\gamma_-}-\sigma^{\gamma_+})_F-(\sigma^{\gamma_-}-\sigma^{\gamma_+})_B
\over (\sigma^{\gamma_-}+\sigma^{\gamma_+})_F+(\sigma^{\gamma_-}+\sigma^{\gamma_+})_B}
 ~~, \label{AFB-polZ-lambdatau}
\eq
which may be defined  for unpolarized $e^{\mp}$ beams,
as well as  separately for $L$ or $R$ electron beams.
It turns out  to be non vanishing in all these three cases.\\

\noindent
\underline{CP conservation}\\
When CP is conserved, one gets from (\ref{CP-conservation})
\bq
A_{FB}(\lambda, -\tau)=-A_{FB}(\lambda, \tau) ~~, \label{CP-conservation-AFBlambdatau}
\eq
which  remains true also for unpolarized $e^\mp$-beams, where one sums over $\lambda=L,R$
obtaining
\bqa
&&  A_{FB}(-\tau)=-A_{FB}(\tau) ~~ \Rightarrow ~~
  A_{FB}(\gamma_+)=-A_{FB}(\gamma_-) ~~. \label{CP-conservation-AFBtau}
\eqa
 If one sums over all final $\gamma$ polarization in (\ref{CP-conservation-AFBtau}), then one obtains
$A_{FB}=0 $.\\

Another consequence of CP conservation concerns the Left-Right asymmetries
in the forward and backward directions
\bq
A^F_{LR}(\tau)=A^B_{LR}(-\tau)\Rightarrow  A^F_{LR}(\gamma_-)=A^B_{LR}(\gamma_+)
~~; \label{CP-conservation-ALRtau}
\eq
compare  (\ref{ALR-tau}).

We also note that  CP conservation implies
\bq
A_F^{{\rm pol} ~\gamma}(\lambda)=-A_B^{{\rm pol}~\gamma}(\lambda) ~~,\label{CP-conservation-Apol}
\eq
for both $\lambda=L,R$ cases, such that the totally integrated
$A^{{\rm pol}~\gamma}(\lambda)$ vanishes; compare (\ref{AFB-polZ-lambdatau}).\\

In the next section we illustrate the above properties for
CP conserving one loop corrections to SM or MSSM models; as well as for the effective, possibly
CP violating  Higgs couplings  in the BSM case (\ref{d-BSM}).
It turns out that  some of the above asymmetries are particularly sensitive
to the dynamical details,  and may be very useful for disentangling
SM from MSSM or BSM corrections.\\

\noindent
\underline{Transverse $e^{\pm}$ polarization}\\
We next turn to  the possibility of $e^-e^+$ collisions with
transversally polarized beams. It has been known since a long time, that
this can reveal in a clear way the presence of various types of BSM effects;
see e.g. \cite{Chiappetta, Renard, Rindani}. In our case, this  is
particularly motivated by the
presence of a relatively important imaginary part in the
$e^-e^+\to  \gamma H$ amplitudes, which may  produce an important  $\sin2\phi$
azimuthal dependence in the transition probability. Restricting to (\ref{e-massless}),
we then obtain
\bq
R=R_0+2P_TP'_T[\cos2\phi R_{\cos2\phi}+\sin2\phi R_{\sin2\phi}] ~~, \label{R-form}
\eq
with the unpolarized part being
\bq
R_0=|F_{--}|^2 +|F_{-+}|^2 +|F_{+-}|^2+|F_{++}|^2  ~~, \label{R0-form}
\eq
$(P_T, P'_T$) being the $e^{\mp}$ degrees of transverse polarization, and
the two azimuthal dependent terms being
\bqa
 R_{\sin2\phi}& = & { \rm Im}F_{--}{\rm Re}F_{+-}
 -{\rm Re}F_{--}{\rm Im}F_{+-}+{\rm Im}F_{-+}{\rm Re}F_{++}-{\rm Re}F_{-+}{\rm Im}F_{++}
~, \label{Rsin-form} \\
 R_{\cos2\phi} &= &{\rm Re}F_{--}{\rm Re}F_{+-}
 +{\rm Im}F_{--}{\rm Im}F_{+-}+{\rm Re}F_{-+}{\rm Re}F_{++}+{\rm Im}F_{-+}{\rm Im}F_{++}
~. \label{Rcos-form}
\eqa
Note that only the $\sin2\phi$-term is proportional to the imaginary parts of the amplitudes,
while the $\cos2\phi$ term is non vanishing even when all amplitudes are purely real.

These $R_0$, $R_{\sin2\phi}$, $R_{\cos2\phi}$ terms can be extracted from the
observed  $R$-form in (\ref{R-form})  by integrating the complete azimuthal distribution
over the  angular  domains
\bqa
 && \Big [ 0, 2\pi  \Big ] ~,~ \nonumber \\
&& {1\over4}\Big [\Big (0,{\pi\over2} \Big )
-\Big ({\pi\over2},\pi \Big )+\Big (\pi,{3\pi\over2} \Big )-\Big ({3\pi\over2},\pi \Big ) \Big ]
~,~ \nonumber \\
&&  {1\over2} \Big [\Big (0,{\pi\over4} \Big )-\Big (0,{3\pi\over4}\Big )
 +\Big (\pi,{5\pi\over4}\Big )-\Big (\pi,{7\pi\over4}\Big ) \Big ] ~. \label{R-form-construction}
\eqa\\

Note that in case of CP conservation, the validity of (\ref{CP-conservation}) makes the three forms
 $R_0$, $R_{\sin2\phi}$, $R_{\cos2\phi}$,  Forward-Backward symmetric.

In the  illustrations  of these transverse terms, we present
the ratios of the polarized terms to the unpolarized one:
\bq
T_{\sin2\phi}={2R_{\sin2\phi}\over R_0}~~, ~~ T_{\cos2\phi}={2R_{\cos2\phi}\over R_0} ~~.
\label{T-terms}
\eq \\

\section{Numerical analysis}

For the  MSSM illustrations, we use the  S1 benchmark  of \cite{bench}, where
the EW  scale values of the various parameters (with masses in TeV) are
\bqa
&&  \mu =0.4~,~  m_{A^0}=0.5 ~,~M_1=0.25 ~,~ M_2=0.5 ~,~ M_3=2 ~, \nonumber \\
&& \tan\beta=20 ~,~    m_{\tilde q}=2 ~,~ m_{\tilde l}= A_\tau= 0.5 ~, ~ A_t=A_b=2.3 ~.~
\label{bench-param}
\eqa
Such a benchmark is   consistent with all present LHC constraints \cite{bench}.

\subsection{Comparison of basic amplitudes}

In Fig.1-4, we give the energy dependencies of the 4 $e^-e^+\to  \gamma H$ amplitudes
at a fixed angle $\theta=60^\circ$, and the angular dependencies at fixed energies of 1 and 5 TeV,
successively for $(H_{SM}$ and $h^0$, in SM and the S1 MSSM benchmark mentioned above
\cite{bench}. In order to not increase the number of figures we do not show the $H^0$ amplitudes.
They are an order of magnitude smaller than the $h^0$ ones because of the benchmark choice \cite{bench},
 where the $\alpha, \beta$ parameters lead to small $H^0$ couplings. Consequently
the $H^0$ cross section is probably not observable.

Because of the HCns theorem \cite{heli1, heli2}, all these HV amplitudes are  suppressed at high energy,
albeit somewhat slowly, because of logarithmic enhancements partially canceling   the naively expected
$M/\sqrt{s}$ suppression.

The left-handed $F_{-\mp}$ amplitudes are (at least 10 times) larger than
the right-handed $F_{+\mp}$; this is due to the
left-handed charged W contribution in triangles and boxes.

The imaginary parts are often non negligible, and in fact comparable to that of the real parts.
They arise from the possibility of on shell intermediate states in various diagrams.

The real parts of the $h^0$ amplitudes are close to the $H_{SM}$ ones, but
the imaginary parts differ because of contributions from virtual
spartner exchanges leading to typical threshold effects.

We observe that globally the sim approximation is quickly good for $H_{SM}$
and $h^0$. In the $H^0$ case it would require higher energies.

Note that, the one loop SM and MSSM amplitudes satisfy (\ref{CP-conservation}),
since SM and the MSSM benchmark  we are considering, respects CP conservation \cite{bench}.

In the SM Figs.1,2  we also include  the possibility of an  effective BSM
involving anomalous couplings between  $H_{SM}$ and the gauge bosons. Since in this example,
 determined by   (\ref{Weff}, \ref{Beff}), CP is also violated, the BSM contributions
  violate the restriction (\ref{CP-conservation}).\\

\subsection{Unpolarized differential cross sections}

In the left panels of Fig.5 we give the unpolarized differential cross sections
for $H_{SM}$  in SM. Correspondingly in the right panels of Fig.5 we present
the  $h^0$ production in S1 MSSM \cite{bench}.
In the various panels we show the energy and angular dependencies, as in Figs.1-4.

Note that  the angular peaks in the forward and backward directions
(coming from $t,u$ channel triangles and boxes)
and the CP invariance of the one loop contributions lead
to forward-backward symmetries, possibly violated by anomalous couplings.

\subsection{$A_{LR}$ Asymmetries}

We first discuss $A_{LR}$ defined in  (\ref{ALR-unpolZ})  for the case where
the polarization of the final photon is not measured.
In Fig.6 we give $A_{LR}$ for $H_{SM}$ (left panels) and the MSSM $h^0$ results (right panels).
The one loop contributions $A_{LR}$ get large values. The weak energy and angle
dependencies come from the dominance of the L amplitudes as seen above.
The $H_{SM}$ and $h^0$ cases are rather similar.\\

The BSM contributions from W eff and B eff  (\ref{Weff}, \ref{Beff}),
 are also large and rather flat.\\

 We next turn $A_{LR}(\tau)$ defined in (\ref{ALR-tau}) for the cases
 the polarization of the final photon is chosen to have specific values
 $\tau=\pm1$.
Fig.7 show $A_{LR}$ for $\tau=+1$, and Fig.8 for $\tau=-1$.

The angular dependencies are particularly interesting.
For the one loop amplitudes which respect CP-invariance, they reflect
the helicity properties of (\ref{CP-conservation-ALRtau}).
Consequently,  the two $\tau$-helicity one-loop $A_{LR}(\tau)$
 satisfy an F/B interchange rule.
But this is strongly violated by the chosen BSM anomalous couplings,
which do not respect CP invariance.\\

\subsection{$A^{{\rm pol}~\gamma}$ Asymmetries}

For unpolarized $e^{\pm}$ beams, these are defined in (\ref{A-polgamma}) and shown in
 Figs.9 for $H_{SM}$ (left panels) and $h^0$ (right panels).

As seen there, the energy dependence is not negligible and the angular dependencies
reflect also the photon helicity properties of the one loop terms.
CP invariance imposes the F/B antisymmetry,
again possibly perturbed by anomalous coupling contributions.\\

The corresponding results for $e^-_L$ and  $e^-_R$-beams, defined in (\ref{Alambda-polgamma}),
are shown in Fig.10 for $H_{SM}$ (left panels) and $h^0$ (right panels).

These 2 cases of electron  polarization lead to very different results;
they can even differ by a sign. As the L amplitudes are
larger than the R ones, this explains why this case leads to results
closer to the unpolarized ones.

The CP forward-backward symmetry relations of (\ref{CP-conservation-Apol})
are well illustrated and again easily violated
by anomalous couplings.

\subsection{Azimuthal dependence}

As described in Sect.5, when transversely polarized beams are available,
the azimuthal dependence of the differential cross section is controlled
by the  coefficients $T_{\cos2\phi}$, $T_{\sin2\phi}$ in (\ref{T-terms}), of the
$\cos 2\phi$, $\sin 2\phi$ terms in (\ref{R-form}).    Fig.11 then shows the energy
and angular dependencies for
$H_{SM}$ (left panels) and $h^0$ (right panels).

These coefficients get non negligible values which means that
there is a significant azimuthal dependence. As already mentioned
$T_{\sin2\phi}$ is particularly interesting because it is governed by the
imaginary parts of the amplitudes which are important in the process
under consideration.
This should constitute an additional source of tests
of the underlying dynamics and of the Higgs couplings.
With CP invariance the angular dependence of these coefficients
is forward-backward symmetric.

In the left panels of Fig.11, we also show the contributions
of the anomalous couplings of W and B types (\ref{Weff}, \ref{Beff}).\\

\section{Conclusions}

We have analyzed the specificity of the process $e^-e^+\to  \gamma H$
as compared to $e^-e^+\to  ZH$.
The new feature is that this process has no Born term and is therefore immediately
sensitive to one loop effects and the underlying Higgs dynamics. This arises
in particular through the contributions of the imaginary parts of the amplitudes.

Our aim in doing this, is to see the differences between SM, MSSM and another (possibly CP violating)
BSM, and to check  especially whether this is observable when using the supersimple approximation.

We have insisted on two important aspects of the supersimple description.
Its ability  to allow the immediate reading of the dynamical contents
of the various standard and non standard contributions.
And in addition, the fact that these supersimple expressions quite accurately  reproduce
the exact one loop effects at high energies.

For achieving this, we have computed the exact one loop amplitudes
and cross sections, as well as
their sim approximations, and compared them numerically.
The sim and exact one loop  amplitudes agree at high energy, but differ at low energies
due to neglected terms behaving like $m^2/s$, possibly modified by  logarithmic corrections,
and also due to threshold effects caused by  virtual contributions  particularly
visible in the SUSY cases.

At 1 TeV, the agreement between sim and the exact one loop results,
is already good in the $H_{SM}$ case, but there are still
some non negligible differences  in the $h^0$ case, which disappear at higher energies.
The $H^0$ case would need even higher energies for achieving such an accuracy; but
$H^0$ production is probably unobservable with the considered
benchmark parameters \cite{bench}.

In addition to the unpolarized differential cross section (angular distribution
and forward-backward asymmetry) we have also considered several other observables
with initial and final polarization asymmetries: $A_{LR}$, $A_{LR}(\tau=\mp1)$,
 $A^{{\rm pol}~\gamma}$, $A^{{\rm pol}~\gamma}_{L}$ , $A^{{\rm pol}~\gamma}_{R}$;
as well as the coefficients of the $\cos2\phi$ and $\sin2\phi$ azimuthal
dependencies (\ref{T-terms}), when the $e^{\pm}$ beams are transversally polarized.

We have illustrated how sensitive all these observables are to the underlying dynamics by comparing
 the SM predictions to the  MSSM   ones, and a simple  type of BSM  physics involving  anomalous
Higgs couplings to gauge bosons, possibly containing also CP violation.

We hope that this work will motivate further studies studies of the
$e^-e^+\to  \gamma H$ process, using in particular  polarized beams
and containing also measurements  of the final $\gamma$ polarization.

\vspace*{2cm}


\renewcommand{\thesection}{A}
\renewcommand{\theequation}{A.\arabic{equation}}
\setcounter{equation}{0}

{\Large   \bf Appendix:  Sim expressions for the $e^-e^+\to \gamma H$ amplitudes in SM and  MSSM}\\

From the decomposition of the invariant amplitude over the CP-conserving invariant forms
defined in (\ref{I1I3I4-forms}) and leading to (\ref{CP-conservation}),
we write  the four HV amplitudes in the form
\bqa
F_{--}& = & {\alpha^2m_W\over\beta_{\gamma}}
\left [{u\sqrt{2}\over\sqrt{s}}N^L_1+{ut\over\sqrt{2s}}(N^L_3-N^L_4) \right ] ~~,  \nonumber \\
F_{+-} &=&  {\alpha^2m_W\over\beta_{\gamma}}
\left [{t\sqrt{2}\over\sqrt{s}}N^R_1-{ut\over\sqrt{2s}}(N^R_3-N^R_4) \right ] ~~,  \nonumber \\
F_{-+} & = & {\alpha^2m_W\over\beta_{\gamma}}
\left [{t\sqrt{2}\over\sqrt{s}}N^L_1-{ut\over\sqrt{2s}}(N^L_3-N^L_4) \right ] ~~,  \nonumber \\
F_{++} &= &  {\alpha^2m_W\over\beta_{\gamma}}
\left [{u\sqrt{2}\over\sqrt{s}}N^R_1+{ut\over\sqrt{2s}}(N^R_3-N^R_4) \right ] ~~, \label{F-amp1}
\eqa
where  $H=H_{SM},h^0,H^0$  and the kinematics are defined in (\ref{kinematics}).

The $N^{L,R}_i$ contributions in (\ref{F-amp1}) are obtained from
the exact 1 loop computation in terms of PV functions \cite{PV}. These  are then expanded
using the high energy forms given in \cite{asPV}.
We thus obtain the so called supersimple (sim)  results written below in SM and MSSM.

We next turn to the forms entering the sim expressions.
These consist of the linear log augmented Sudakov form and the
forms involving ratios of Mandelstam variables, as in \cite{simZH, super, WW}.
These are
 \bq
\overline{\ln s_{ij}(a)}  \equiv  \ln{-s-i\epsilon \over m_im_j}+b_0^{ij}(m^2_a)-2 ~~,
\label{Sud-ln-form}
 \eq
 where  $b_0^{ij}(m^2_a)$ is  given in e.g.   Eqs.(A.6) of \cite{WW}, with  (i,j) denoting
 internal exchanges and  $a$    an  on-shell particle, such that the $aij$ tree
 level coupling is non vanishing. And the forms
 \bq
\overline{\ln^2r_{xy}}=\ln^2r_{xy}+ \pi^2 ~~~,~~~ \ln r_{xy} ~~~.~~ \label{lnr-form}
\eq
where
\bq
   r_{xy} \equiv \frac{-x-i\epsilon}{-y-i\epsilon} ~~, \label{rxy-form}
 \eq
 with $x,y$ standing for the Mandelstam-variables $s,t,u$.

For $e^-e^+\to \gamma H$ though, the additional  augmented quadratic Sudakov logarithms though,
are  different from those in \cite{simZH, super, WW}. The forms we meet here are
\bqa
&&\overline{\ln^2 x_{WW}}=\ln^2{-x- i \epsilon \over m^2_W}+2L_{\gamma WW}+2L_{HWW}~~, \nonumber\\
&& \overline{\ln^2x_{W}}=\ln^2{-x - i \epsilon \over m^2_W}+2L_{eW\nu}+L_{\gamma WW}+L_{HWW}~~,
\nonumber\\
&& \overline{\ln^2x_{Z}}=\ln^2{-x- i \epsilon \over m^2_Z}+2L_{eZe}+L_{HZZ}~~, \nonumber\\
&& \overline{\ln^2x_{t}} = \ln^2{-x- i \epsilon \over m^2_t}+2L_{\gamma tt}+2L_{Htt}  ~~, \nonumber\\
&& \overline{\ln^2x_{b}} =\ln^2{-x- i \epsilon \over m^2_b}+2L_{\gamma tt}+2L_{Hbb}  ~~,
\label{quadratic-Sud}
\eqa
and the forms involving charginos or neutralinos  described by the indices $(i,j,k)$ and given by
\bqa
&& \overline{\ln^2x_j}=\ln^2{-x-i\epsilon \over M^2_j}+2L_{Hji}+2L_{ej\tilde{l}_L}~~
{\rm ~ with} ~~~ \tilde{l}_L=\tilde{\nu}_L~,~\tilde{e}_L ~, \nonumber \\
&& \overline{\ln^2s_{ii}}=\ln^2{s\over M^2_i}+2L_{\gamma ii}+2L_{Hii}~~, ~~
\overline{\ln^2s_{ji}}=\ln^2{s\over M^2_j}+2L_{\gamma jj}+2L_{Hji}  ~, \label{quadratic-Sud-chi}
\eqa
where $L_{aij}$ in (\ref{quadratic-Sud}, \ref{quadratic-Sud-chi}) are given
in   Eqs.(22) of \cite{asPV}. Again  (i,j) denote
 internal exchanges and  $a$    an  on-shell external particle, such that the $aij$ tree
 level coupling is non vanishing. \\

To  describe the sim expressions, we also need the  constants of Table A.1,
the  $(t,b)$ and sfermion couplings respectively given by
\bqa
&& C^L_t={-3-2s^2_W\over3s^2_Wc^2_W} ~,~ C^L_b={-3+2s^2_W\over6s^2_Wc^2_W} ~,~
C^R_t={-10\over3c^2_W} ~,~ C^R_b={-1\over3c^2_W} ~~, \label{tb-couplings}  \\[0.1cm]
&& f_{Z\tilde{f}} =  -{(I^3_{\tilde{f}}-s^2_WQ_{\tilde{f}}) \over s_Wc_W}~,~
f_{H\tilde{f}}={m_W (I^3_{\tilde{f}}-s^2_WQ_{\tilde{f}})\over s_Wc^2_W}C^+_H
-{m^2_f\over s_Wm_W}f_{fH} ~, \label{sfermion}
\eqa
as well as the chargino and neutralino couplings
\bqa
&& A^L_i(\tilde{e}_L)=-~{e\over s_W}Z^-_{1i} ~~, ~~A^L_i(\tilde{\nu}_L)=-~{e\over s_W}Z^+_{1i}
\nonumber \\
&& A^{0L}_i(\tilde{e}_L)={e\over \sqrt{2}s_Wc_W}(Z^N_{1i}s_W+Z^N_{2i}c_W)
~~, ~~ A^{0R}_i(\tilde{e}_R)=-{e\sqrt{2}\over c_W}Z^{N*}_{1i} \nonumber \\
&&  O^L_{Zij}=-{1\over2s_Wc_W}(Z^{+*}_{1i}Z^{+}_{1j}+\delta_{ij}(c^2_W-s^2_W)) ~, \nonumber \\
&& O^R_{Zij}=-{1\over2s_Wc_W}(Z^{-}_{1i}Z^{-*}_{1j}+\delta_{ij}(c^2_W-s^2_W)) ~, \nonumber \\
&& c^L_{h^0ij}=-{1\over\sqrt{2}s_W}(-\sin\alpha Z^{-}_{2i}Z^{+}_{1j}
+\cos\alpha Z^{-}_{1i}Z^{+}_{2j})~~, ~~c^R_{h^0ij}=c^{L*}_{h^0ji}~, \nonumber \\
&& O^{0L}_{Zij}={1\over2s_Wc_W}(Z^{N*}_{4i}Z^{N}_{4j}-Z^{N*}_{3i}Z^{N}_{3j})
~~, ~~O^R_{Zij}={1\over2s_Wc_W}(Z^{N}_{3i}Z^{N*}_{3j}-Z^{N}_{4i}Z^{N*}_{4j})~, \nonumber \\
&& c^{0L}_{h^0ij}={1\over2s_Wc_W}[(-\sin\alpha Z^{N}_{3i}-\cos\alpha Z^{N}_{4j})
(Z^{N}_{1i}s_W-Z^{N}_{2i}c_W))+(i\to j)]~, ~ \nonumber \\
&& c^{0R}_{h^0ij}=c^{0L*}_{h^0ji} , \label{ino-couplings}
\eqa
\noindent
given in terms of the usual mixing matrices \cite{Rosiek}.

\begin{table}[tb]
\begin{center}
{ Table A.1: Parameters for the HV amplitudes in SM and MSSM  }\\
  \vspace*{0.3cm}
\begin{small}
\begin{tabular}{||c|c|c|c||}
\hline \hline
   & $H_{SM}$ & $h^0$ & $H^0$  \\
 \hline
 $C^-_H$ & 1 & $\sin(\beta-\alpha)$  &  $\cos(\beta-\alpha)$   \\
  $C^+_H$ & 0 & $\sin(\beta+\alpha)$ & $-\cos(\beta+\alpha)$      \\
   $f_{GGH}$ &  $-{m^2_H\over2s_Wm_W}$& ${m_W\over2s_Wc^2_W}\cos(2\beta)\sin(\beta+\alpha)$
   &  - ${m_W\over2s_Wc^2_W} \cos(2\beta)\cos(\beta+\alpha)]$  \\
    $f_{HHH}$ & 0 & $-{m_W\over s_W}\left [{\cos(2\beta)\sin(\beta+\alpha)\over2c^2_W}
+\sin(\beta-\alpha) \right ]$ & ${m_W\over s_W}
\left [{\cos(2\beta)\cos(\beta+\alpha)\over2c^2_W}-\cos(\beta-\alpha) \right ]$       \\
   $f_{tH}$ & 1 & ${\cos\alpha\over\sin\beta}$ &  ${\sin\alpha\over\sin\beta}$      \\
     $f_{bH}$ & 1 & $-{\sin\alpha\over\cos\beta}$ &  ${\cos\alpha\over\cos\beta}$      \\
 \hline  \hline
\end{tabular}
 \end{small}
\end{center}
\end{table}

The  $H^0$-case is subsequently obtained from the above $h^0$-one
through the replacement
\bq
(~ h^0 \Rightarrow H^0 ~) \mapsto ( \sin\alpha \Rightarrow -\cos\alpha  ~~~,~~~
\cos\alpha \Rightarrow  \sin\alpha ) ~~.  \label{H0-replacement}
\eq\\

Using  the above forms and couplings, we give below the  sim results for the $N^{L,R}_i$
contributions to the HV amplitudes in (\ref{F-amp1}). These include an SM part
which is easily identified, and the MSSM SUSY contributions to the $h^0$ case,
arising from sfermion and chargino-neutralino exchanges.

In the expressions below, the
SM and sfermion contributions are always included explicitly in the $N^{L,R}_i$-forms,
while the chargino-neutralino contributions are collected in  $T$ and $B$ forms
entering them.  We thus have
\bqa
 && sN^L_1=C^-_H \left \{{\overline{\ln^2s_{WW}}\over2c^2_Ws^3_W}
-{1+4c^2_W\over4s^3_Wc^2_W}[\overline{\ln s_{WW}(H)}+\overline{\ln s_{WW}(\gamma)}]
+{1\over2s^3_Wc^2_W} \right \}
\nonumber\\
&& -{(f_{GGH}+f_{HHH})\over2c^2_Ws^2_Wm_W} \Big [ 1-{1\over2}[ \overline{\ln s_{WW}(H)}
+\overline{\ln s_{WW}(\gamma)}] \Big ] \nonumber\\
&& + {m^2_t\over s_Wm^2_W}C^L_tf_{tH} \left [{\overline{\ln^2s_{t}}\over4}-
{1\over2} \Big (\overline{\ln s_{tt}(H)} +\overline{\ln s_{tt}(\gamma)} \Big )+1 \right ]\nonumber\\
&&+{m^2_b\over s_W m^2_W}C^L_bf_{bH} \left [{\overline{\ln^2s_{b}}\over4}-
{1\over2} \Big (\overline{\ln s_{bb}(H)}
+\overline{\ln s_{bb}(\gamma)} \Big ) +1 \right ] \nonumber\\
&&-\frac{s C^-_H }{s^3_W} \left \{ {\overline{\ln t_{WW}(H)}\over t}+{\overline{\ln u_{WW}(H)}\over u}
+ {(2s^2_W-1)^2\over 2 c^3_W}
\left [{\overline{\ln t_{ZZ}(H)}\over t}+{\overline{\ln u_{ZZ}(H)}\over u} \right ] \right \}\nonumber\\
&& -{2\over m_W}\Sigma_{\tilde{f}} Q_{\tilde{f}} f_{H\tilde{f}} \left [Q_{\tilde{f}}
+{(2s^2_W-1)\over 2s_Wc_W}f_{Z\tilde{f}} \right ]
\left [1-{1\over2} \left (\overline{\ln s_{\tilde{f}\tilde{f}}}(H)
+\overline{\ln s_{\tilde{f}\tilde{f}}}(\gamma) \right ) \right ]\nonumber\\
&& +{s C^+_{H} \over 2c^2_Ws_W^3} \left \{ {\overline{\ln t_{\tilde{\nu}\tilde{\nu}}}(H)\over t}
+{\overline{\ln u_{\tilde{\nu}\tilde{\nu}}}(H)\over u}
+{(2s^2_W-1)\over 2c^2_W}
\left [{\overline{\ln t_{\tilde{e_L}\tilde{e_L}}}(H)\over t}
+{\overline{\ln u_{\tilde{e_L}\tilde{e_L}}}(H)\over u} \right ] \right \}\nonumber\\
&&+\frac{C^-_H}{2s^3_W} \Bigg \{ {s\over t}( \overline{\ln^2t_{W}}-\overline{\ln^2r_{us}})
+{s\over u}( \overline{\ln^2u_{W}} -\overline{\ln^2r_{ts}} ) \nonumber \\
&&  +{(1-2s^2_W)^2\over 2 c^4_W} \Big [{s\over u}\overline{\ln^2u_{Z}}+{s\over t}\overline{\ln^2t_{Z}}
+2\overline{\ln^2r_{tu}} \Big ] \Bigg \}
\nonumber\\
&&-{(1-2s^2_W)C^+_{H}\over8s^3_Wc^4_W}
\left [{s\over u}\overline{\ln^2r_{ts}}+{s\over t}\overline{\ln^2r_{us}} \right ]
+T^{L1}_{\chi\chi}+T^{L1}_{\chi\chi\chi}+B^{L1}_{\chi\chi}+B^{L1}_{\chi\chi\chi} ~, \label{NL1}
\eqa
with the chargino-neutralino contributions being
\bqa
&&T^{L1}_{\chi\chi}=\sum_{\chi^+_i\chi^+_j}A^{L*}_j(\tilde{\nu}_L)
A^{L}_i(\tilde{\nu}_L)\Big [{M_j\over m_W}c^R_{Hji}{s\over2t}
(\overline{\ln^2t_{j}}-2\overline{\ln t_{ji}(H)})-{M_i\over m_W}c^L_{Hji}{s\over t}
(\overline{\ln t_{ji}(H)})\nonumber\\
&&+{M_j\over m_W}c^L_{Hij}{s\over2u}
(\overline{\ln^2u_{j}}-2\overline{\ln u_{ji}(H)})-{M_i\over m_W}c^R_{Hij}{s\over u}
(\overline{\ln u_{ji}(H)}) \Big ]\nonumber\\
&&+\sum_{\chi^0_i\chi^0_j}A^{0L*}_j(\tilde{e}_L)A^{0L}_i(\tilde{e}_L)\Big [{M_j\over m_W}c^{0R}_{Hji}
{s\over2t} (\overline{\ln^2t_{j}}-2\overline{\ln t_{ji}(H)})-{M_i\over m_W}c^{0L}_{Hji}{s\over t}
(\overline{\ln t_{ji}(H)})\nonumber\\
&&+{M_j\over m_W}c^{0L}_{Hij}{s\over2u}
(\overline{\ln^2u_{j}}-2\overline{\ln u_{ij}(H)})-{M_i\over m_W}c^{0R}_{Hij}{s\over u}
(\overline{\ln u_{ij}(H)}) \Big ]~, \nonumber \\[0.2cm]
&&T^{L1}_{\chi\chi\chi}=-\sum_{\chi^+_i\chi^+_j}{4\over m_W} \Bigg \{2Q_ec^L_{Hii}M_i
\left [{1\over4}\overline{\ln^2s_{ii}}-\overline{\ln s_{ii}(H)}+1 \right ]\nonumber\\
&& +g^L_{Ze}(O^R_{Zij}+O^L_{Zij})c^L_{Hji}M_i
\left [{1\over4}\overline{\ln^2s_{ji}}-\overline{\ln s_{ij}(H)}+1 \right ] \Bigg \}
~, \nonumber \\[0.2cm]
&&B^{L1}_{\chi\chi}=\sum_{\chi^0_i\chi^0_j}A^{L0*}_j(\tilde{e}_L)A^{L0}_i(\tilde{e}_L)
\left [{M_i\over 2m_W}c^{0L}_{Hji}
+{M_j\over 2m_W}c^{0R}_{Hji} \right ]\overline{\ln^2r_{tu}} ~, \nonumber \\
&&B^{L1}_{\chi\chi\chi}=\sum_{\chi^+_i\chi^+_k}{M_k\over m_W}
c^{R}_{Hki}A^{L*}_k(\tilde{\nu}_L)A^{L}_i(\tilde{\nu}_L)
\left [\overline{\ln^2s_{ki}}+{s\over u}\overline{\ln^2r_{ts}}+{s\over t}\overline{\ln^2r_{us}} \right ]
~, \label{TL1BL1-forms}
\eqa

and
\bqa
&& N^L_3-N^L_4=-C^-_{H}\Bigg \{{2\over s^3_W}\Big [{(t-u)\over2tu}\overline{\ln^2s_{WW}}
+{\overline{\ln^2t_{W}}\over 4t}-{\overline{\ln^2u_{W}}\over 4u}\nonumber\\
&& +{s\overline{\ln r_{tu}}\over ut}+{(2u^2-2t^2+ut)\over4tu^2}\overline{\ln^2r_{ts}}
+{(2u^2-2t^2-ut)\over4ut^2}\overline{\ln^2r_{us}}\Big ]\nonumber\\
&& +{1\over 2s^3_W}\Big [{\overline{\ln^2u_{W}}\over u}-{\overline{\ln^2t_{W}}\over t}
+{\overline{\ln^2r_{us}}\over t}-{\overline{\ln^2r_{ts}}\over u}\Big ]\nonumber\\
&&+ {(2s_W^2-1)^2\over 2c^4_Ws^3_W}
\Big [{s\over ut}(\overline{\ln^2t_{Z}}-\overline{\ln^2u_{Z}})+{2s\over ut}(\overline{\ln u_{ZZ}(H)}
-\overline{\ln t_{ZZ}(H)})\nonumber\\
&&+{\overline{\ln^2r_{tu}}\over u}-{\overline{\ln^2r_{tu}}\over t}\Big ] \Bigg \}
+{{(2s^2_W-1)\over 4s^3_Wc^4_W}}C^+_{H}
\Big [{s\overline{\ln^2r_{ts}}\over u^2}
-{s\overline{\ln^2r_{us}}\over t^2}
+2s{\overline{\ln r_{ts}}-\overline{\ln r_{us}}\over ut} \Big ]\nonumber\\
&& +{1\over s^3_Wc^2_W}C^+_{H}{s\overline{\ln r_{tu}}\over tu}
+B^{L34}_{\chi\chi}+B^{L34}_{\chi\chi\chi} ~~, \label{NL3NL4}
\eqa
with
\bqa
&&B^{L34}_{\chi\chi}=\sum_{\chi^0_i\chi^0_j}A^{L0*}_j(\tilde{e}_L)A^{L0}_i(\tilde{e}_L)
\Bigg \{{M_i\over m_W}c^{0L}_{Hji}
\Big [-{s\overline{\ln^2u_i}\over ut}-2s{\overline{\ln r_{tu}}\over ut}-{\overline{\ln^2r_{tu}}\over t}
\Big ] \nonumber\\
&&+{M_j\over m_W}c^{0R}_{Hji} \Big [{s\overline{\ln^2t_j}\over ut}-2s{\overline{\ln r_{tu}}\over ut}
+{\overline{\ln^2r_{tu}}\over u}\Big ] \Bigg \} ~, \nonumber \\
&&B^{L34}_{\chi\chi\chi}=\sum_{\chi^+_i\chi^+_k}{M_k\over m_W}c^{R}_{Hki}
A^{L*}_k(\tilde{\nu}_L)A^{L}_i(\tilde{\nu}_L)
\Bigg [{\overline{\ln^2s_{ik}}\over t}-{\overline{\ln^2s_{ki}}\over u}
+{4s\overline{\ln r_{ut}}\over ut} \nonumber\\
&&  +{(2t^2-u^2+tu)\over tu^2}\overline{\ln^2r_{ts}}
-~{(2u^2-t^2+tu)\over ut^2}\overline{\ln^2r_{us}} \Bigg ]~, \label{BL34-forms}
\eqa

and
\bqa
&& sN^R_1=C^-_H \Big [{\overline{\ln^2s_{WW}}
+1-{1\over2}(\overline{\ln s_{WW}(H)}+\overline{\ln s_{WW}(\gamma)})\over s_Wc^2_W} \Big ] \nonumber\\
&& -{(f_{GGH}+f_{HHH})\over c^2_Wm_W} \Big [1-{1\over2}(\overline{\ln s_{WW}(H)}
+\overline{\ln s_{WW}(\gamma)})\Big ] \nonumber\\
&&+ {m^2_t\over s_Wm^2_W}C^R_tf_{tH} \Big [{\overline{\ln^2s_{t}}\over4}-
{1\over2}(\overline{\ln s_{tt}(H)}
+\overline{\ln s_{tt}(\gamma)})+1 \Big ]\nonumber\\
&& +{m^2_b\over s_Wm^2_W}C^R_bf_{bH} \Big [{\overline{\ln^2s_{b}}\over4}-
{1\over2}(\overline{\ln s_{bb}(H)} +\overline{\ln s_{bb}(\gamma)})+1 \Big ]
\nonumber\\
&&+C^-_H \Big \{ {-2s_W\over c^4_W}
\Big [{s\overline{\ln t_{ZZ}(H)}\over t}+{s\overline{\ln u_{ZZ}(H)}\over u} \Big ]
+{s_W\over c^4_W} \Big [{s\overline{\ln^2t_{W}}\over t}+{s\overline{\ln^2u_{W}}\over u}
+2\overline{\ln^2r_{tu}} \Big ] \Big \}\nonumber\\
&& -{2s_W\over c^4_W}C^+_{H}
\Big [{s\overline{\ln t_{\tilde{e_R}\tilde{e_R}}}(H)\over t}
+{s\overline{\ln u_{\tilde{e_R}\tilde{e_R}}}(H)\over u} \Big ]\nonumber\\
&&-{2\over m_W}\Sigma_{\tilde{f}} Q_{\tilde{f}}f_{H\tilde{f}} \Big [Q_{\tilde{f}}
+{s_W\over c_W}f_{Z\tilde{f}} \Big] \Big [1-{1\over2}(\overline{\ln s_{\tilde{f}\tilde{f}}}(H)
+\overline{\ln s_{\tilde{f}\tilde{f}}}(\gamma)) \Big ]\nonumber\\
&& -{s_W\over c^4_W}C^+_{H}\Big [{s\over u}\overline{\ln^2r_{ts}}+{s\over t}\overline{\ln^2r_{us}} \Big ]
+T^{R1}_{\chi\chi}+T^{R1}_{\chi\chi\chi}+B^{R1}_{\chi\chi} ~~, \label{NR1}
\eqa
with
\bqa
&&T^{R1}_{\chi\chi}=\sum_{\chi^0_i\chi^0_j}A^{0R*}_j(\tilde{e}_R)A^{0R}_i(\tilde{e}_R)
\Bigg [{M_j\over m_W}c^{0L}_{Hji}{s\over2t}
(\overline{\ln^2t_{j}}-2\overline{lnt_{ji}(H)})-{M_i\over m_W}c^{0R}_{Hji}{s\over t}
(\overline{\ln t_{ji}(H)})\nonumber\\
&&+{M_j\over m_W}c^{0R}_{Hij}{s\over2u}
(\overline{\ln^2u_{j}}-2\overline{\ln u_{ij}(H)})-{M_i\over m_W}g^L_{Hij}{s\over u}
(\overline{\ln u_{ij}(H)}) \Bigg ]  ~, \nonumber \\
&&T^{R1}_{\chi\chi\chi}=-\sum_{\chi^+_i\chi^+_j}{4\over m_W}\Bigg \{2Q_ec^L_{Hii}M_i
\Big [{1\over4}\overline{\ln^2s_{ii}}-\overline{\ln s_{ii}(H)}+1 \Big ]\nonumber\\
&& +g^R_{Ze}(O^R_{Zij}+O^L_{Zij})c^L_{Hji}M_i
\Big [{1\over4}\overline{\ln^2s_{ji}}-\overline{\ln s_{ij}(H)}+1 \Big ] \Bigg \} ~, \nonumber \\
&&B^{R1}_{\chi\chi}=\sum_{\chi^0_i\chi^0_j}A^{0R*}_j(\tilde{e}_R)A^{0R}_i(\tilde{e}_R)
\Big [{M_i\over 2m_W}c^{0R}_{Hji}+{M_j\over 2m_W}c^{0L}_{Hji} \Big ]\overline{\ln^2r_{tu}}
~, \label{TR1BR1-forms}
\eqa

and
\bqa
&& N^R_3-N^R_4={2s_W\over c^4_W}C^-_{H}\Bigg [{s(\overline{\ln^2t_{Z}}-\overline{\ln^2u_{Z}})\over ut}
+2s\Big [ {\overline{\ln u_{ZZ}(H)}-\overline{\ln t_{ZZ}(H)}\over ut}\Big ]
\nonumber\\
&& +{\overline{\ln^2r_{tu}}\over u}-{\overline{\ln^2r_{tu}}\over t} \Bigg ]
-{2s_W\over c^4_W}C^+_{H} \Big [{s\overline{\ln^2r_{ts}}\over u^2}
-{s\overline{\ln^2r_{us}}\over t^2}
+2s{\overline{\ln r_{ts}}-\overline{\ln r_{us}}\over ut} \Big ]
\nonumber\\
&& +B^{R34}_{\chi\chi} ~~, \label{NR3NR4}
\eqa
with
\bqa
&&B^{R34}_{\chi\chi}=\sum_{\chi^0_i\chi^0_j}A^{0R*}_j(\tilde{e}_R)A^{0R}_i(\tilde{e}_R)
\Bigg \{{M_i\over m_W}c^{0R}_{Hji}
\Big [-{s\overline{\ln^2u_i}\over ut}
-2s{\overline{\ln r_{tu}}\over ut}-{\overline{\ln^2r_{tu}}\over t}]\nonumber\\
&& +{M_j\over m_W}c^{0L}_{Hji}[{s\overline{\ln^2t_j}\over ut}
-2s{\overline{\ln r_{tu}}\over ut}+{\overline{\ln^2r_{tu}}\over u}\Big ] \Bigg \} ~~. \label{BR34-forms}
\eqa

\vspace*{1cm}
A  rough  approximation for the chargino and neutralino contributions in
(\ref{TL1BL1-forms}, \ref{BL34-forms}, \ref{TR1BR1-forms}, \ref{BR34-forms}) is respectively given by
 neglecting the various mass differences in the summations over virtual states,
 using common "average" masses.
From the unitarity of the mixing matrices one then obtains the results
\bqa
T^{L1}_{\chi\chi} &\simeq &\Big \{- {<M^+_{12}>\sin\alpha\over \sqrt{2}s^3_Wm_W}
+{(1-2s^2_W)\over4s^3_Wc^3_W} {<M^0_{L+}>\over m_W} \Big \}
\Big [{s\overline{\ln^2t}\over 2t} \nonumber \\
 && -2{s\overline{\ln t}\over t}
+{s\overline{\ln^2u}\over 2u}-2{s\overline{\ln u}\over u} \Big  ]  ~, \nonumber \\
T^{L1}_{\chi\chi\chi} & \simeq &-{4\over m_W}\Big [{1\over4}\overline{\ln^2s}-\overline{\ln s}+1 \Big ]
 \Big \{{\sqrt{2}\over s_W} +{(2s^2_W-1)(1-4c^2_W)\over4\sqrt{2}s^3_Wc^2_W} \Big \} \nonumber \\
 && \cdot \Big (<M^+_{21}>\cos\alpha-<M^+_{12}>\sin\alpha \Big )  ~, \nonumber \\
 B^{L1}_{\chi\chi} &\simeq &  {(1-2s^2_W)\over4s^3_Wc^3_W}{<M^0_{L+}>\over m_W}\overline{\ln^2r_{tu}}
 ~, \nonumber \\
 B^{L1}_{\chi\chi\chi} &\simeq &
{<M^+_{12}>\sin\alpha \over\sqrt{2}s^3_Wm_W}\left [\overline{\ln^2s}+{s\over u}\overline{\ln^2r_{ts}}
+{s\over t}\overline{\ln^2r_{us}} \right ] ~~, \nonumber \\
B^{L34}_{\chi\chi} & \simeq & {(1-2s^2_W)\over4s^3_Wc^3_W} {<M^0_{L+}>\over m_W}
\left [{s\overline{\ln^2t}\over ut}-{s\overline{\ln^2u}\over ut}
  -4s{\overline{\ln r_{tu}}\over ut}+{\overline{\ln^2r_{tu}}\over u}
-{\overline{\ln^2r_{tu}}\over t} \right ]  ~, \nonumber \\
B^{L34}_{\chi\chi\chi} &\simeq & -{<M^+_{12}>\sin\alpha \over\sqrt{2}s^3_Wm_W}
\Big [{\overline{\ln^2s}\over t}
-{\overline{\ln^2s}\over u}+{4s\overline{\ln r_{ut}}\over ut}
+{2t^2-u^2+tu\over tu^2}\overline{\ln^2r_{ts}}\nonumber\\
&&-~{2u^2-t^2+tu\over ut^2}\overline{\ln^2r_{us}} \Big ] ~, \nonumber \\
T^{R1}_{\chi\chi} &\simeq &
-~{1\over c^3_W}{<M^0_{R-}>\over m_W}\Big [{s\overline{\ln^2t}\over 2t}
-2{s\overline{\ln t}\over t}
+{s\overline{\ln^2u}\over 2u}-2{s\overline{\ln u}\over u}\Big ] ~, \nonumber \\
T^{R1}_{\chi\chi\chi} &\simeq &-{4\over m_W}\Big \{{\sqrt{2}\over s_W}
+{1-4c^2_W\over2\sqrt{2}s_Wc^2_W} \Big \}
\Big [ {1\over4}\overline{\ln^2s}-\overline{\ln s}+1 \Big ] (<M^+_{21}>\cos\alpha \nonumber \\
&& -<M^+_{12}>\sin\alpha)
~, \nonumber \\
B^{R1}_{\chi\chi} & \simeq & -~{<M^0_{R+}>\over c^3_Wm_W} \overline{\ln^2r_{tu}} ~, \nonumber \\
B^{R34}_{\chi\chi} & \simeq & -{<M^0_{R+}>\over c^3_Wm_W}
\left [{s\overline{\ln^2t}\over ut}-{s\overline{\ln^2u}\over ut}
-4s{\overline{\ln r_{tu}}\over ut}+{\overline{\ln^2r_{tu}}\over u}
-{\overline{\ln^2r_{tu}}\over t}\right ] ~~\label{TB-approximate}
\eqa
with
\bqa
&& <M^+_{12}>={m_W\sqrt{2}\over\sqrt{1+\tan^2\beta}}
~~, ~~<M^+_{21}>={m_W\sqrt{2}\tan\beta\over\sqrt{1+\tan^2\beta}}~~, \nonumber \\
&& <M^0_{L+}>=\sin\alpha(s_W<M^N_{13}>+c_W<M^N_{23}>)+\cos\alpha(s_W<M^N_{14}>+c_W<M^N_{24}>)~~, \nonumber \\
&& <M^0_{L-}>=-\sin\alpha(s_W<M^N_{13}>+c_W<M^N_{23}>)+\cos\alpha(s_W<M^N_{14}>+c_W<M^N_{24}>)~~, \nonumber \\
&& <M^0_{R-}>=\sin\alpha <M^N_{13}>-\cos\alpha <M^N_{14}> ~~, \nonumber \\
&& <M^0_{R+}>=\sin\alpha <M^N_{13}>+\cos\alpha <M^N_{14}>~~, \nonumber \\
&& <M^{'0+}_{L-}>=\sin\alpha(s_W<M^N_{13}>-c_W<M^N_{23}>)+\cos\alpha(s_W<M^N_{14}>-c_W<M^N_{24}>)~~, \nonumber \\
&& <M^N_{13}>={-m_Ws_W\over c_W\sqrt{1+\tan^2\beta}}
~~,~~<M^N_{23}>={m_W\over \sqrt{1+\tan^2\beta}}~~, \nonumber \\
&& <M^N_{14}>={m_Ws_W\tan\beta\over c_W\sqrt{1+\tan^2\beta}}
~~,~~<M^N_{24}>={-m_W\tan\beta\over \sqrt{1+\tan^2\beta}}~~, \label{chargino-neutralino-param}
\eqa
where   $M^{+\top}$ denotes the  $\tilde \chi^+$ mass matrix, and  $M^N$
the neutralino one.

The quantities $\overline{\ln^2x}$ and $\overline{\ln x}$
in (\ref{TB-approximate}) are  approximated by their pure logarithmic contents
$\ln^2(x/ M^2) $ and $\ln(x/ M^2)$, where $M$ are adequate average sparticle masses.
Correspondingly, the average $<M>$-quantities  in (\ref{chargino-neutralino-param})  denote common
(chargino or neutralino) mass matrix elements.
The values of these average masses could be determined by fitting each B and T
contribution  in (\ref{TL1BL1-forms}, \ref{BL34-forms}, \ref{TR1BR1-forms}, \ref{BR34-forms}),
  to its corresponding (\ref{TB-approximate}, \ref{chargino-neutralino-param}) 
  simplified expression. 
  
  However the numerical results are not  very accurate at low energies, because they do not
reproduce the various threshold structures. In addition the values of the average
masses should be adapted to each specific benchmark choice. 
In fact the use of the  simplified expressions (\ref{TB-approximate}, \ref{chargino-neutralino-param}) 
is only to show in a direct way the nature
of the $B$ and $T$ contributions. In the numerical calculations it is best to use the 
expressions in (\ref{TL1BL1-forms}, \ref{BL34-forms}, \ref{TR1BR1-forms}, \ref{BR34-forms}).\\

\noindent
\underline{Global approximation for the helicity amplitudes}:\\
Finally we give a global fit of the four helicity  amplitudes, which
reproduces at intermediate energies (in the domain 0.6-5. TeV),
their angular and energy dependencies. The fit is rather accurate, being valid at a few percent level. It consists in fitting the one loop contributions, to the forms
\bqa
F_{--} &=& {\alpha^2m_W\over\beta_g}
\left [{u\sqrt{2}\over s\sqrt{s}}(C^L_1+C^{\prime L}_1 \cot^2\theta)
+{(u-t)\over s\sqrt{2s}}C^L_{34} \right ]
~~ \nonumber \\
F_{+-} &= &{\alpha^2m_W\over\beta_g}
\left [{t\sqrt{2}\over s\sqrt{s}}(C^R_1+C^{\prime R}_1 \cot^2\theta)
-{(u-t)\over s\sqrt{2s}}C^R_{34} \right ] ~~ \nonumber \\
F_{-+} &=& {\alpha^2m_W\over\beta_g}
\left [{t\sqrt{2}\over s\sqrt{s}}(C^L_1+C^{\prime L}_1 \cot^2\theta)
-{(u-t)\over s\sqrt{2s}}C^L_{34} \right ] ~~ \nonumber \\
F_{++} &= & {\alpha^2m_W\over\beta_g}
\left [{u\sqrt{2}\over s\sqrt{s}}(C^R_1+C^{\prime R}_1 \cot^2\theta)
+{(u-t)\over s\sqrt{2s}}C^R_{34} \right ] ~~, \label{F-amp2}
\eqa
with
\bqa
C^{L,R}_{1,34} &= & a^{L,R}_{1,34} \ln^2{s\over m^2_W}
+b^{L,R}_{1,34} \ln{s\over m^2_W} +c^{L,R}_{1,34}
\nonumber \\
C^{\prime L,R}_{1,34} &= & a^{\prime L,R}_{1,34} \ln^2{s\over m^2_W}
+b^{\prime L,R}_{1,34} \ln{s\over m^2_W} +c^{\prime L,R}_{1,34} ~~. \label{CLR-1-34}
\eqa
Note the factors $(u-t) $,  in front of the $C_{34}$ coefficients in
(\ref{F-amp2}), which reproduce the fact that these terms, arising only from boxes,
vanish at $\theta=\pi/2$, due to crossing relations.

Note also that the terms
$C^{\prime L,R}_1 \cot^2\theta$ reproduce the angular dependencies coming from t,u channel
triangles and boxes.

The effective constants in (\ref{F-amp2}, \ref{CLR-1-34}) are given in Table A.2.
Note the similarity of the real parts in the $H_{SM}$ and the $h^0$ cases,
and the large differences in the imaginary parts, due to the averaging of the threshold
effects in the virtual contributions of the spartners.

\begin{table}[tb]
\begin{center}
{ Table A.2: Constants  fitting the  amplitudes   (\ref{F-amp2}, \ref{CLR-1-34}),
   for SM and S1 MSSM,\\  in the intermediate anergy domain 0.6-5 TeV \cite{bench} }\\
  \vspace*{0.3cm}
\begin{small}
\begin{tabular}{||c|c|c||}
\hline \hline
\multicolumn{3}{||c||}{ $H_{SM}$ } \\ \hline
 $a^{L}_{1}=-31.4+i5.7 $ & $b^{L}_{1}=150.3-i21.6$  & $c^{L}_{1}=-319.1-i22.4 $     \\
 $a^{L}_{34}= 13.0-i1.8 $ & $b^{L}_{34}= -54.5-i29.9 $ & $c^{L}_{34}= 47.4+i82.6 $    \\
 $a^{R}_{1}= -10.9+i3.1 $  & $b^{R}_{1}= 79.6+i3.5 $ & $c^{R}_{1}= -143.4-i47.4 $     \\
 $a^{R}_{34}= 0.57 $     & $b^{R}_{34}=  -1.26 $  & $c^{R}_{34}= -0.90 $    \\
 $a^{\prime L}_{1}= 17.7-i1.2 $ & $b^{\prime L}_{1}= -88.8+i19.2 $ &  $c^{\prime L}_{1}=149.6-i57.9$ \\
 $a^{\prime R}_{1}=2.7$  & $b^{\prime R}_{1}=-15.3$  & $c^{\prime R}_{1}=24.2$    \\
 \hline
 \multicolumn{3}{||c||}{ $h^0$ S1 MSSM } \\ \hline
 $a^{L}_{1}=-46.1+i55.2$ & $b^{L}_{1}=336.9-i596.4$  & $c^{L}_{1}= -832.0+i1482.3$   \\
 $a^{L}_{34}=14.2-i1.6$ & $b^{L}_{34}=-65.9-i32.2 $ & $c^{L}_{34}= 74.0+i88.4 $    \\
$a^{R}_{1}= -12.9+i7.82 $   & $b^{R}_{1}= 114.6-i43.6 $ &  $c^{R}_{1}= -249.2+i61.8 $   \\
 $a^{R}_{34}= 0.40+i0.42 $   & $b^{R}_{34}=0.42-i3.80 $  & $c^{R}_{34}= -4.86+i8.52 $    \\
 $a^{\prime L}_{1}= 17.69-i1.60 $  & $b^{\prime L}_{1}= -88.1+i23.1$  & $c^{\prime L}_{1}= 146.6-i67.8 $ \\
 $a^{\prime R}_{1}= 2.22+i0.16 $  & $b^{\prime R}_{1}= -11.2-i1.5 $ & $c^{\prime R}_{1}= 15.3+i3.3 $    \\
   \hline
 \multicolumn{3}{||c||}{ $H^0$ S1 MSSM } \\ \hline
 $a^{L}_{1}=8.5+i32.2 $ & $b^{L}_{1}= -61.4-i421.7 $  & $c^{L}_{1}= 102.6+i1176.1 $    \\
 $a^{L}_{34}= -5.0+i2.7 $ & $b^{L}_{34}= 47.4-i23.0 $ & $c^{L}_{34}= -109.7+i49.1 $    \\
 $a^{R}_{1}= -2.1+i1.8 $  & $b^{R}_{1}= 28.9-i27.7 $ &  $c^{R}_{1}= -83.7+i82.0 $   \\
$a^{R}_{34}= 0.16-i0.043 $   & $b^{R}_{34}= -1.46+i0.39$  & $c^{R}_{34}= 3.20-i0.88 $    \\
$a^{\prime L}_{1}=0.18+i0.10 $   & $b^{\prime L}_{1}=-3.36-i2.28 $ & $c^{\prime L}_{1}= 10.7+i7.5 $    \\
$a^{\prime R}_{1}= 0.24-i0.0064 $   & $b^{\prime R}_{1}=-2.26+i0.0070 $ & $c^{\prime R}_{1}= 5.13+i0.073 $    \\
 \hline  \hline
\end{tabular}
 \end{small}
\end{center}
\end{table}

\begin{figure}[h]
\vspace{-1cm}
\[
\epsfig{file=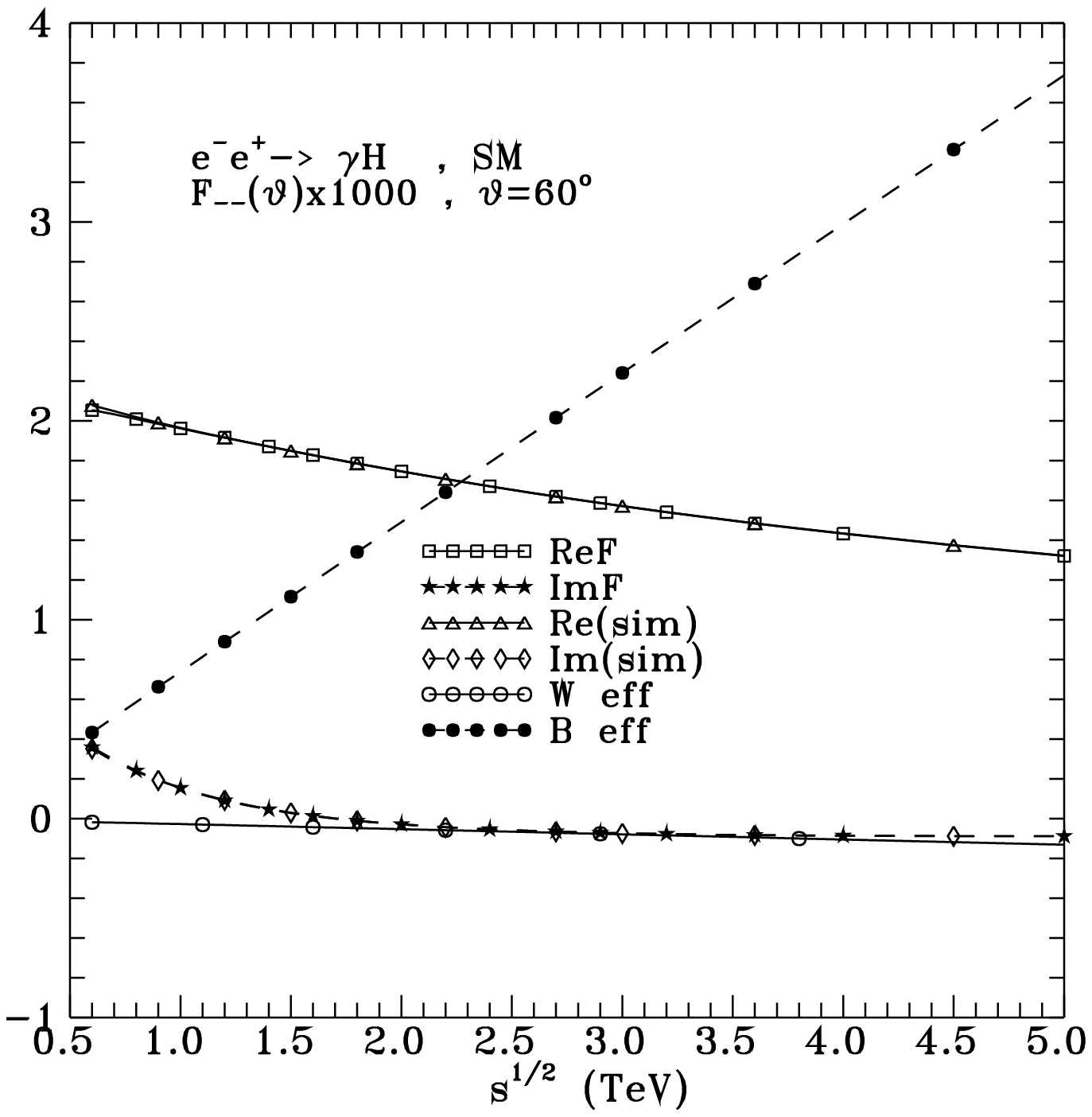, height=6.cm}\hspace{1.cm}
\epsfig{file=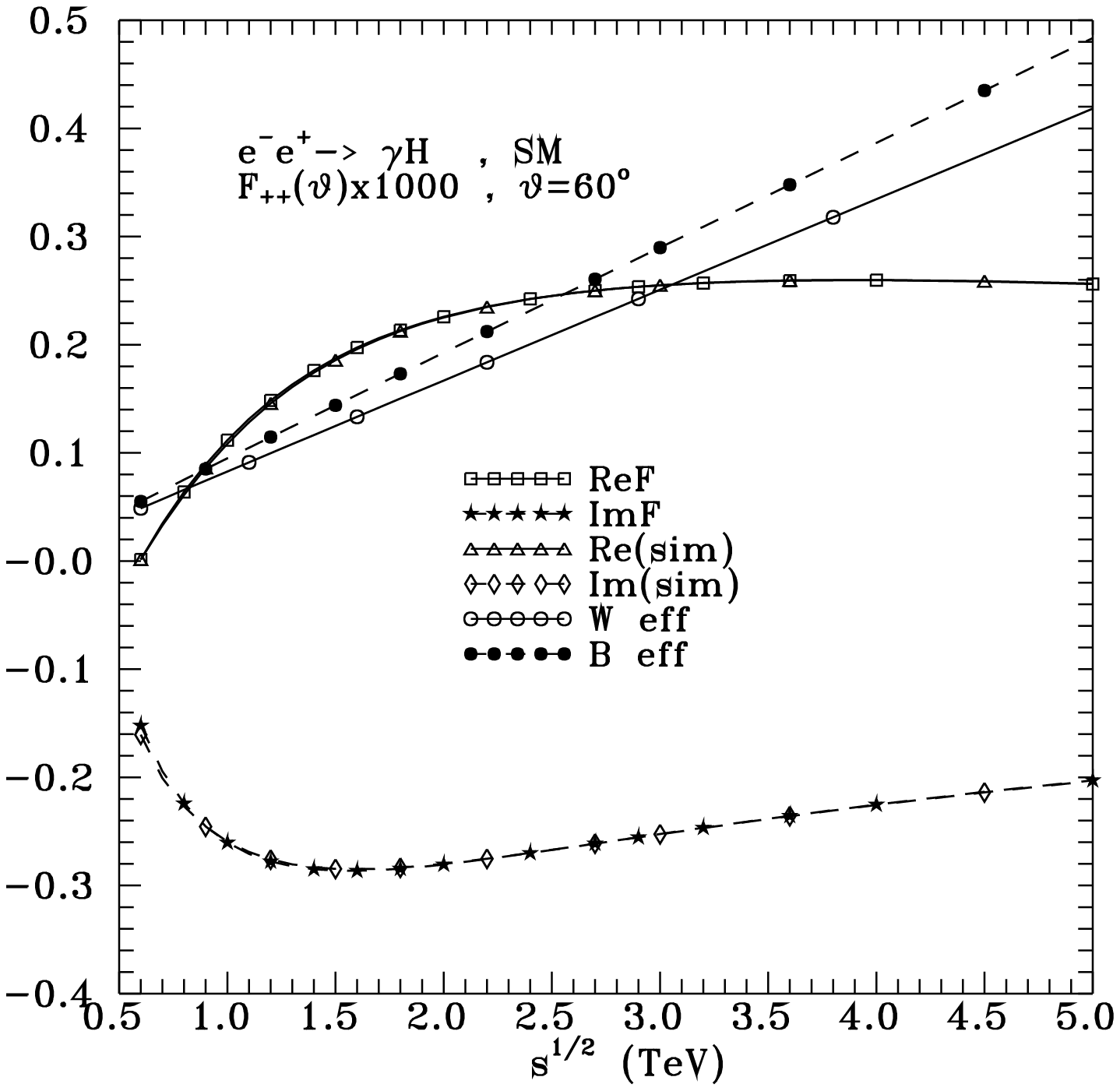,height=6.cm}
\]
\[
\epsfig{file=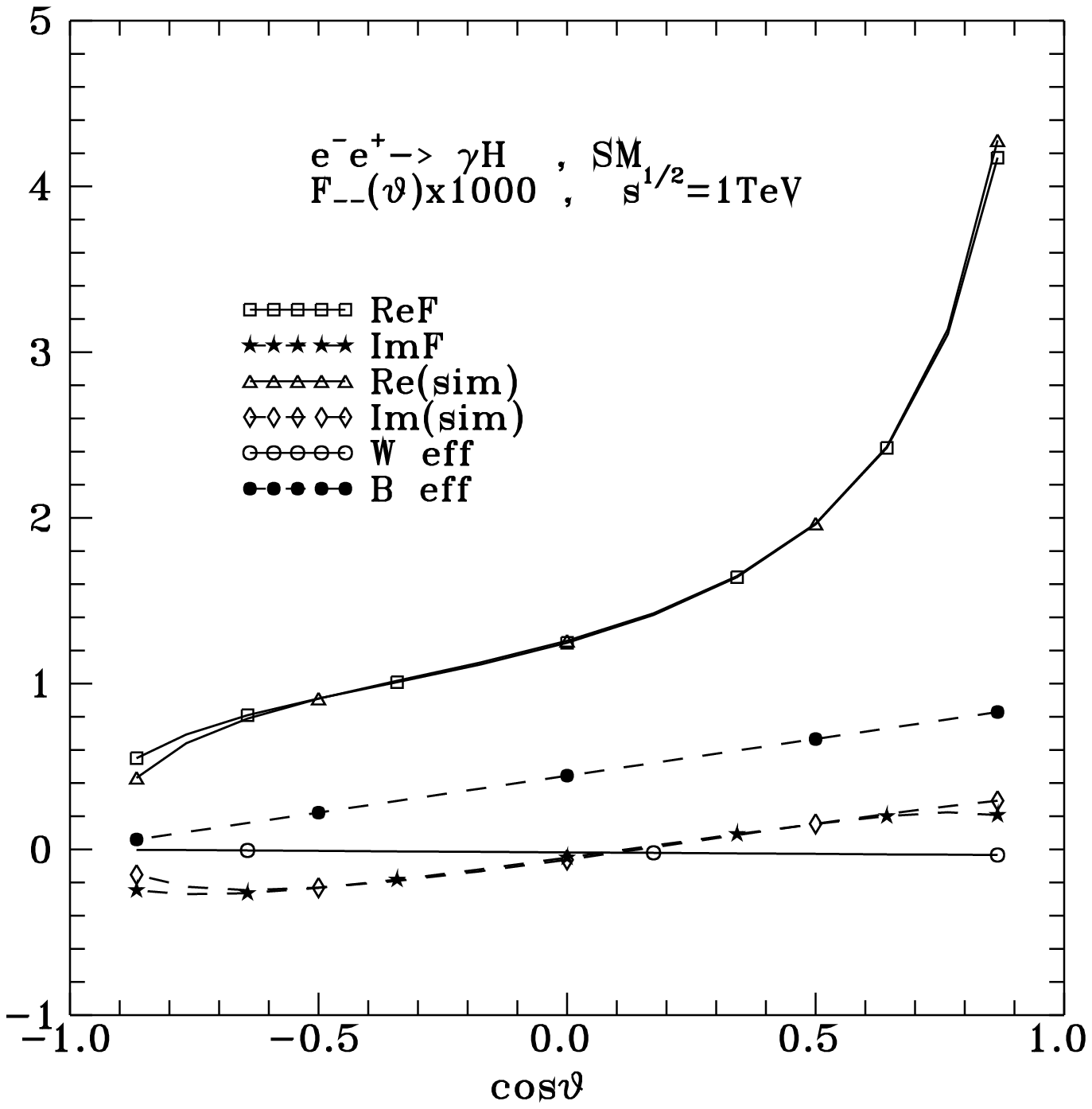, height=6.cm}\hspace{1.cm}
\epsfig{file=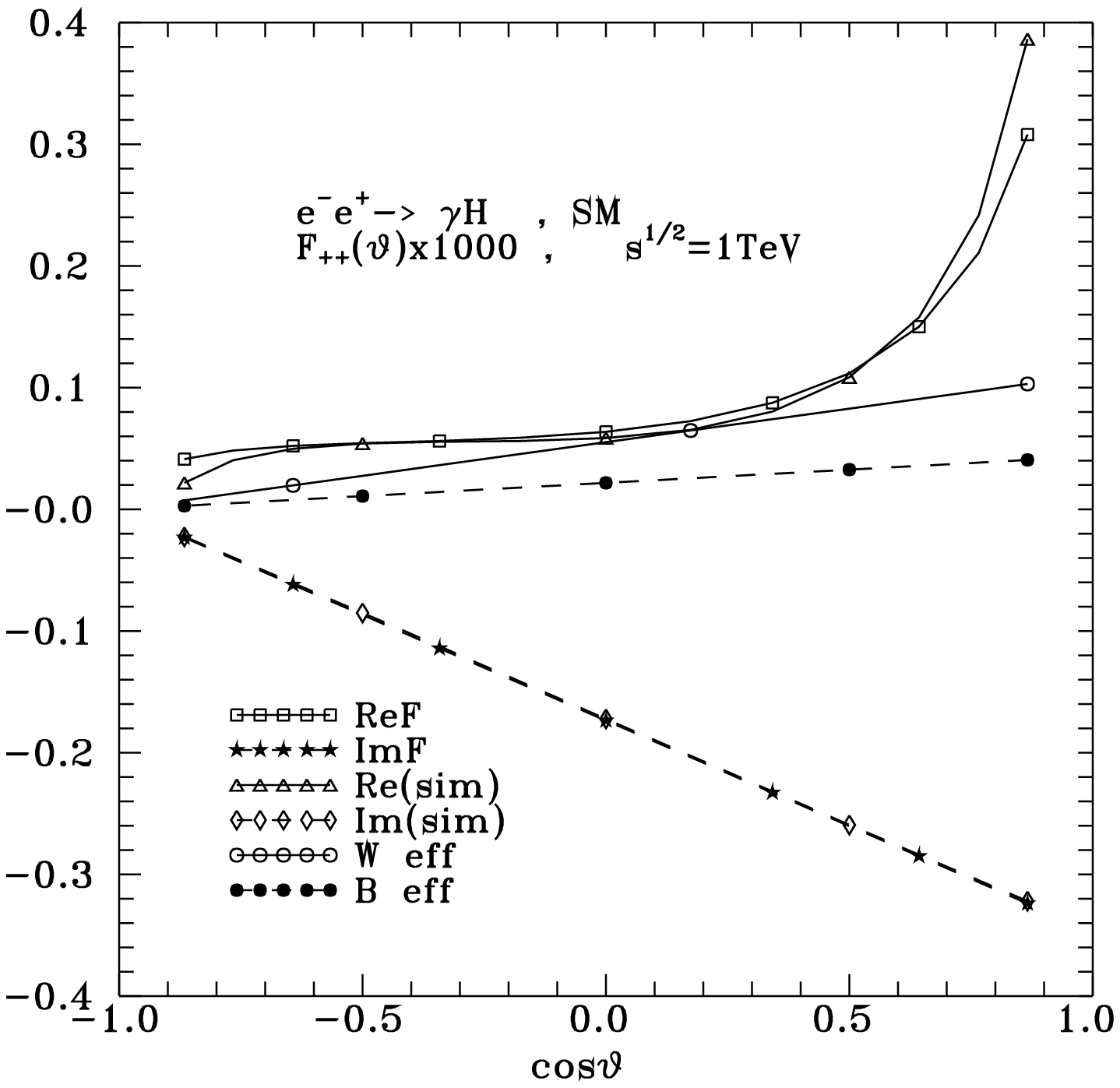,height=6.cm}
\]
\[
\epsfig{file=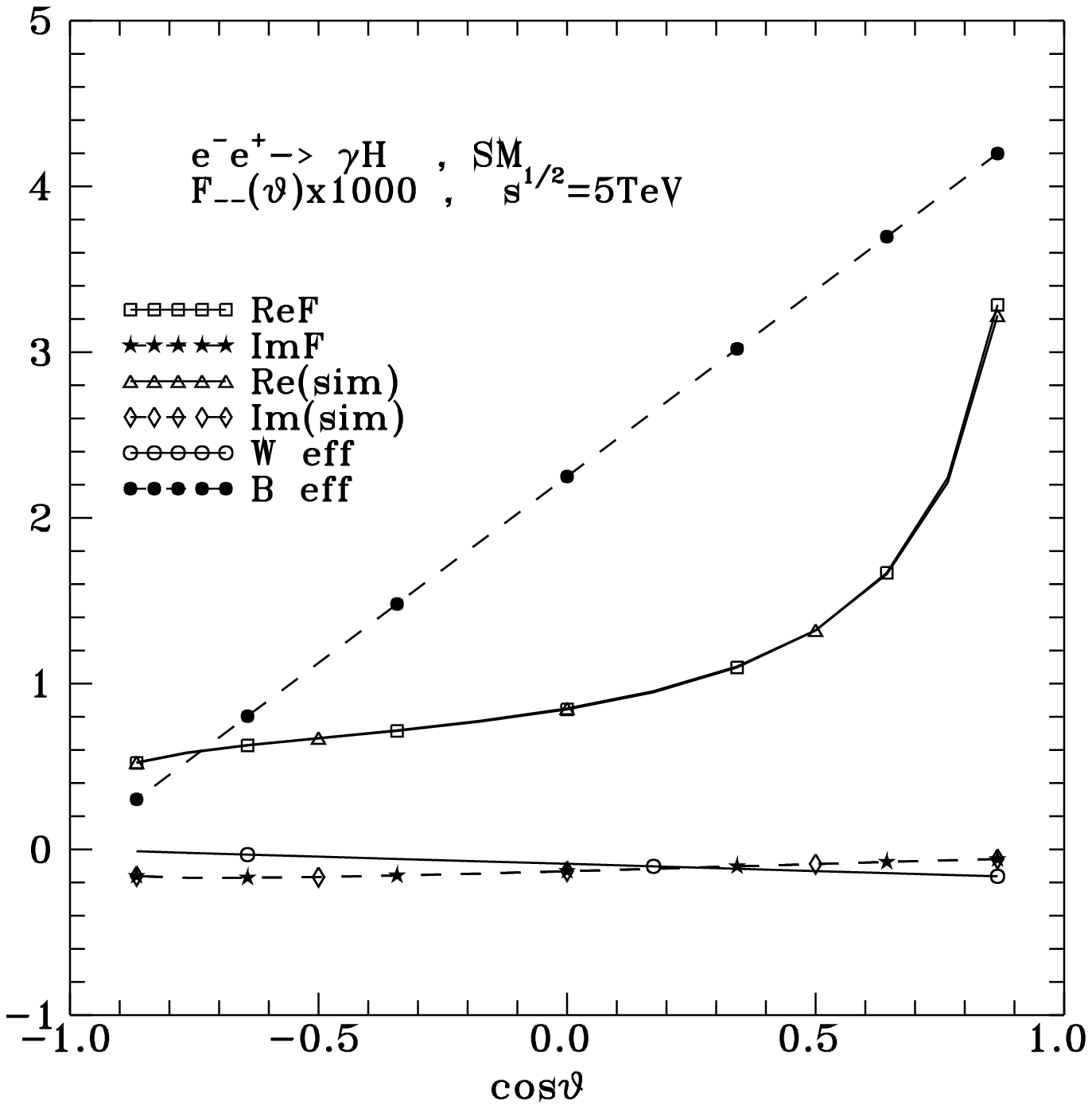, height=6.cm}\hspace{1.cm}
\epsfig{file=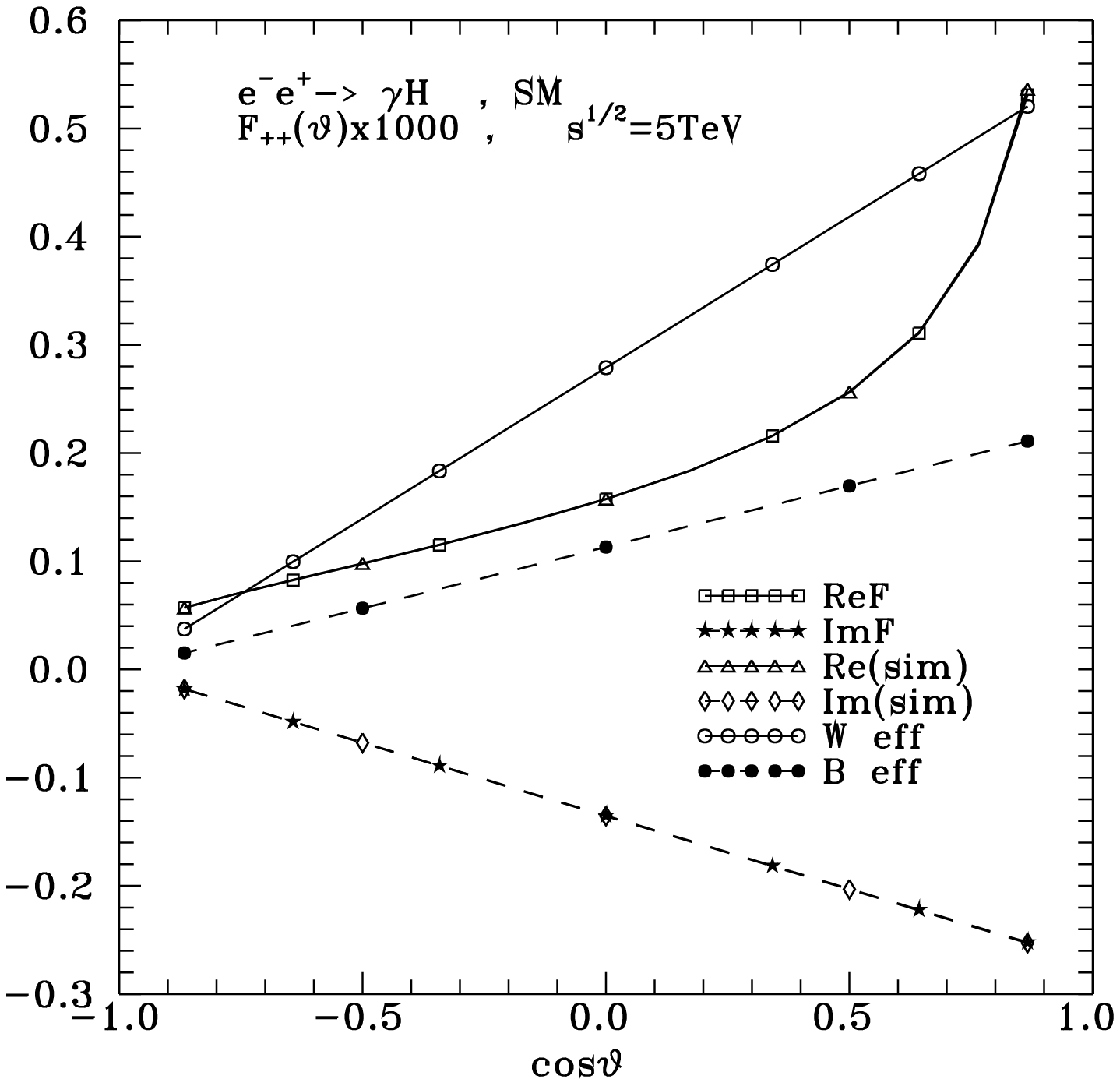,height=6.cm}
\]
\caption[1]{ The $F_{--}$ (left panels) and $F_{++}$ (right panels) amplitudes in SM.
Upper row gives the energy dependence at $60^\circ$, while the middle (lower) row gives the
 angular dependence at 1 (5) TeV. The W eff and  B eff couplings  of the BSM model
are defined in (\ref{Weff} , \ref{Beff}). }
\label{SM1-amp}
\end{figure}

\clearpage

\begin{figure}[h]
\vspace{-1cm}
\[
\epsfig{file=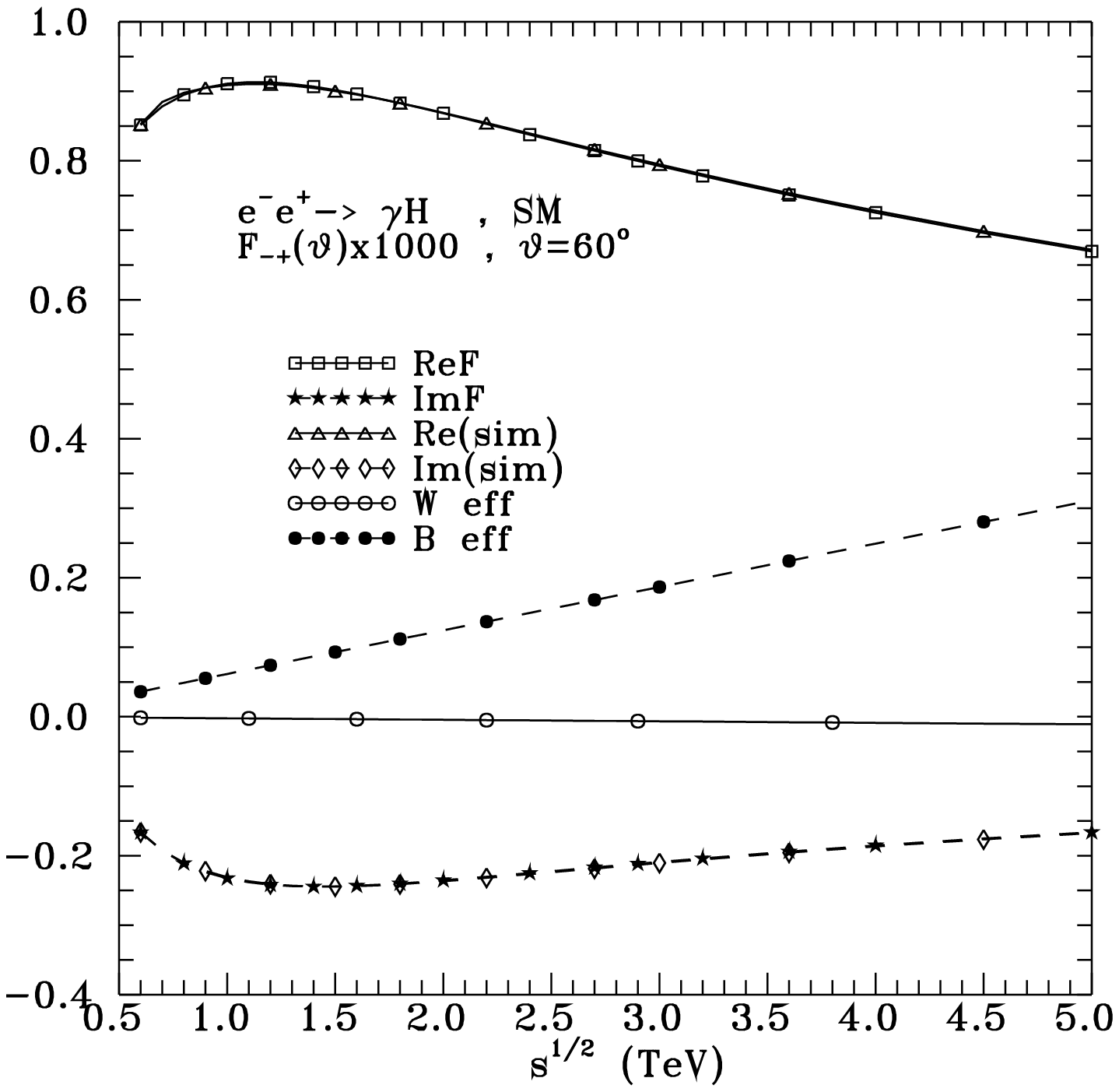, height=6.cm}\hspace{1.cm}
\epsfig{file=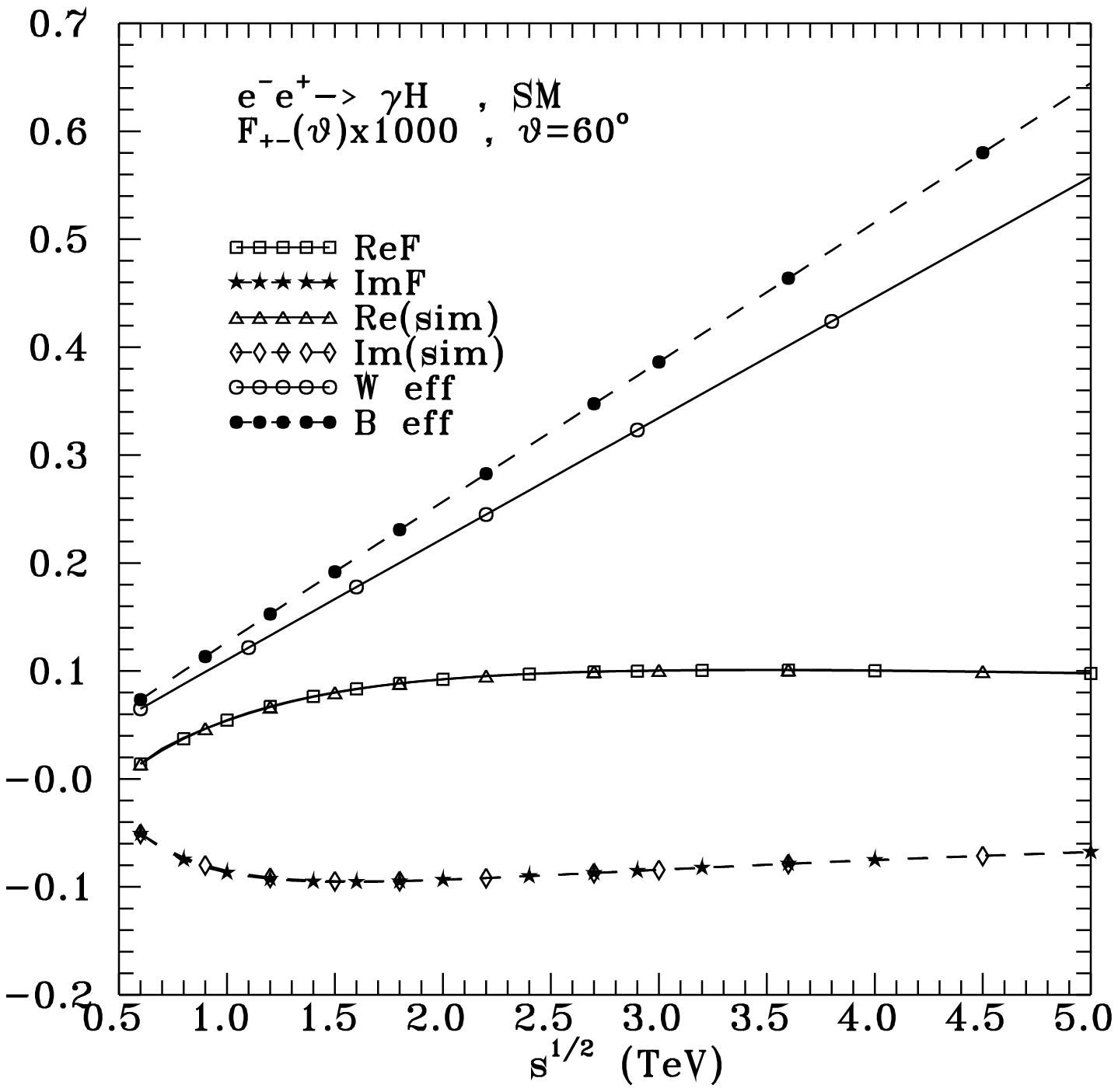,height=6.cm}
\]
\[
\epsfig{file=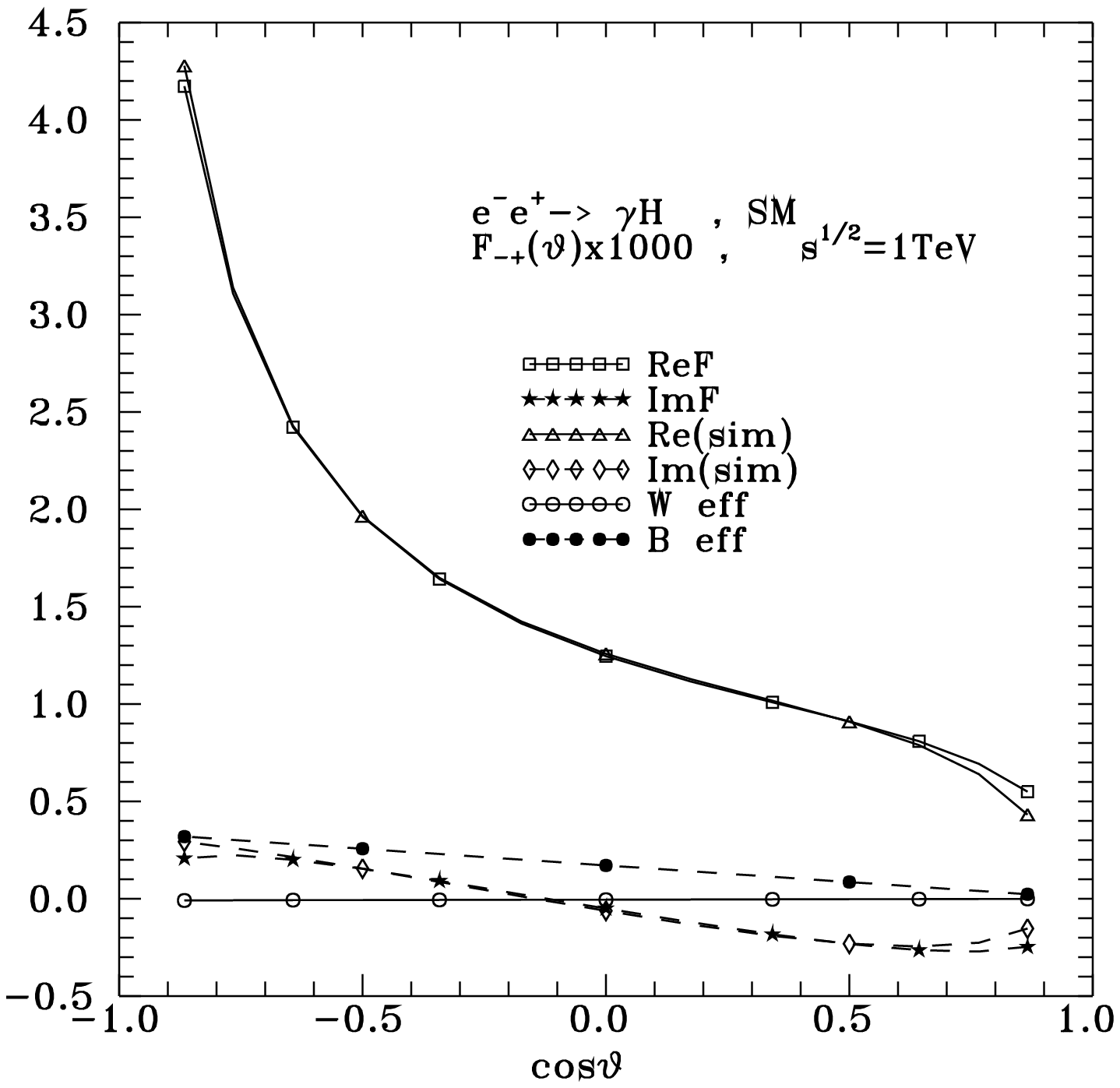, height=6.cm}\hspace{1.cm}
\epsfig{file=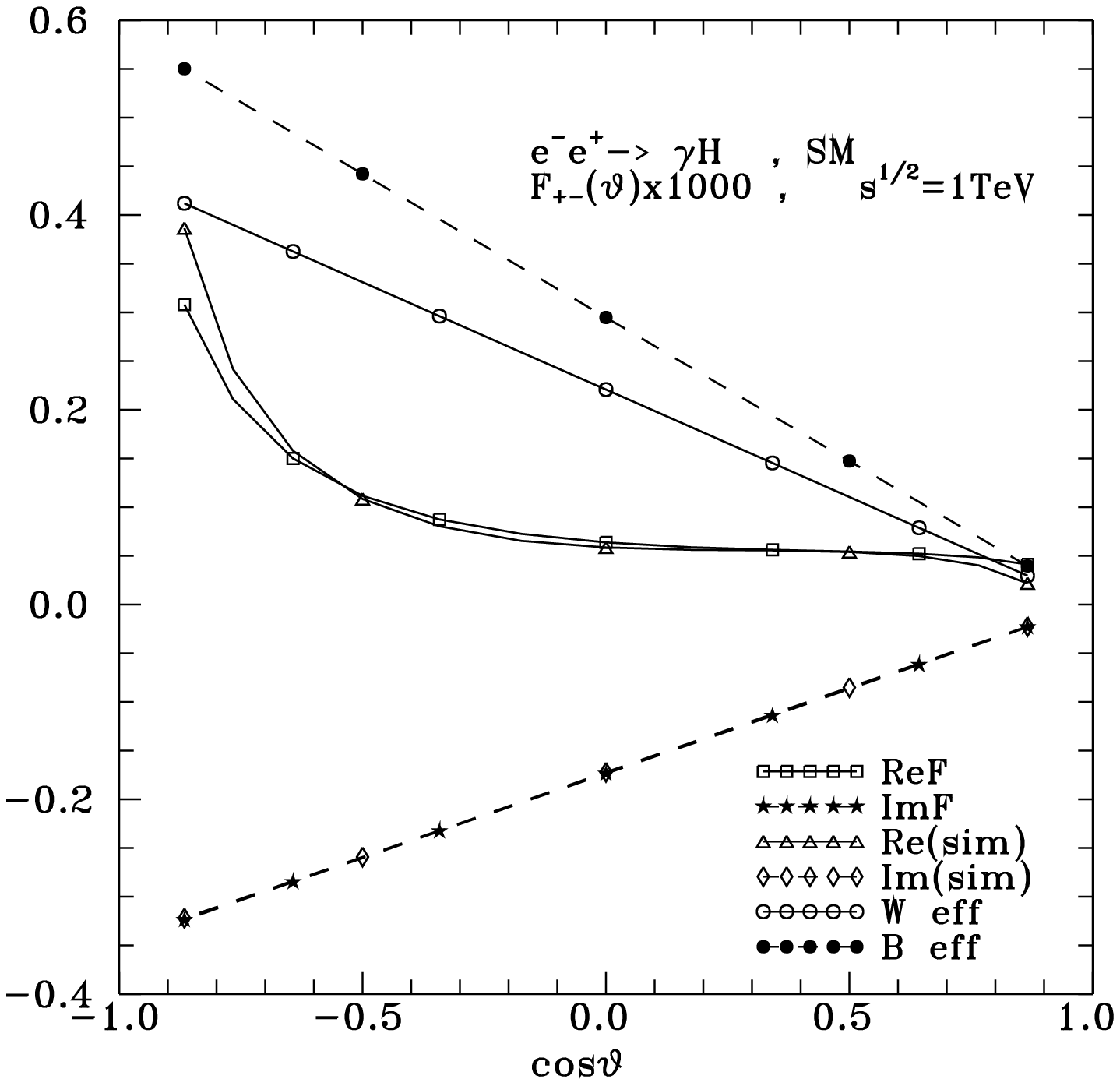,height=6.cm}
\]
\[
\epsfig{file=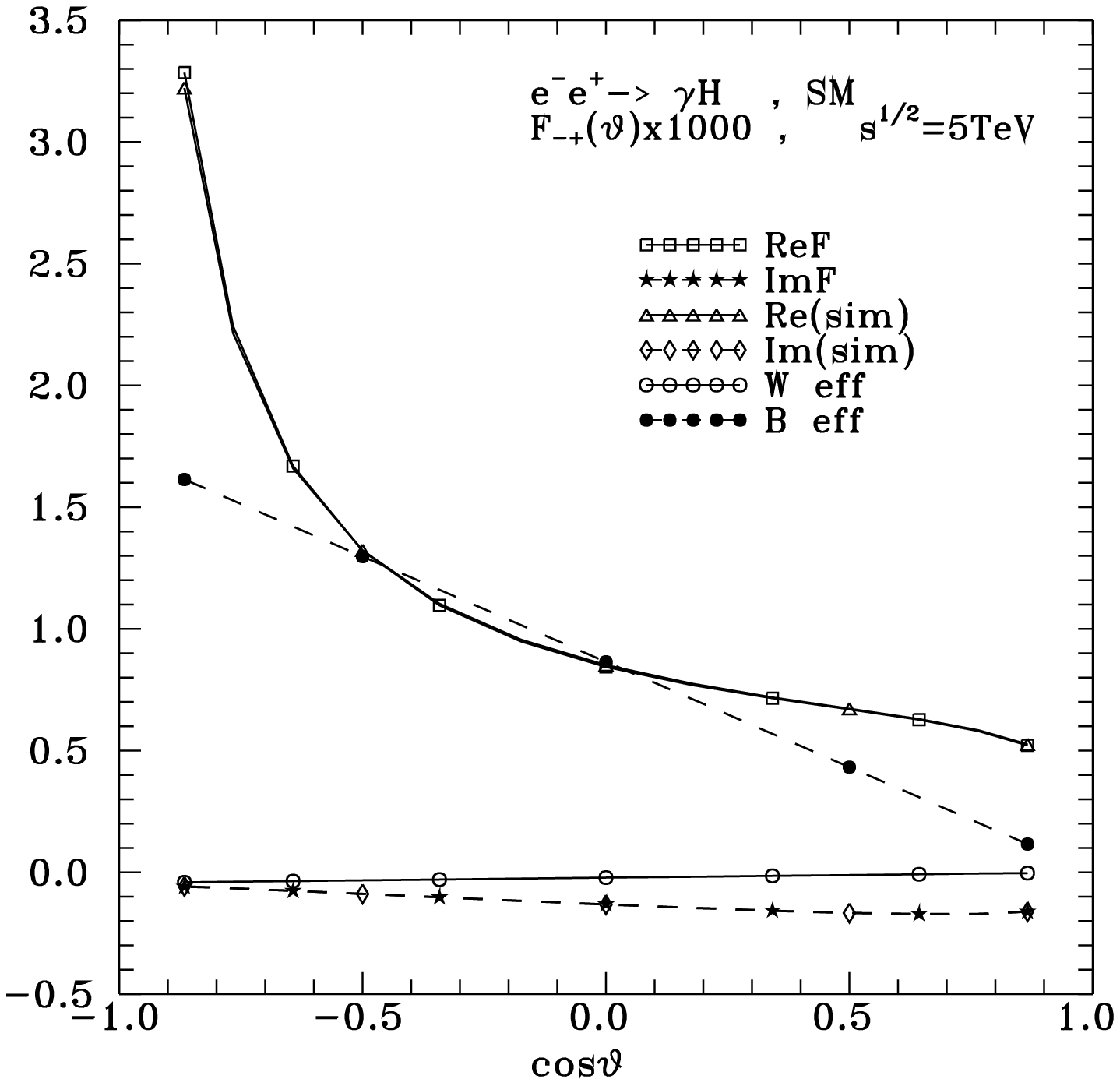, height=6.cm}\hspace{1.cm}
\epsfig{file=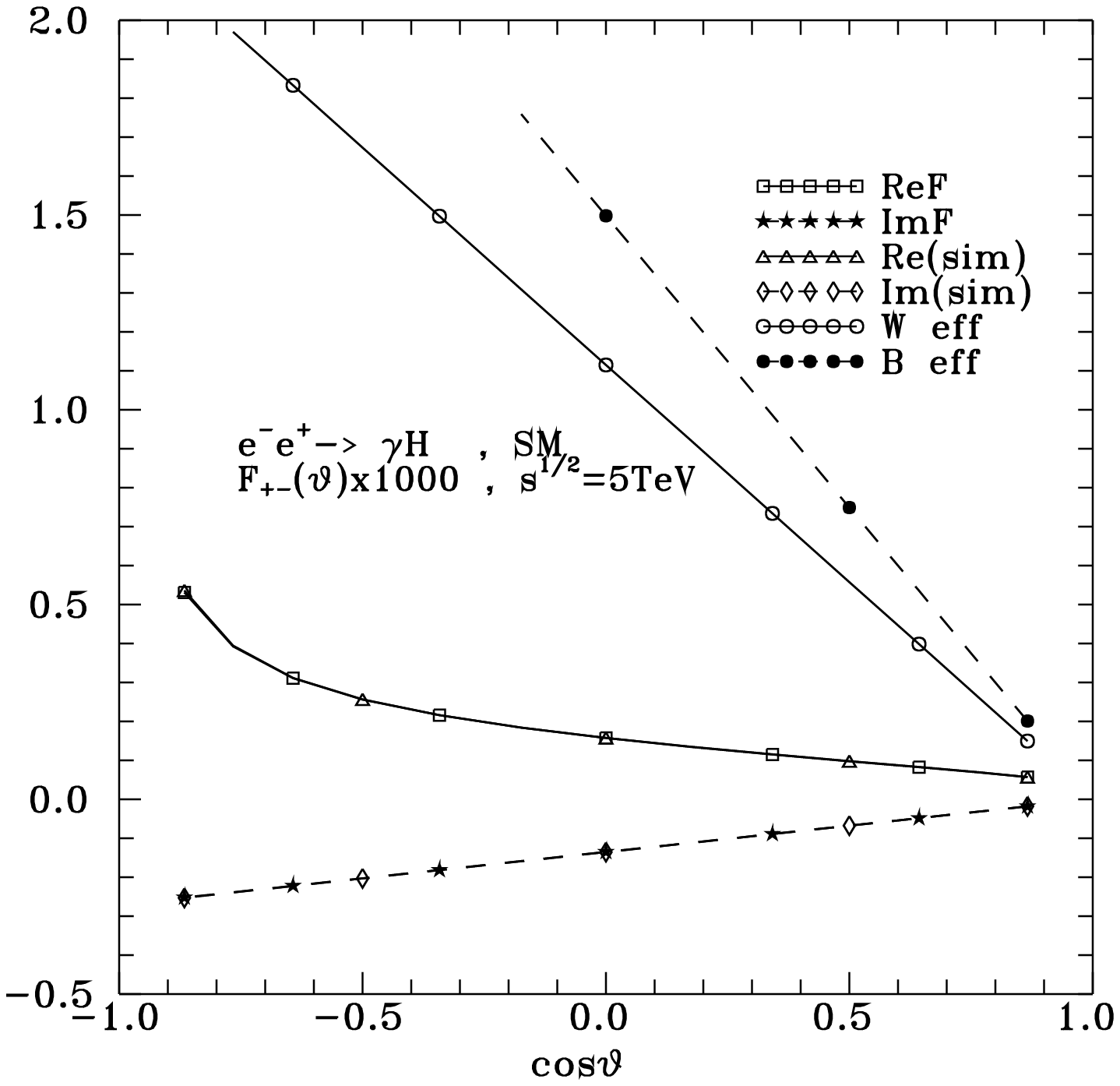,height=6.cm}
\]
\caption[1]{The $F_{-+}$ (left panels) and $F_{+-}$ (right panels) amplitudes in SM.
Panels and BSM couplings as in Fig.\ref{SM1-amp}.}
\label{SM2-amp}
\end{figure}

\clearpage

\begin{figure}[h]
\vspace{-1cm}
\[
\epsfig{file=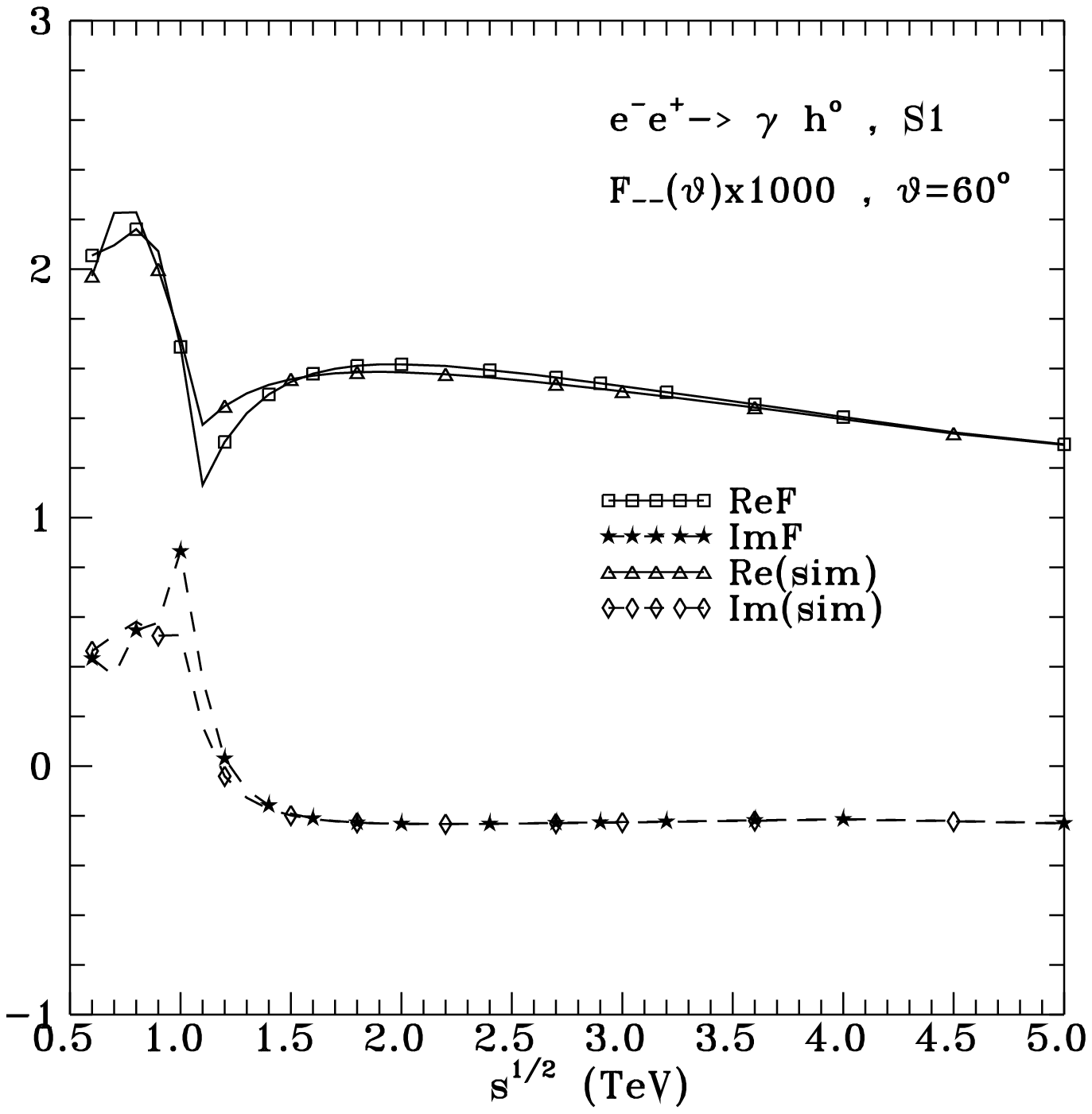, height=6.cm}\hspace{1.cm}
\epsfig{file=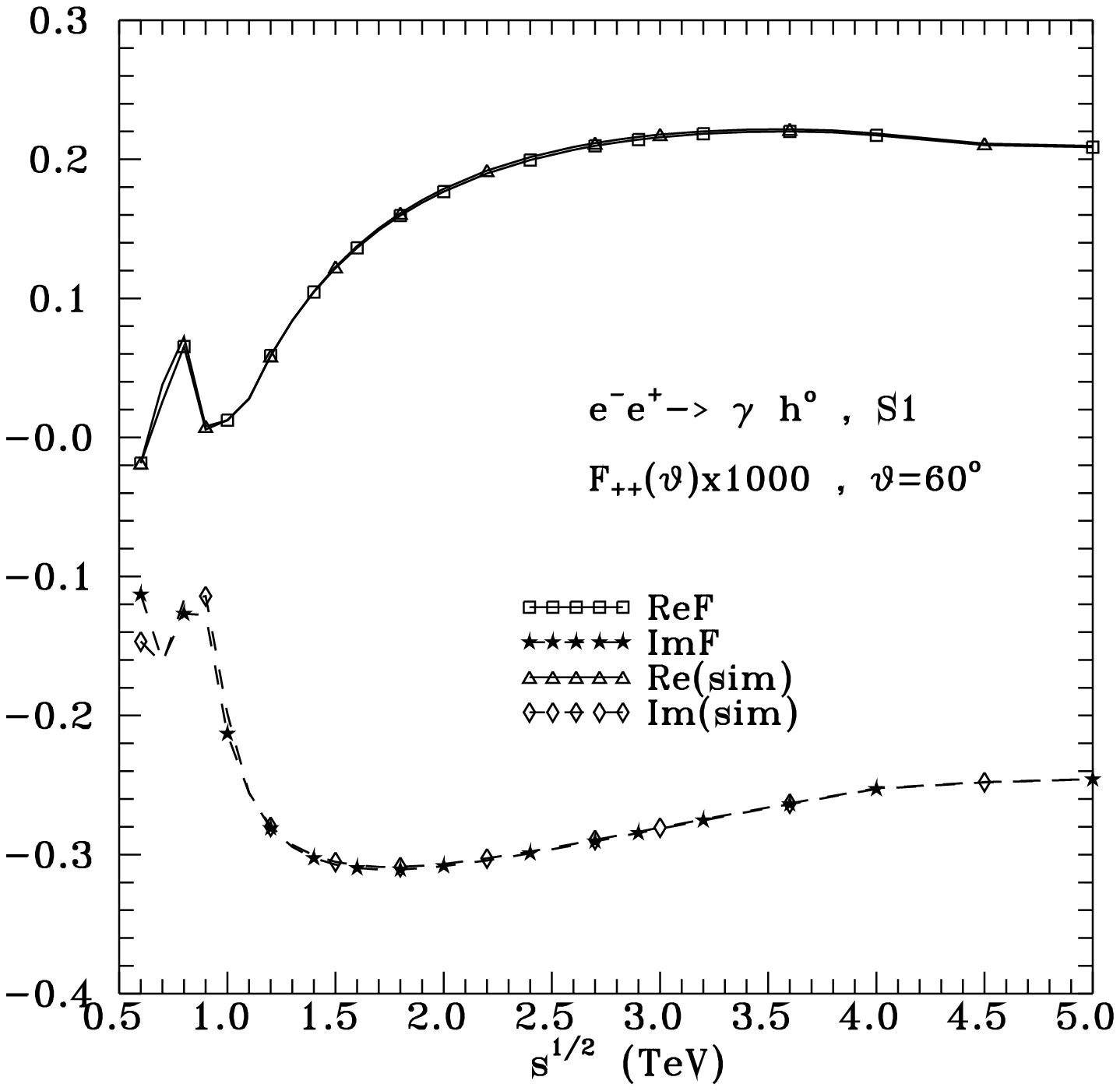,height=6.cm}
\]
\[
\epsfig{file=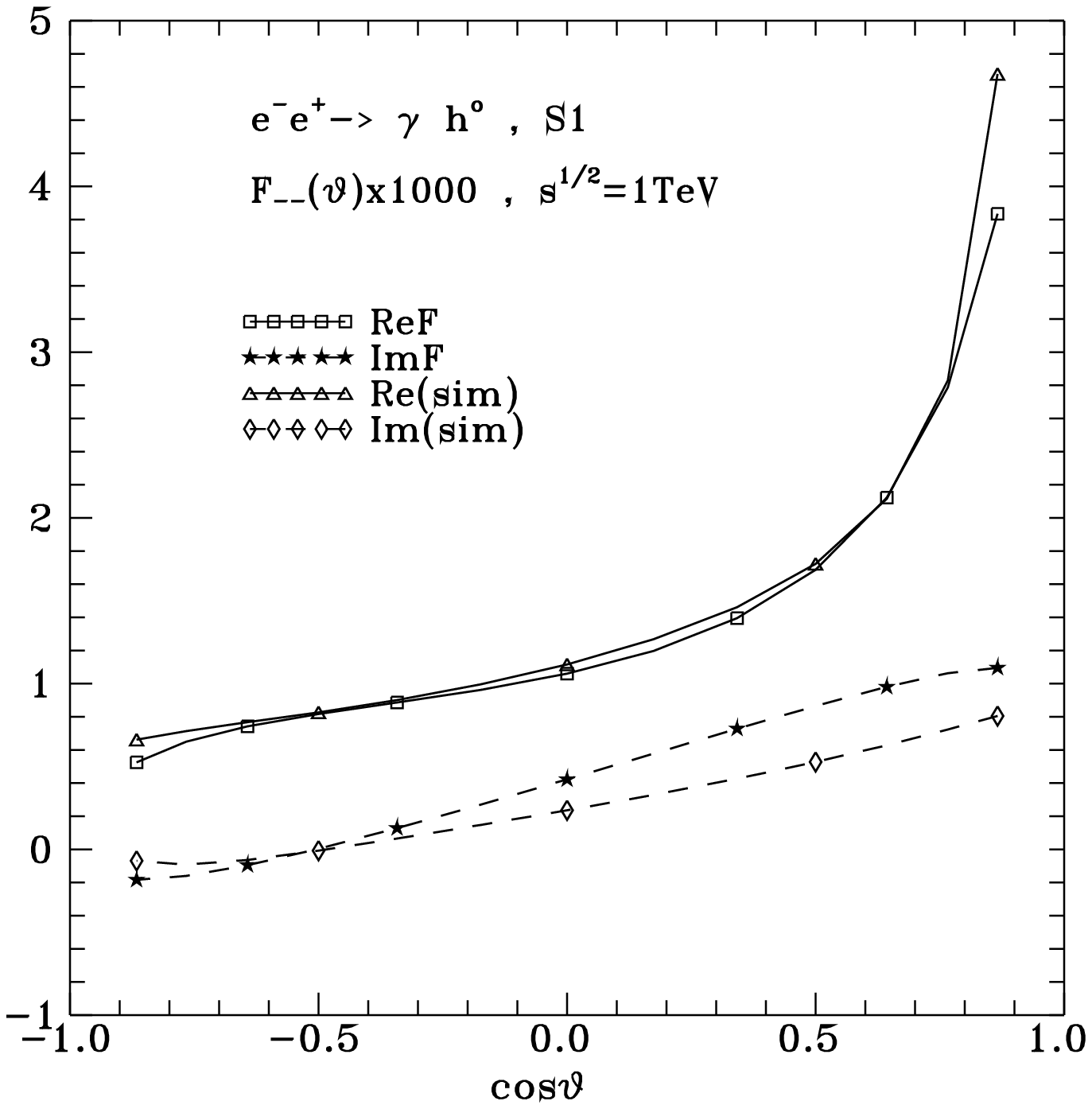, height=6.cm}\hspace{1.cm}
\epsfig{file=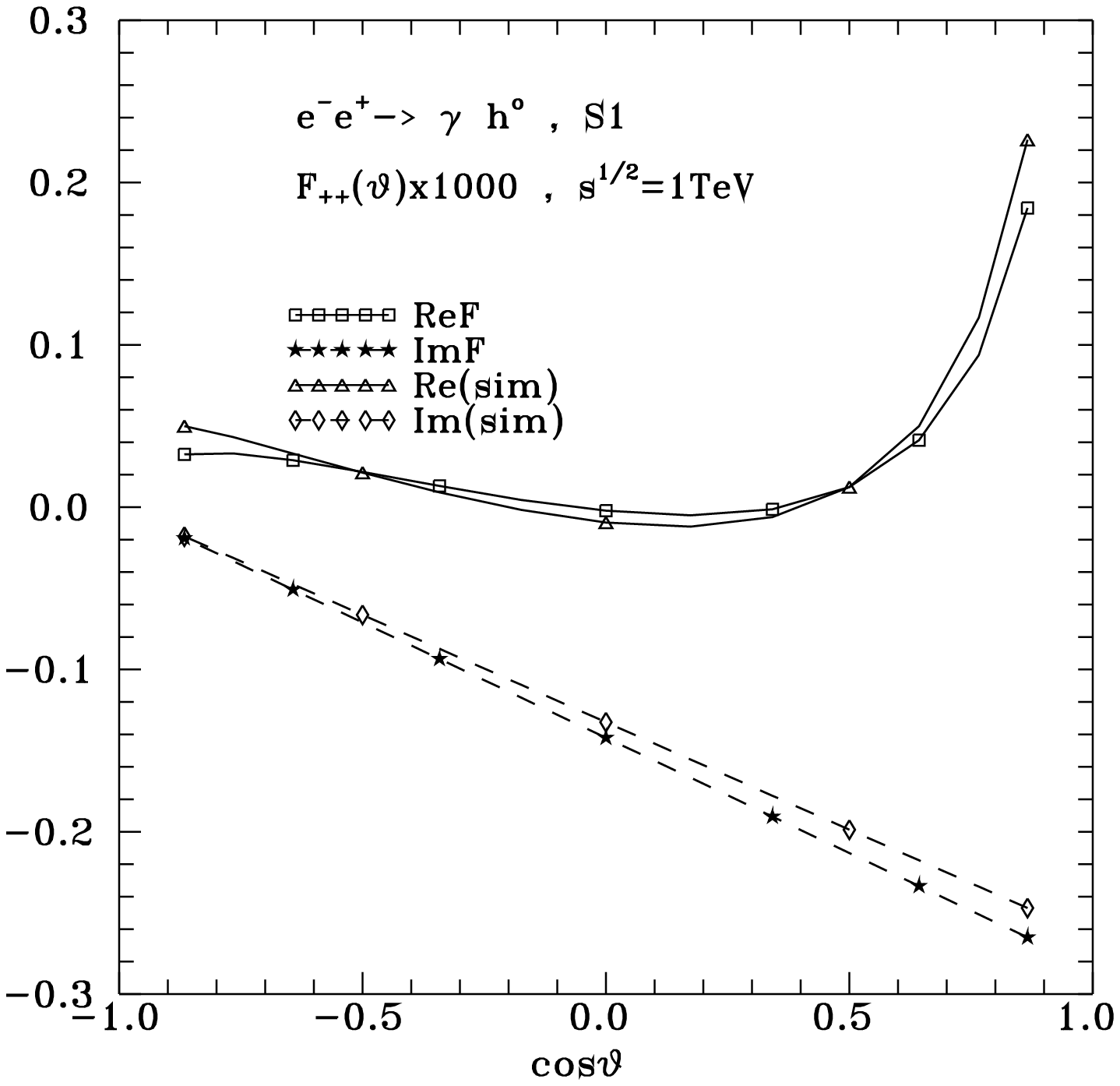,height=6.cm}
\]
\[
\epsfig{file=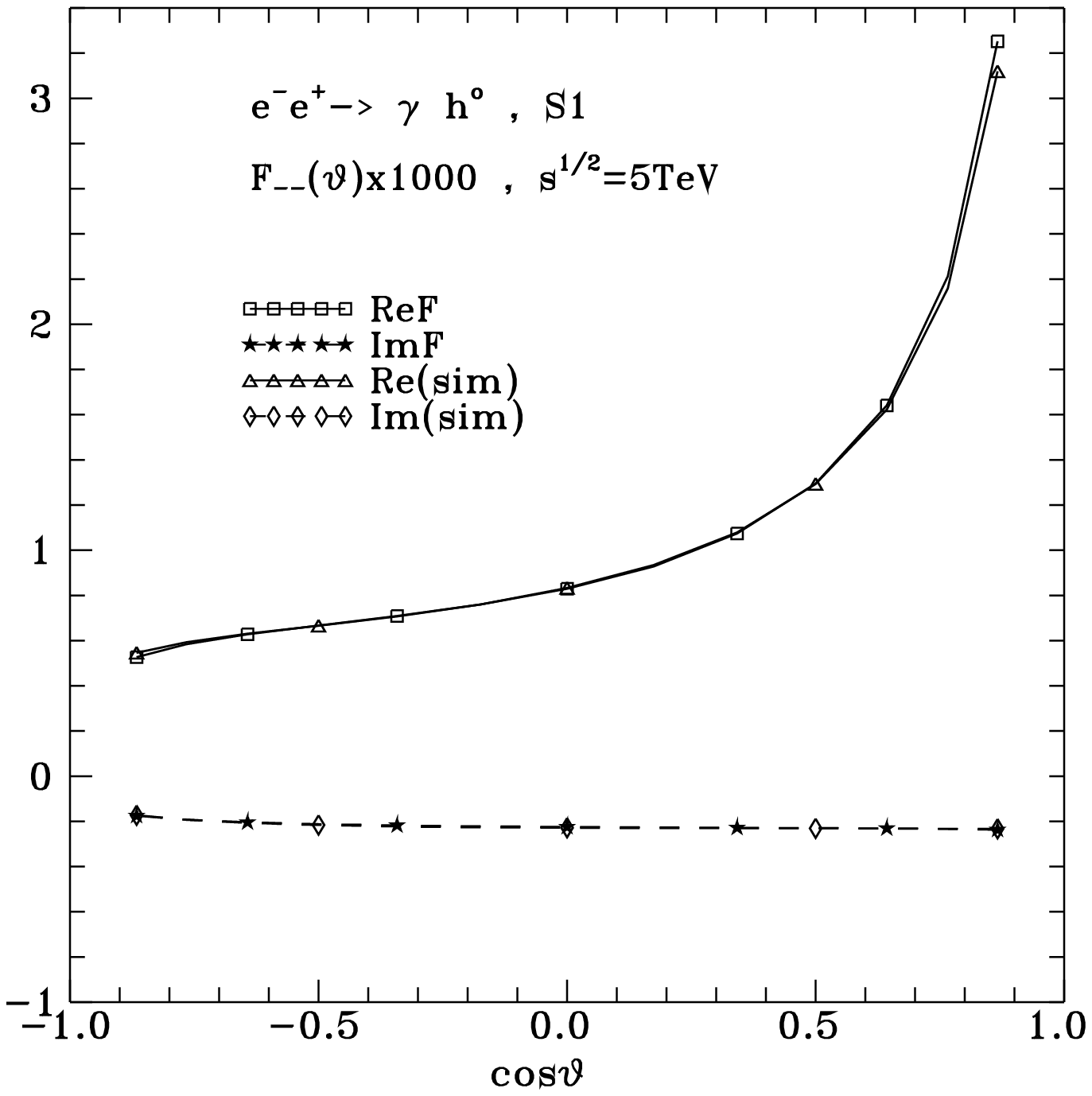, height=6.cm}\hspace{1.cm}
\epsfig{file=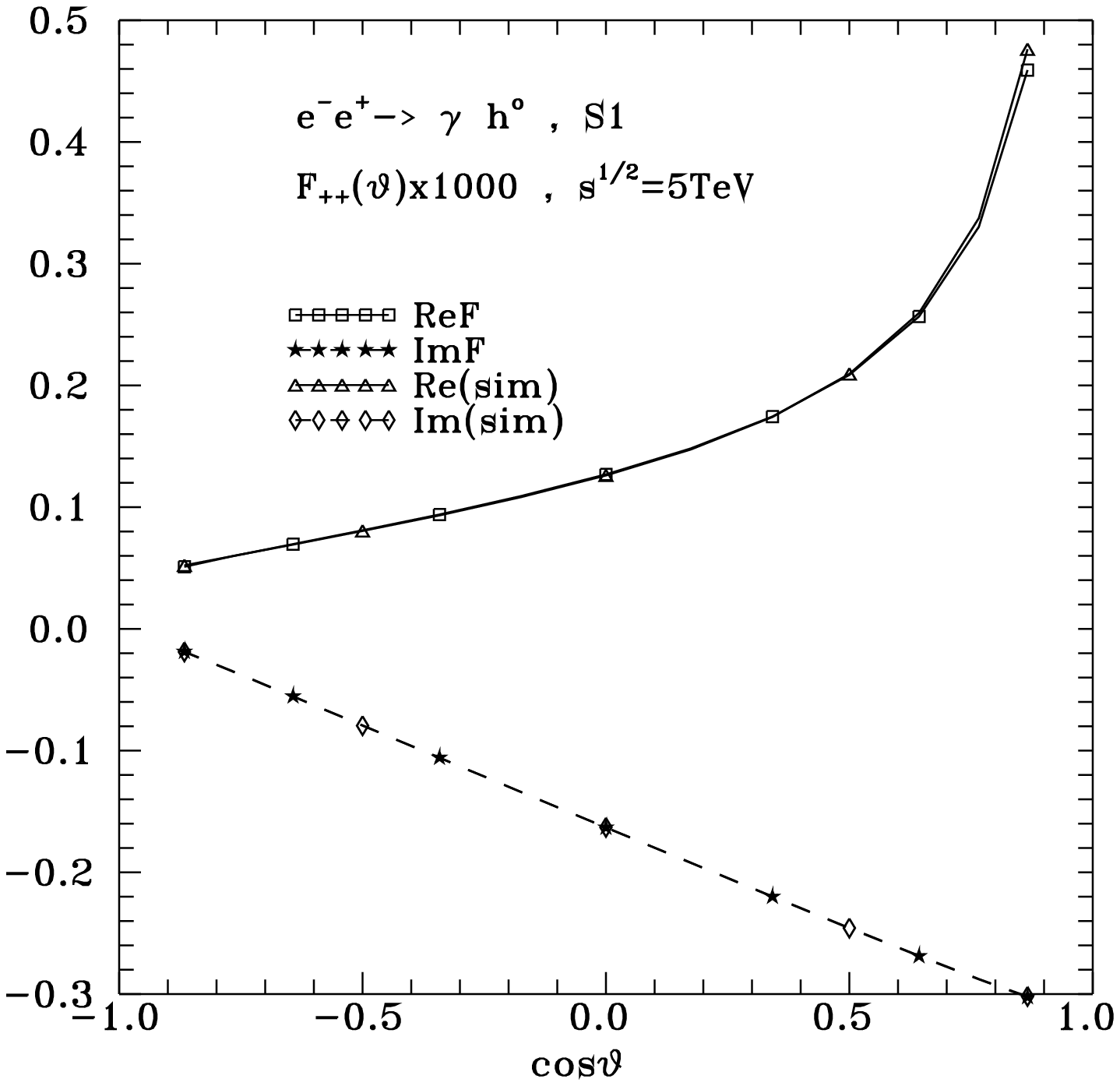,height=6.cm}
\]
\caption[1]{The $F_{--}$ (left panels) and $F_{++}$ (right panels) amplitudes for $h^0$
in S1 MSSM; see (\ref{bench-param}).
Panels as in Fig.\ref{SM1-amp}. }
\label{$h^0$1-amp}
\end{figure}

\clearpage

\begin{figure}[h]
\vspace{-1cm}
\[
\epsfig{file=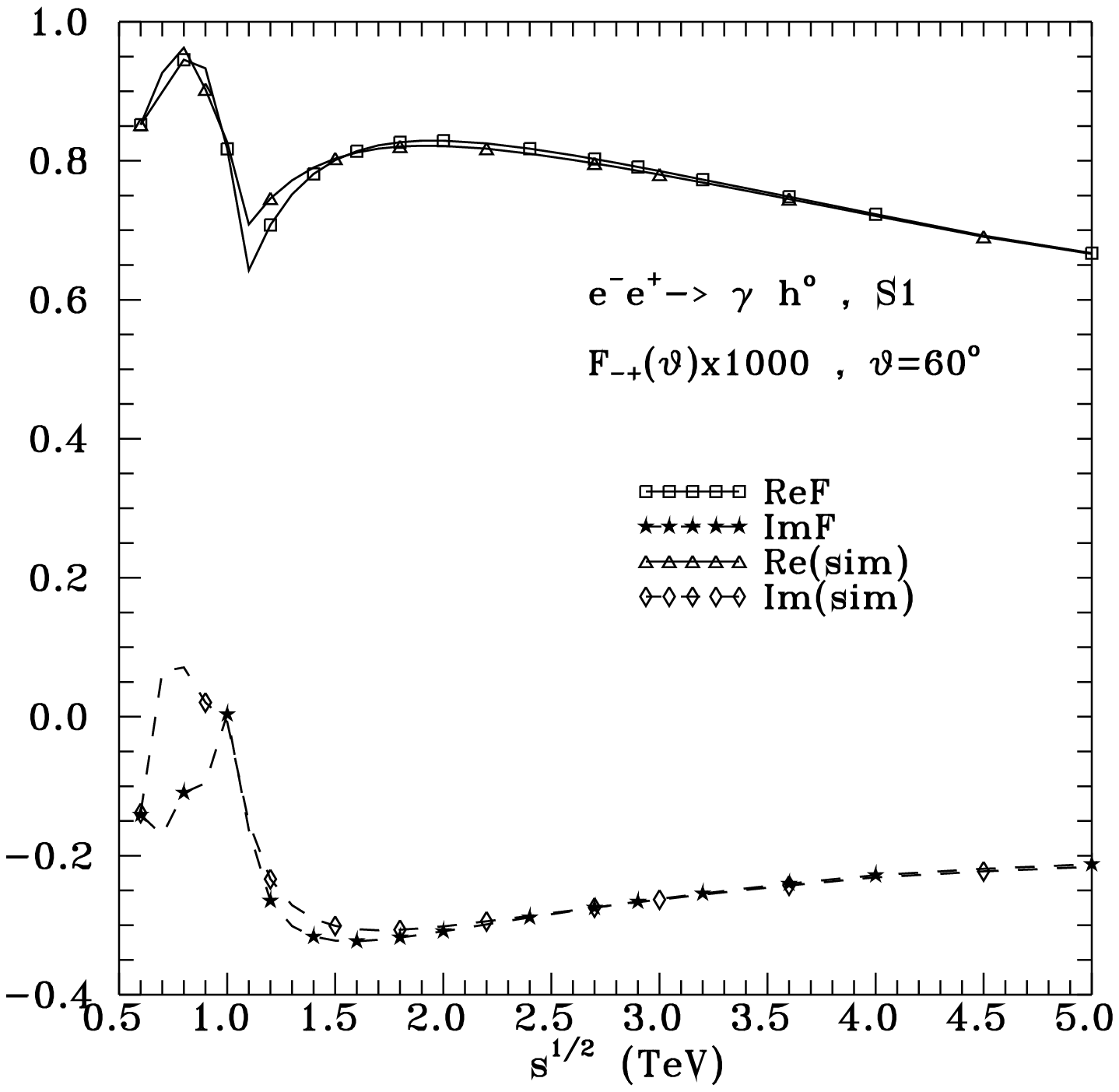, height=6.cm}\hspace{1.cm}
\epsfig{file=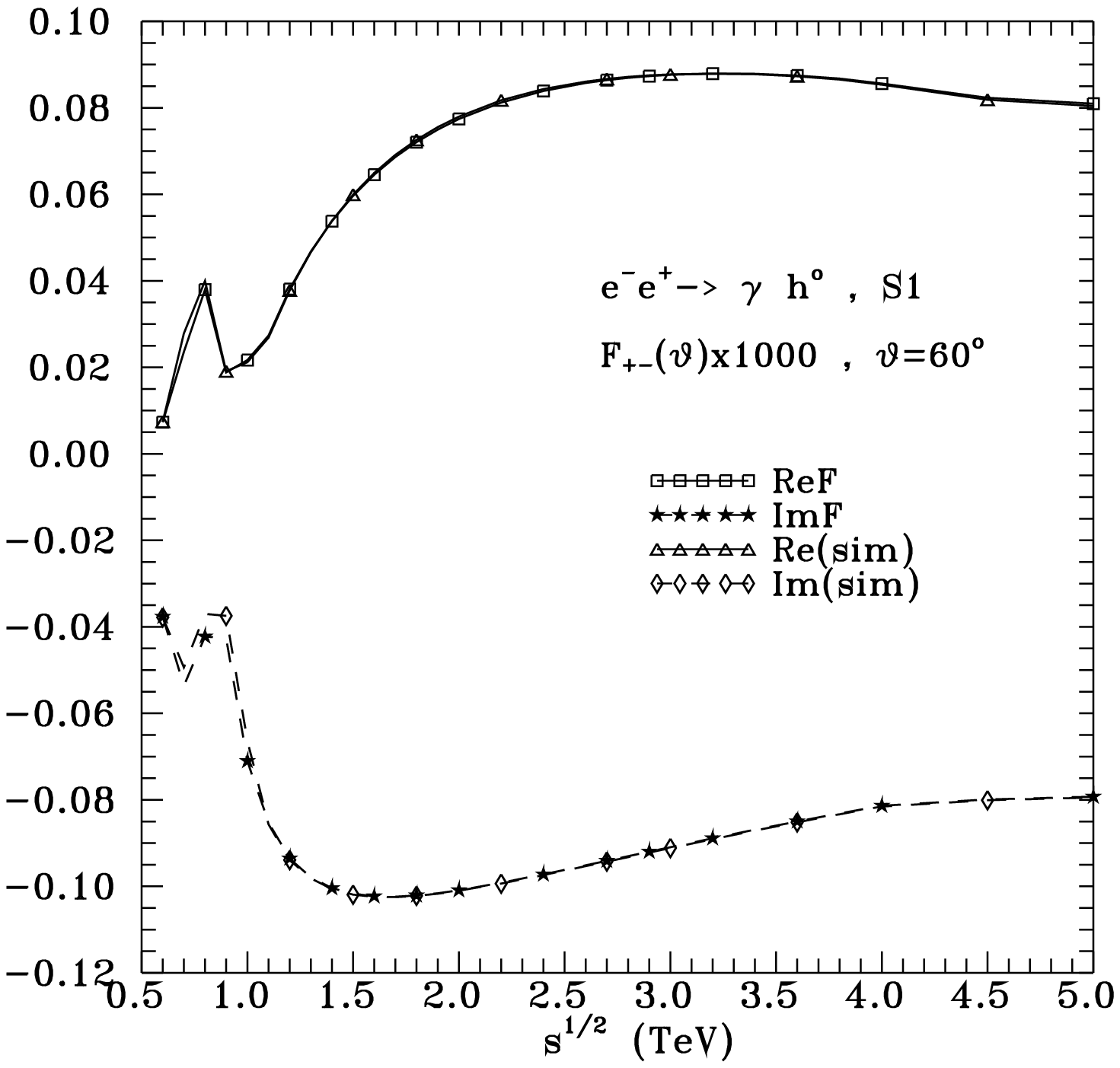,height=6.cm}
\]
\[
\epsfig{file=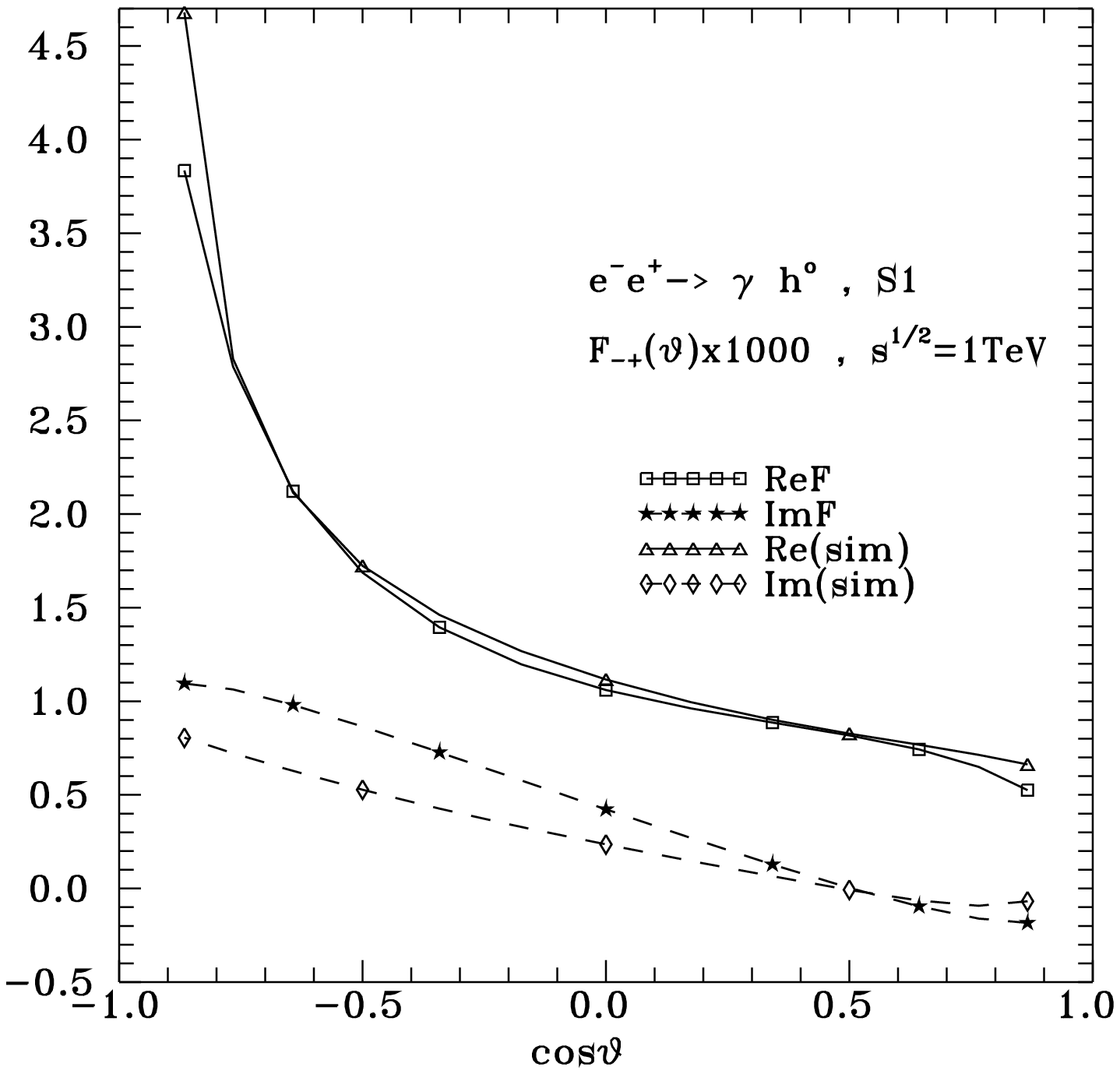, height=6.cm}\hspace{1.cm}
\epsfig{file=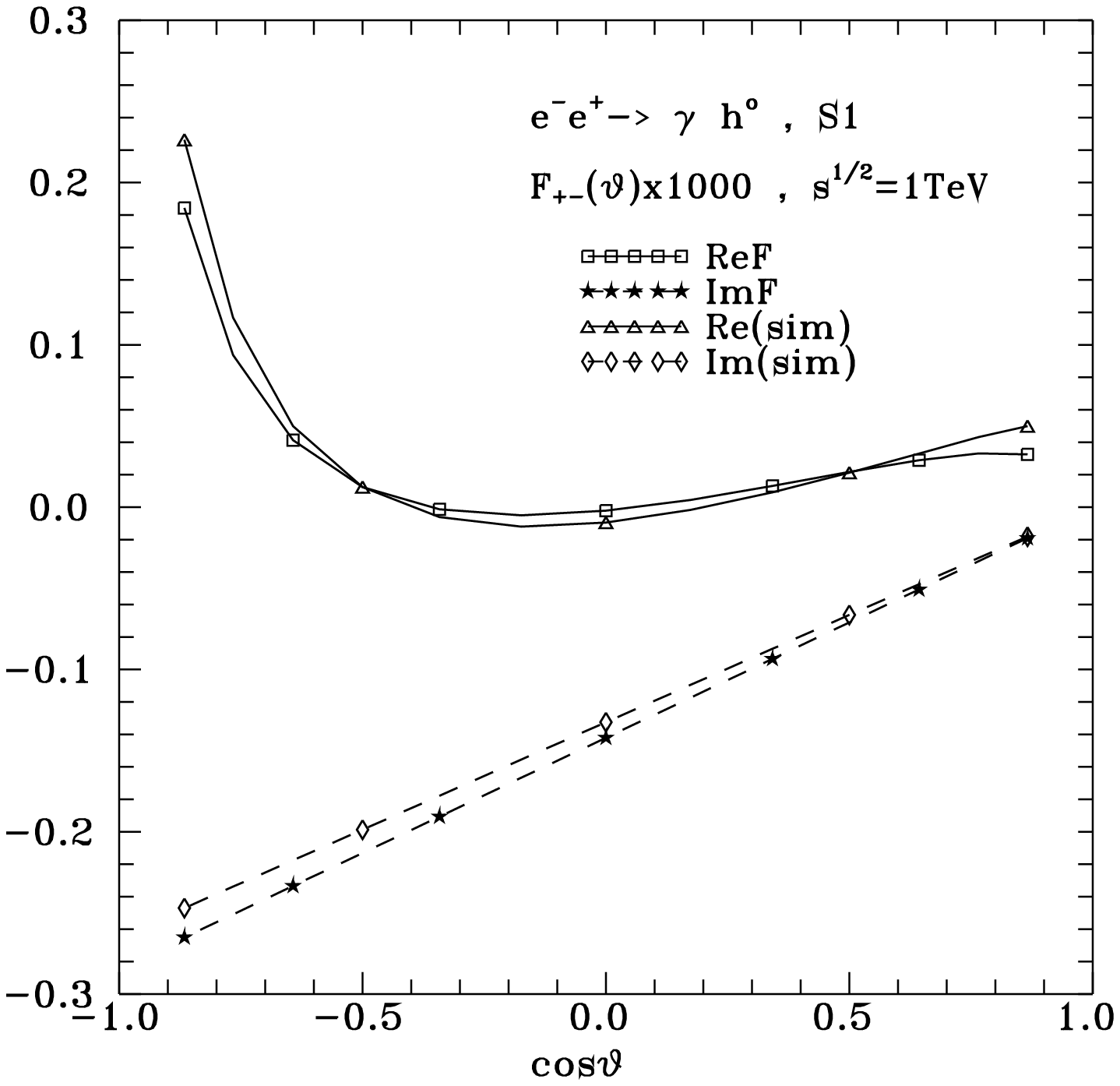,height=6.cm}
\]
\[
\epsfig{file=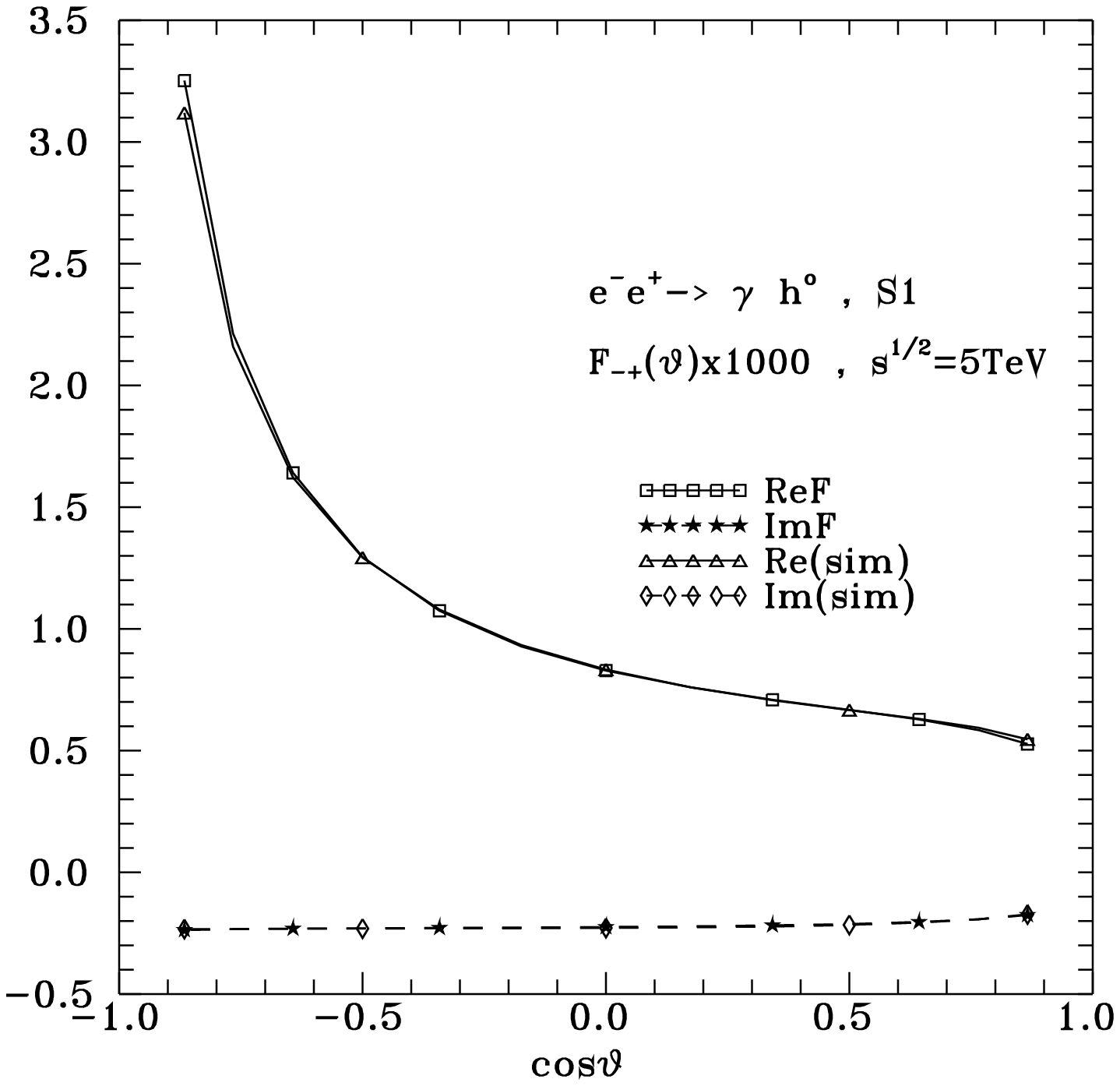, height=6.cm}\hspace{1.cm}
\epsfig{file=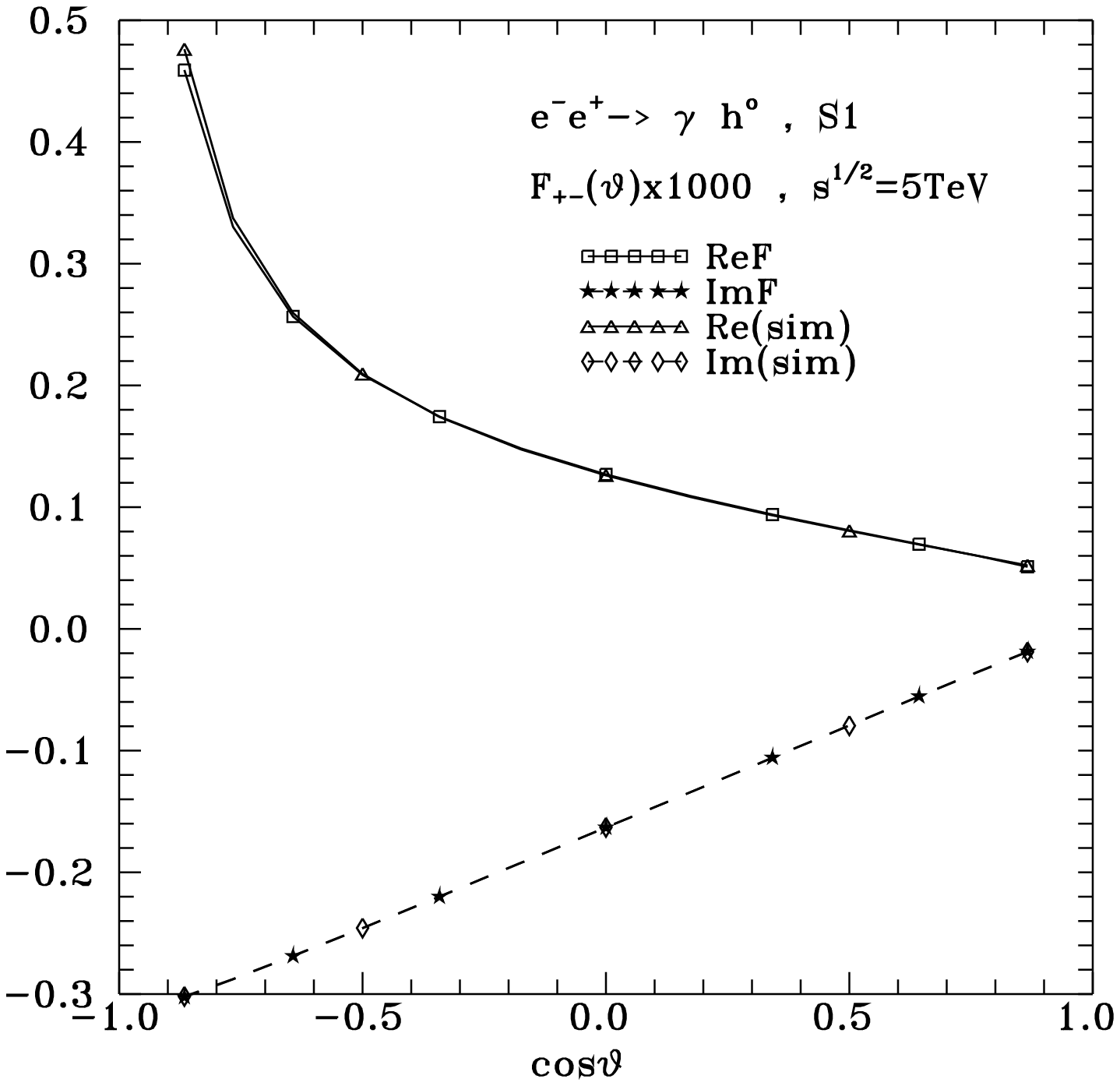,height=6.cm}
\]
\caption[1]{ The $F_{-+}$ (left panels) and $F_{+-}$ (right panels) amplitudes
for $h^0$ in S1 MSSM; see  (\ref{bench-param}).
Panels as in Fig.\ref{SM1-amp}. }
\label{$h^0$2-amp}
\end{figure}

\clearpage

\begin{figure}[h]
\vspace{-1cm}
\[
\epsfig{file=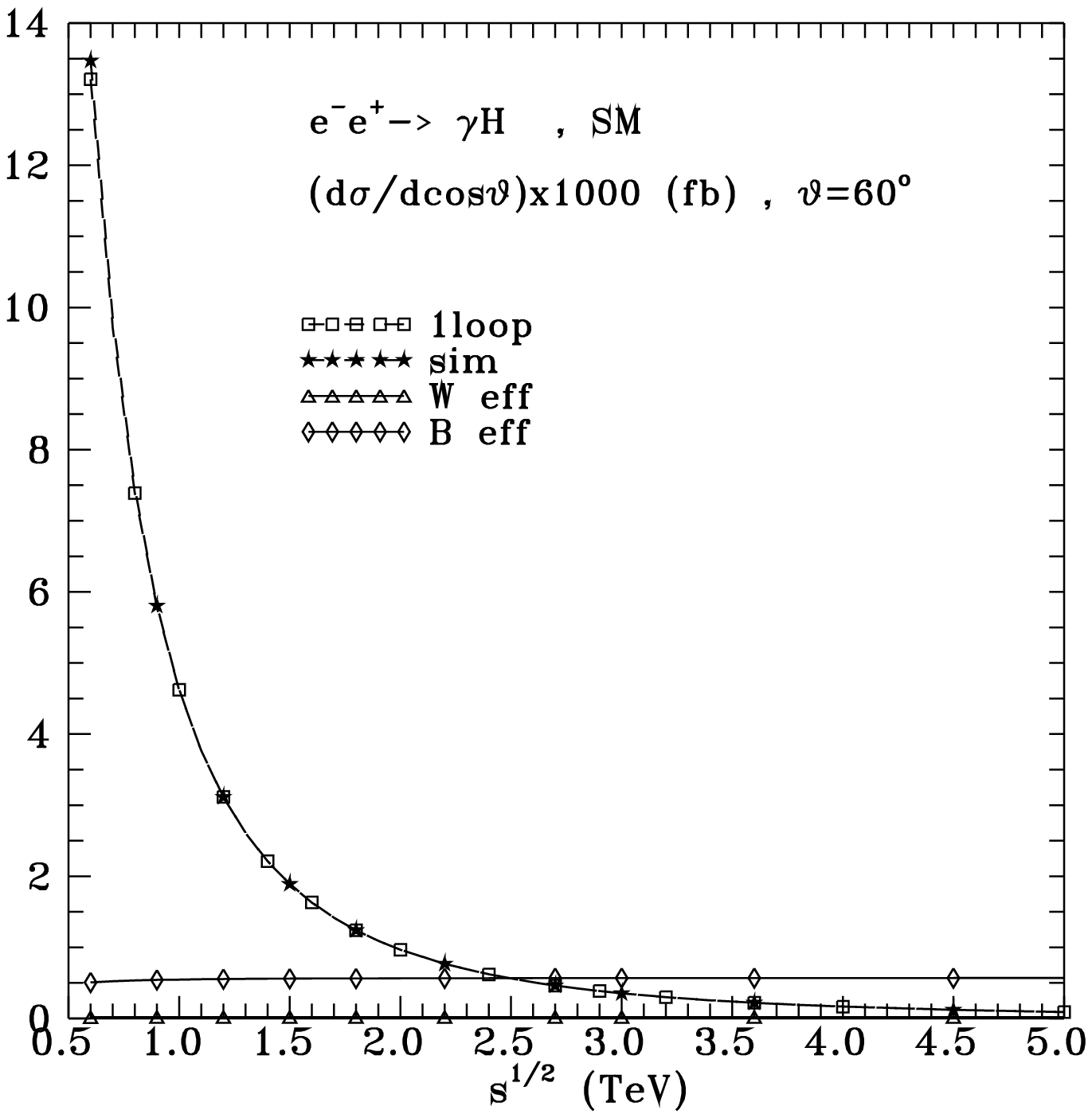, height=6.cm}\hspace{1.cm}
\epsfig{file=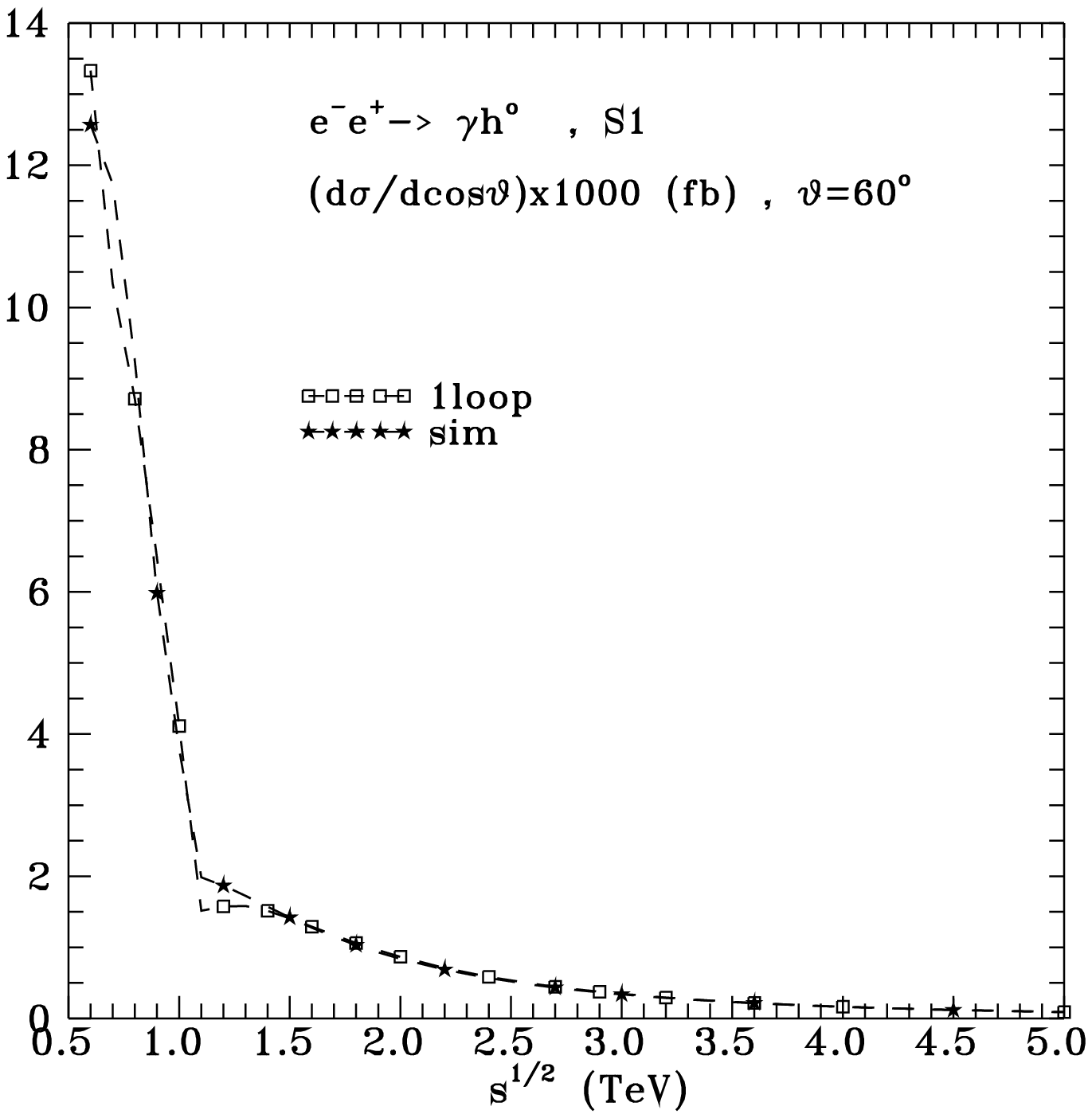,height=6.cm}
\]
\[
\epsfig{file=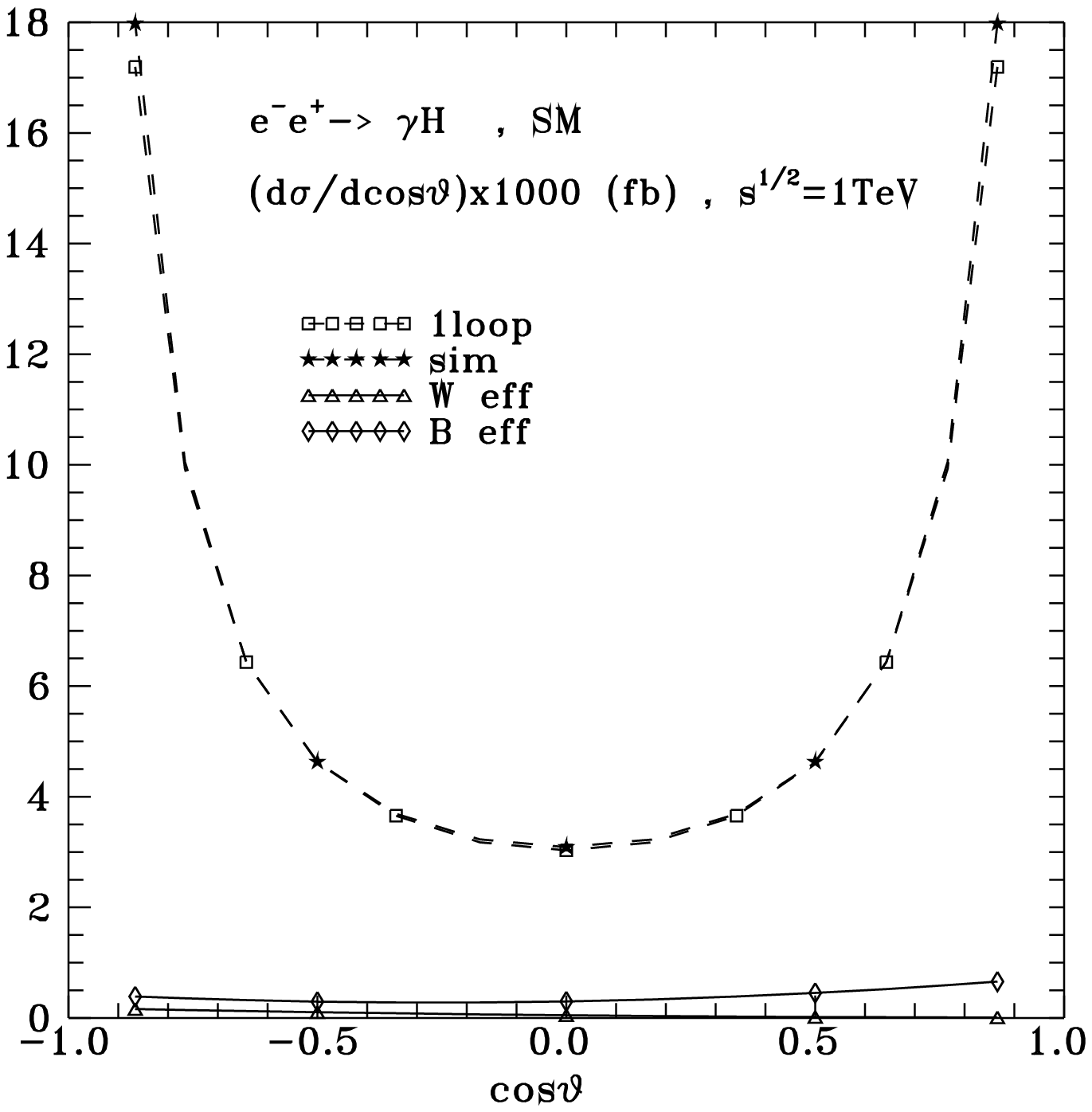, height=6.cm}\hspace{1.cm}
\epsfig{file=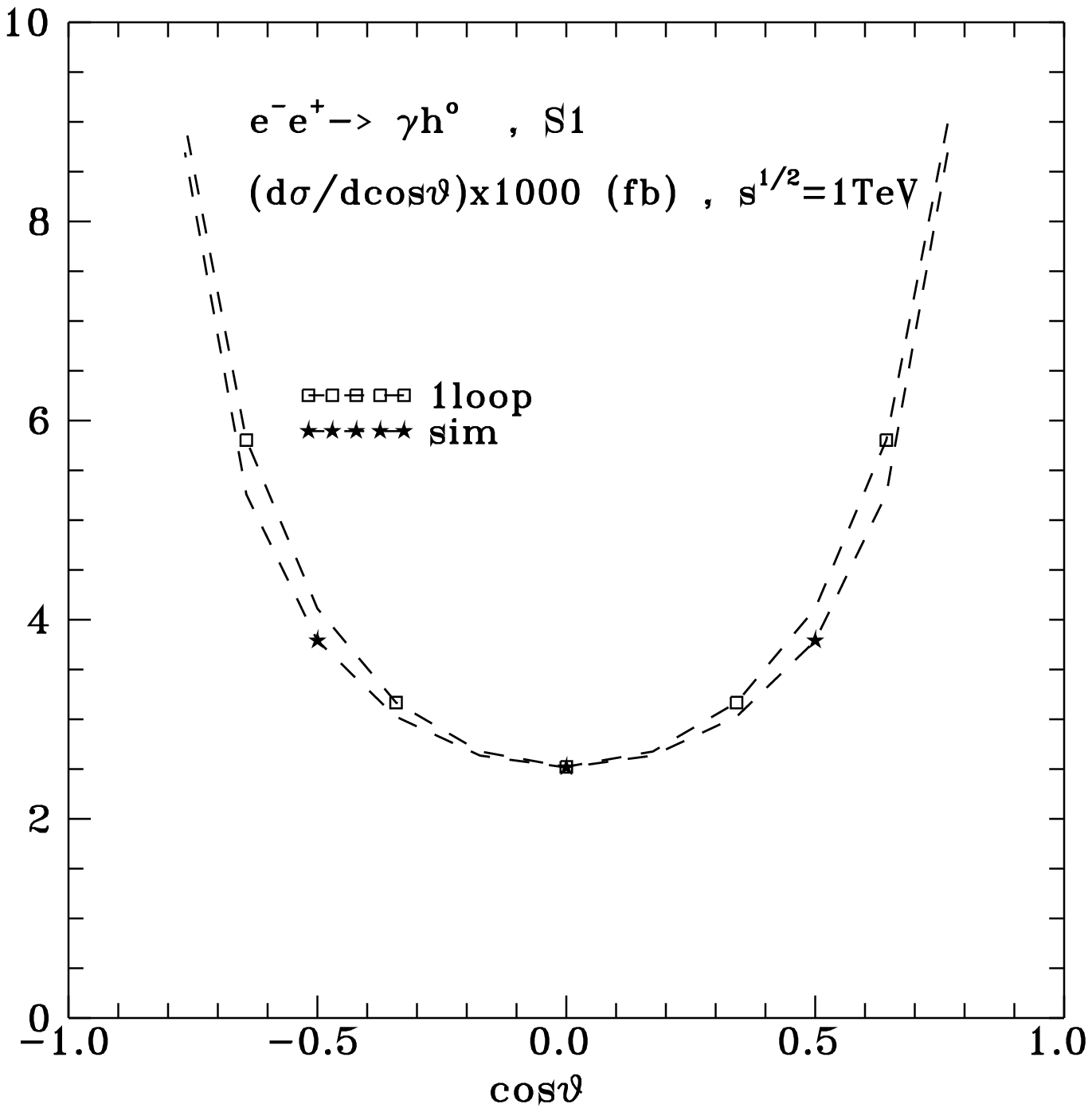,height=6.cm}
\]
\[
\epsfig{file=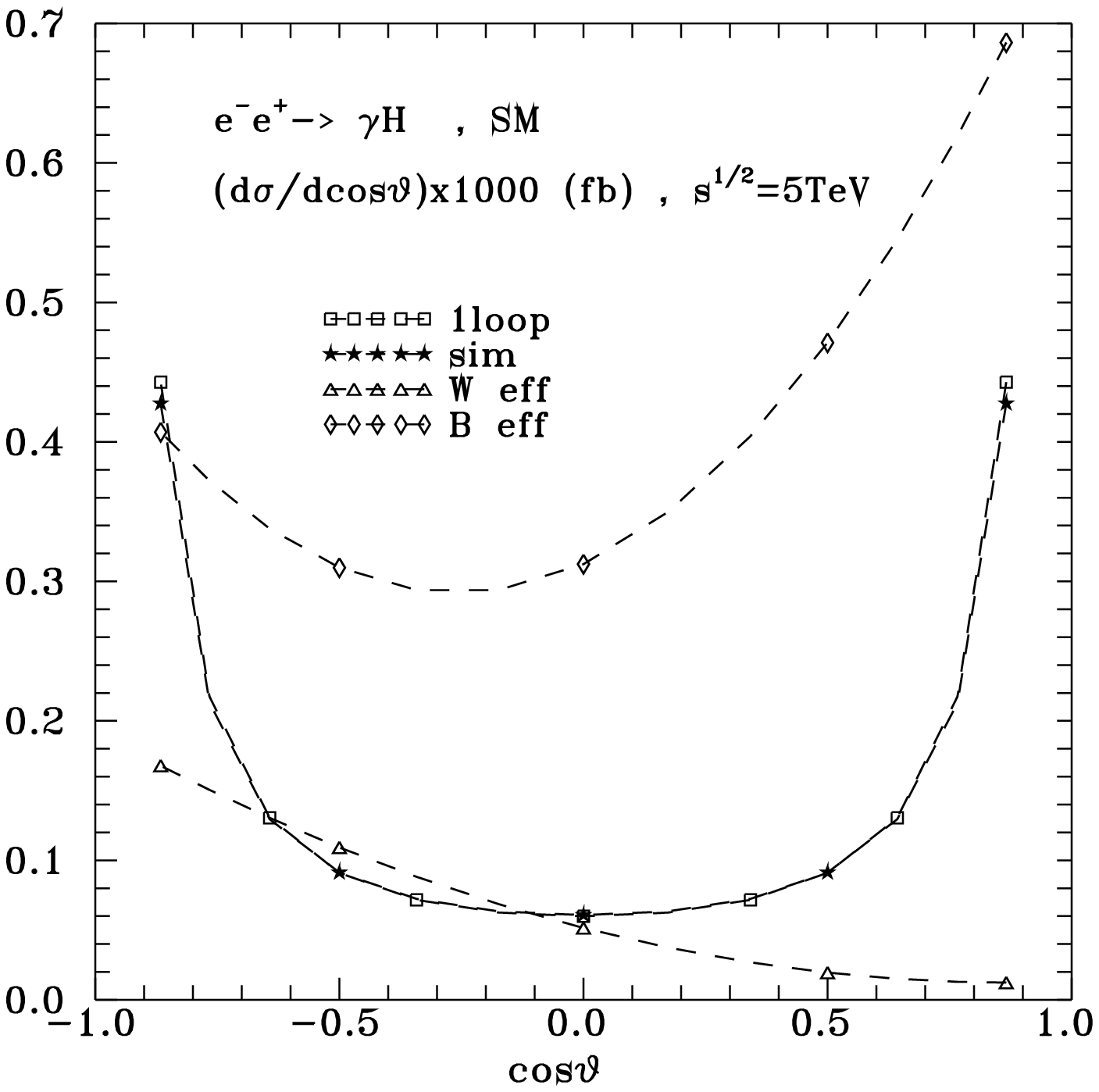, height=6.cm}\hspace{1.cm}
\epsfig{file=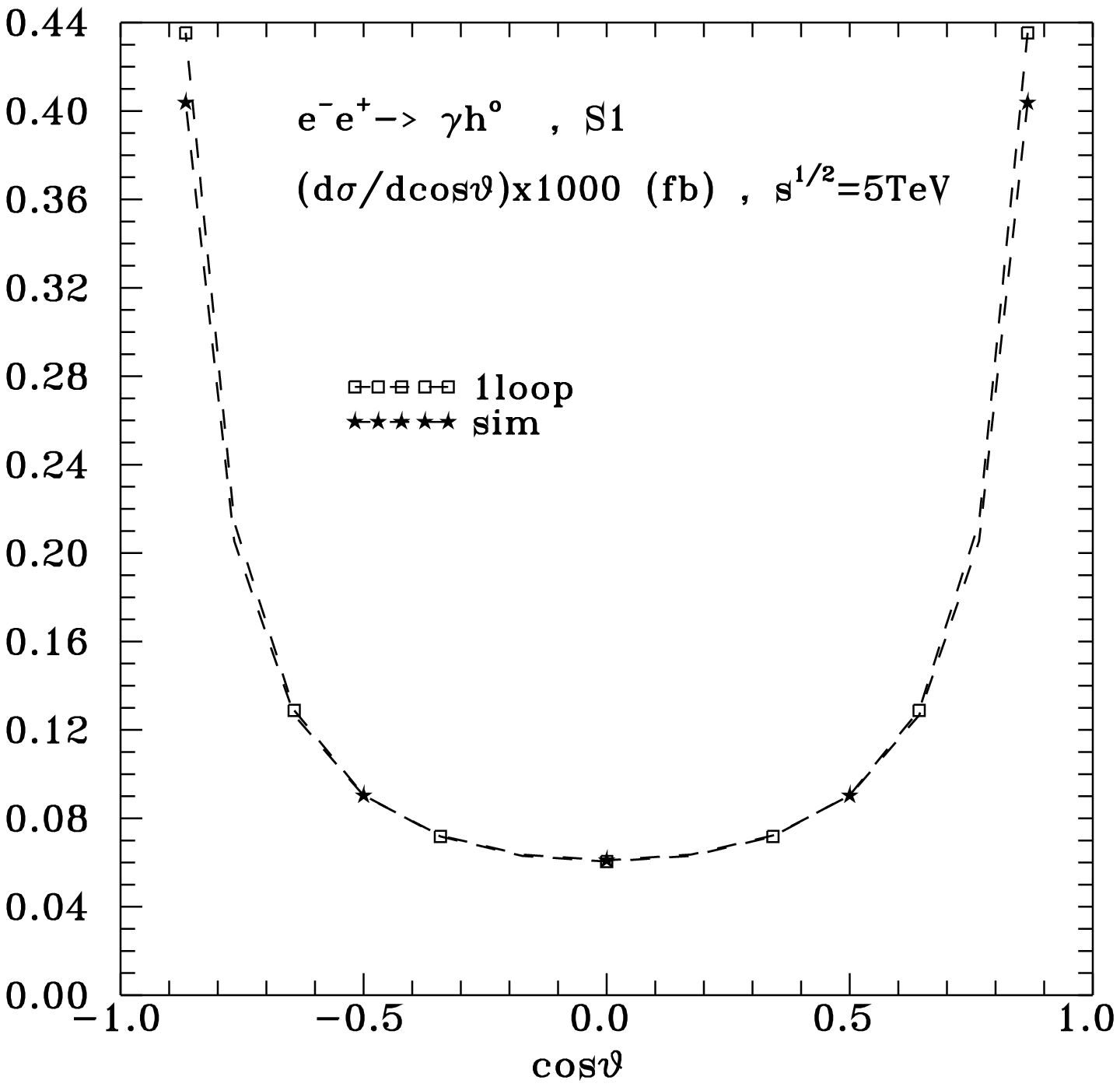,height=6.cm}
\]
\caption[1]{ Cross sections for SM (left panels) and for $h^0$ in S1 MSSM (right panels).
BSM and MSSM parameters as in previous figures.}
\label{sigma-SM-h0}
\end{figure}

\clearpage

\begin{figure}[h]
\vspace{-1cm}
\[
\epsfig{file=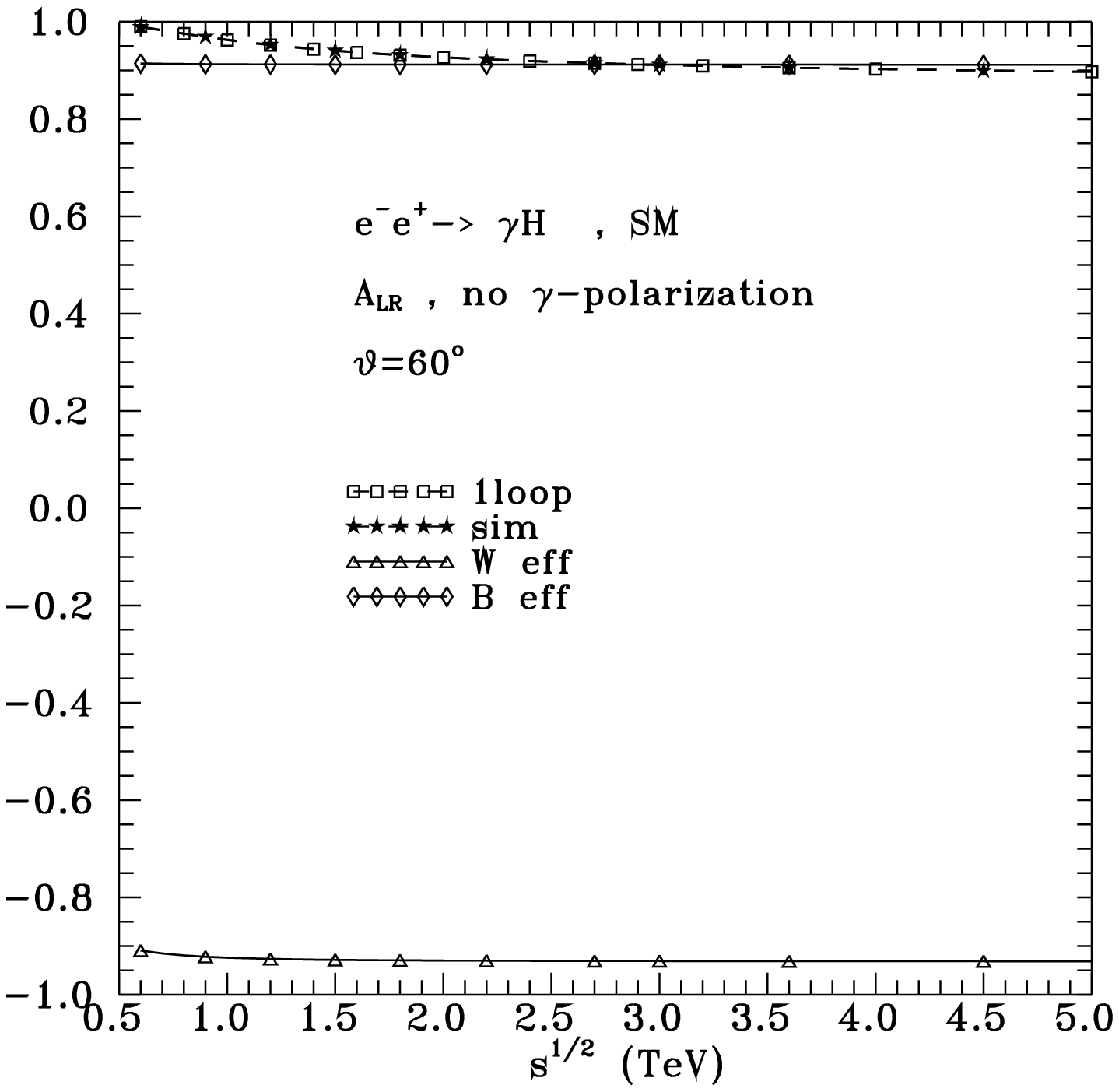, height=6.cm}\hspace{1.cm}
\epsfig{file=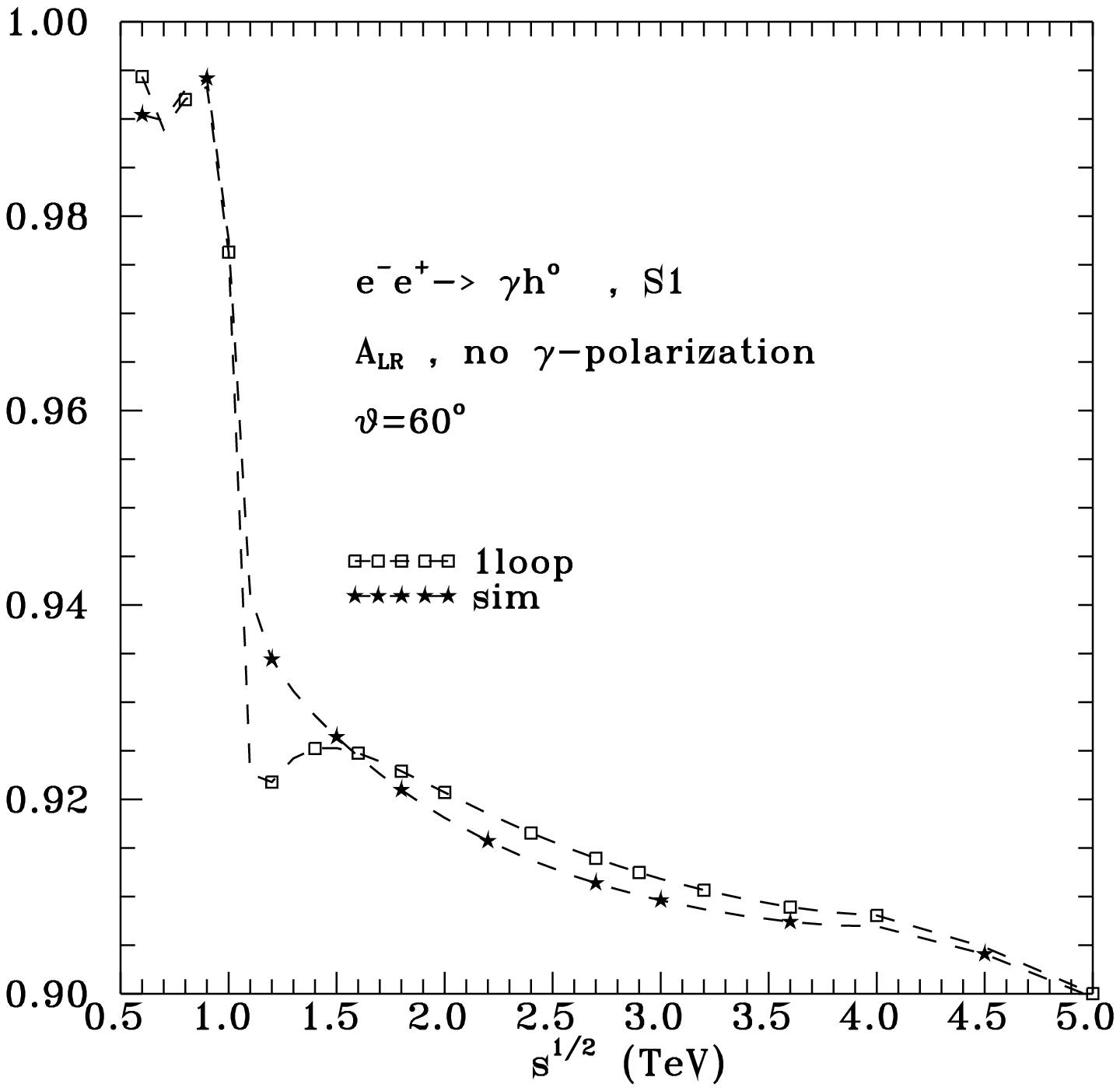,height=6.cm}
\]
\[
\epsfig{file=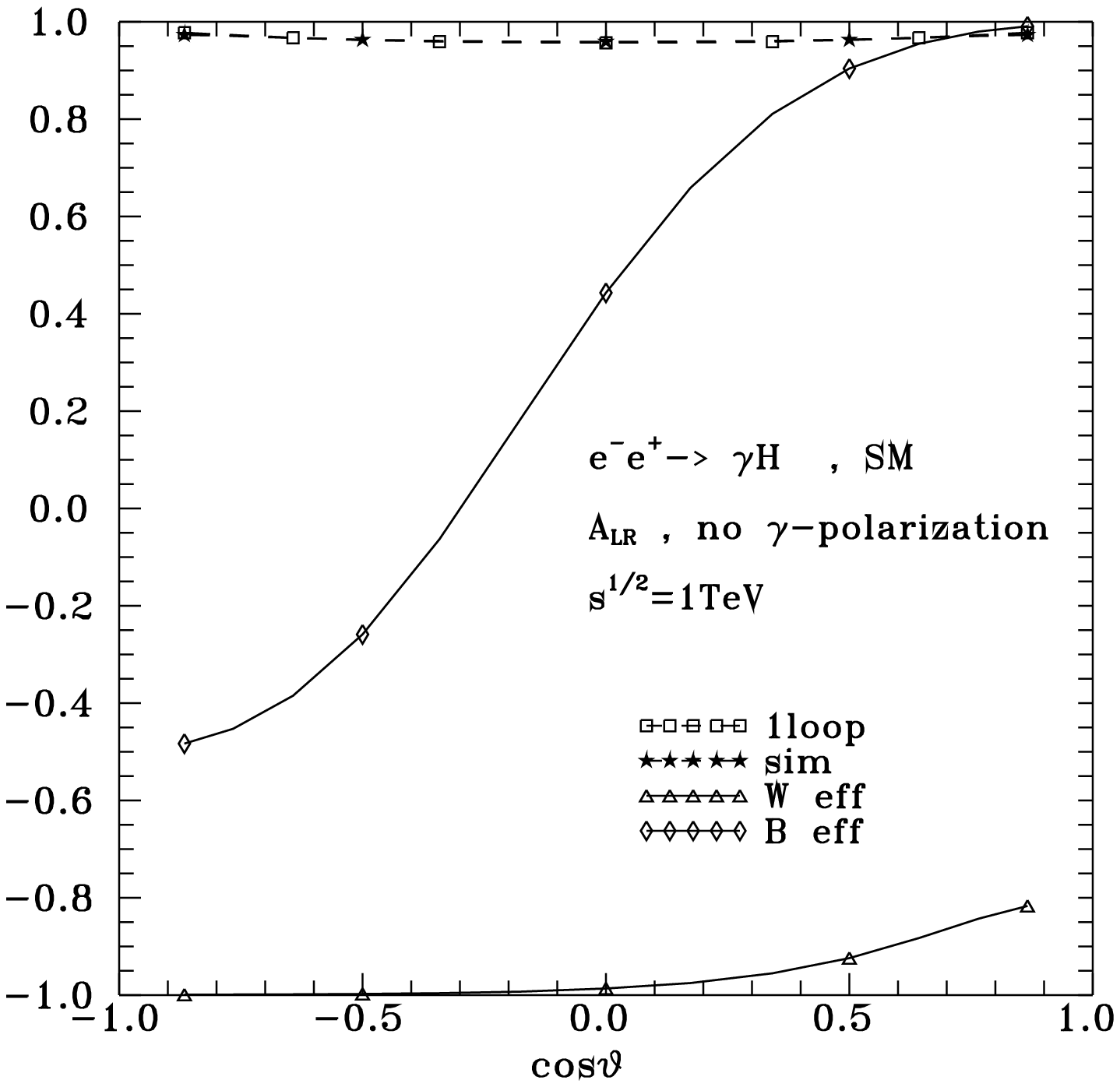, height=6.cm}\hspace{1.cm}
\epsfig{file=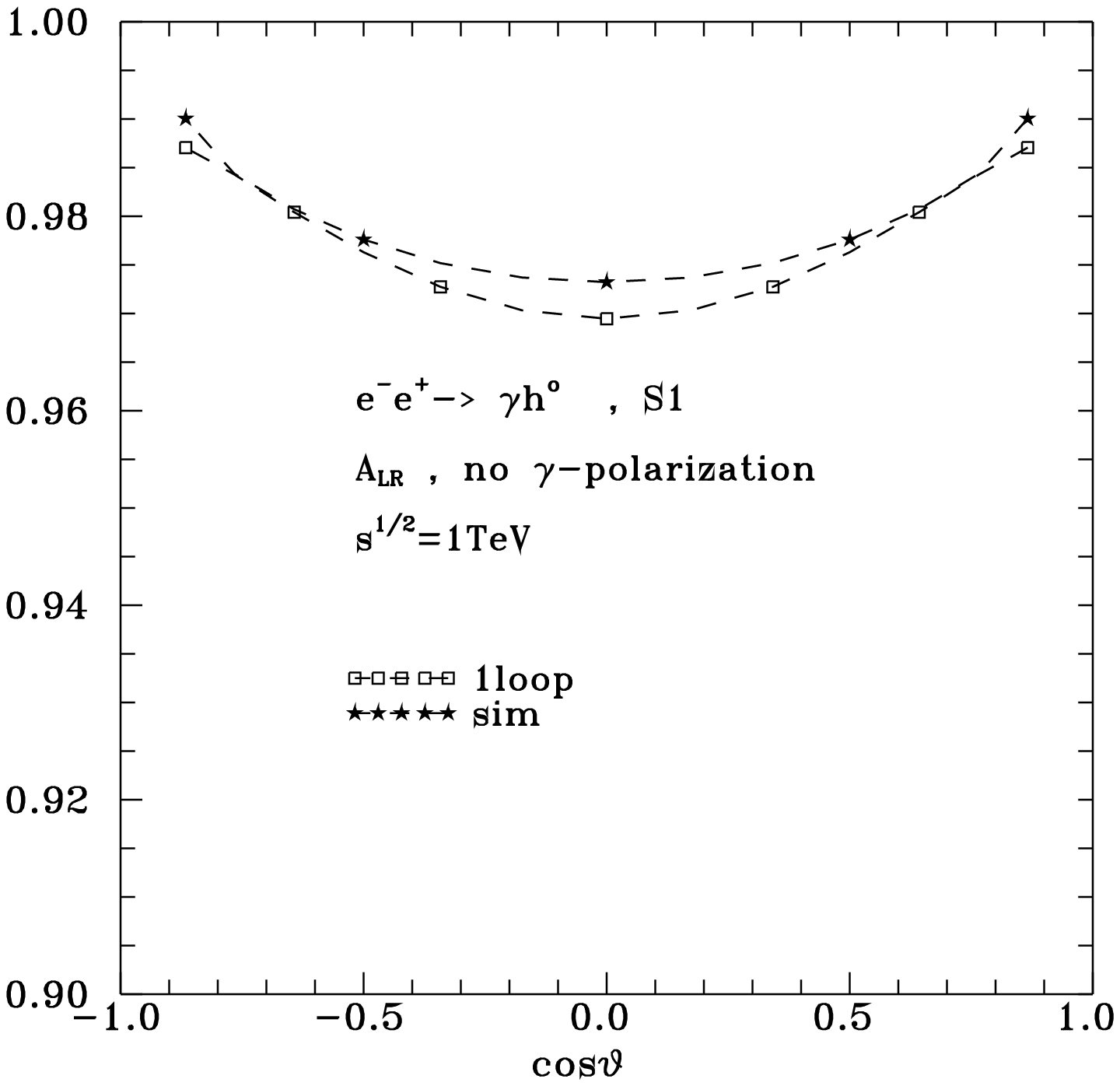,height=6.cm}
\]
\[
\epsfig{file=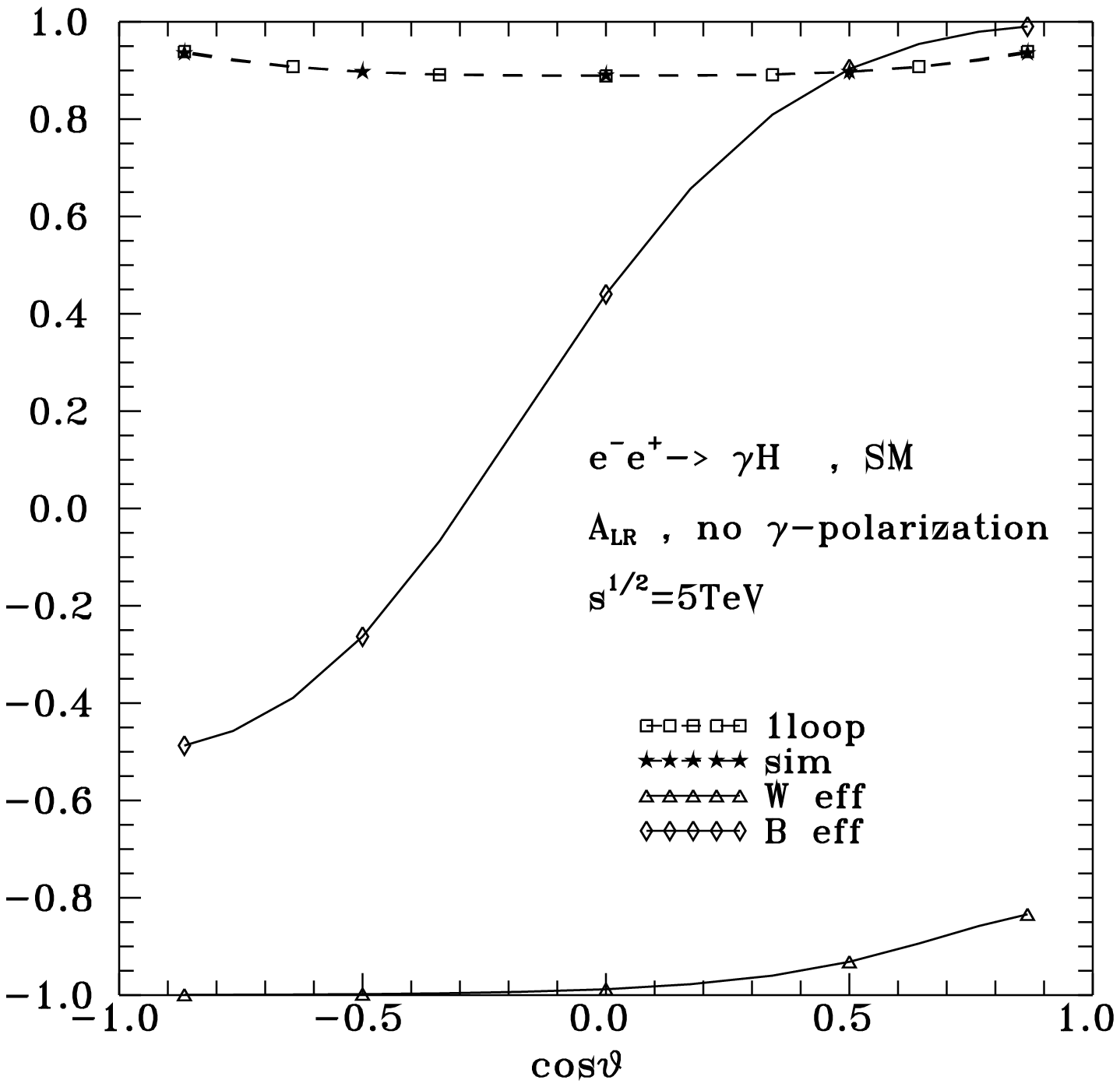, height=6.cm}\hspace{1.cm}
\epsfig{file=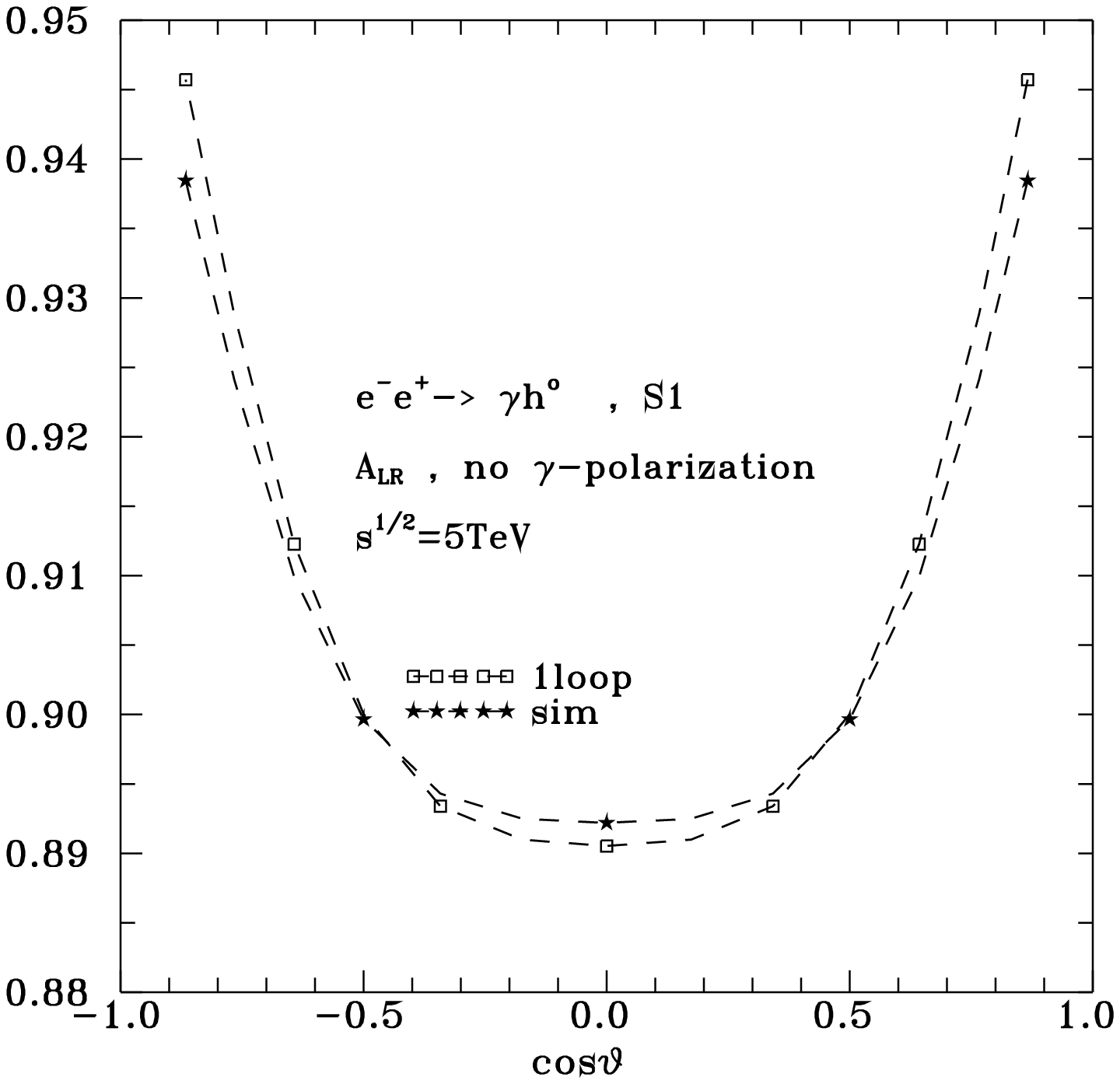,height=6.cm}
\]
\caption[1]{$A_{LR}$ without looking at photon polarization,
for SM (left panels) and for $h^0$ in S1 MSSM (right panels).
BSM and MSSM parameters as in previous figures.}
\label{ALR-nopol-SM-h0}
\end{figure}

\clearpage

\begin{figure}[h]
\vspace{-1cm}
\[
\epsfig{file=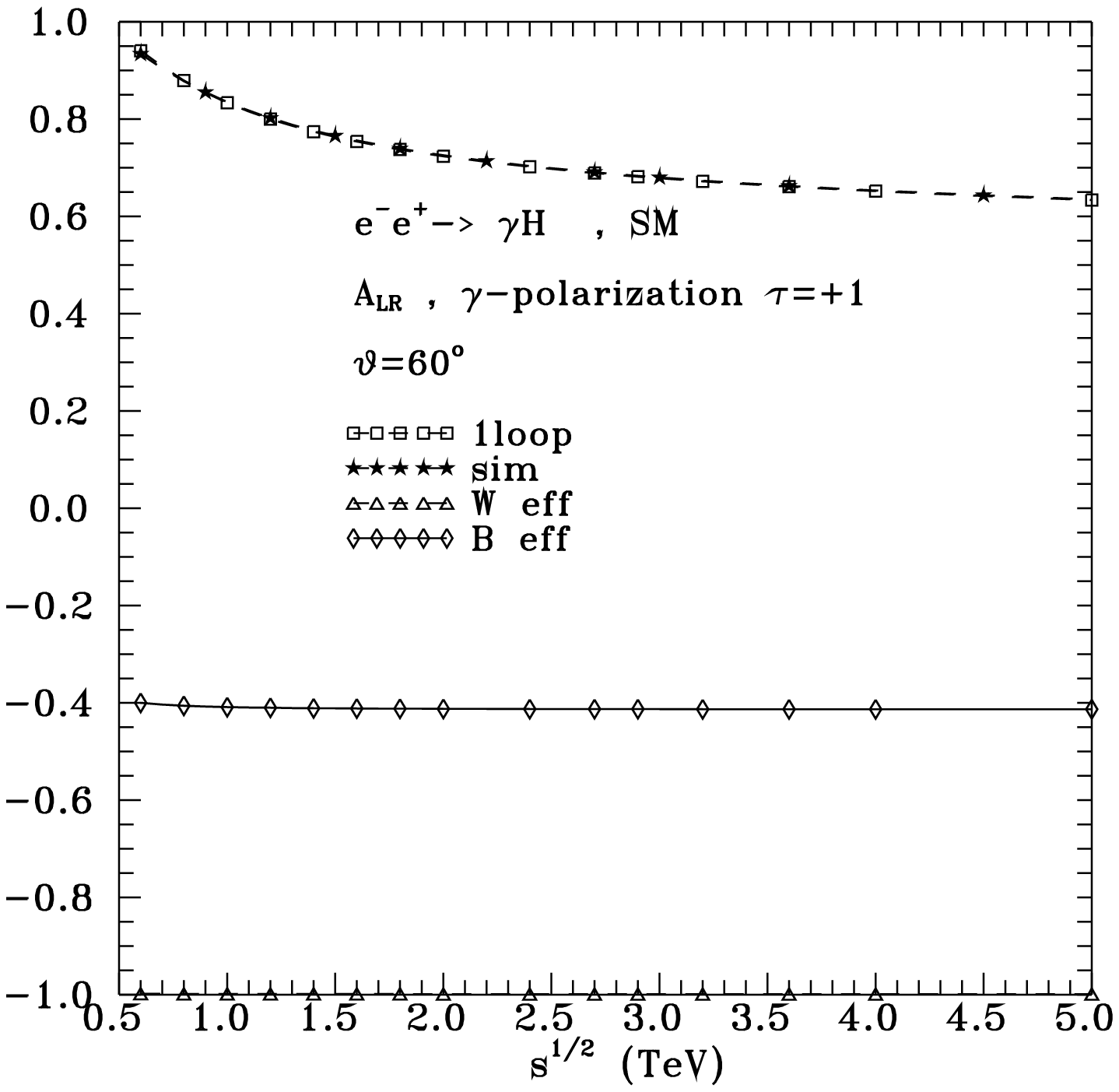, height=6.cm}\hspace{1.cm}
\epsfig{file=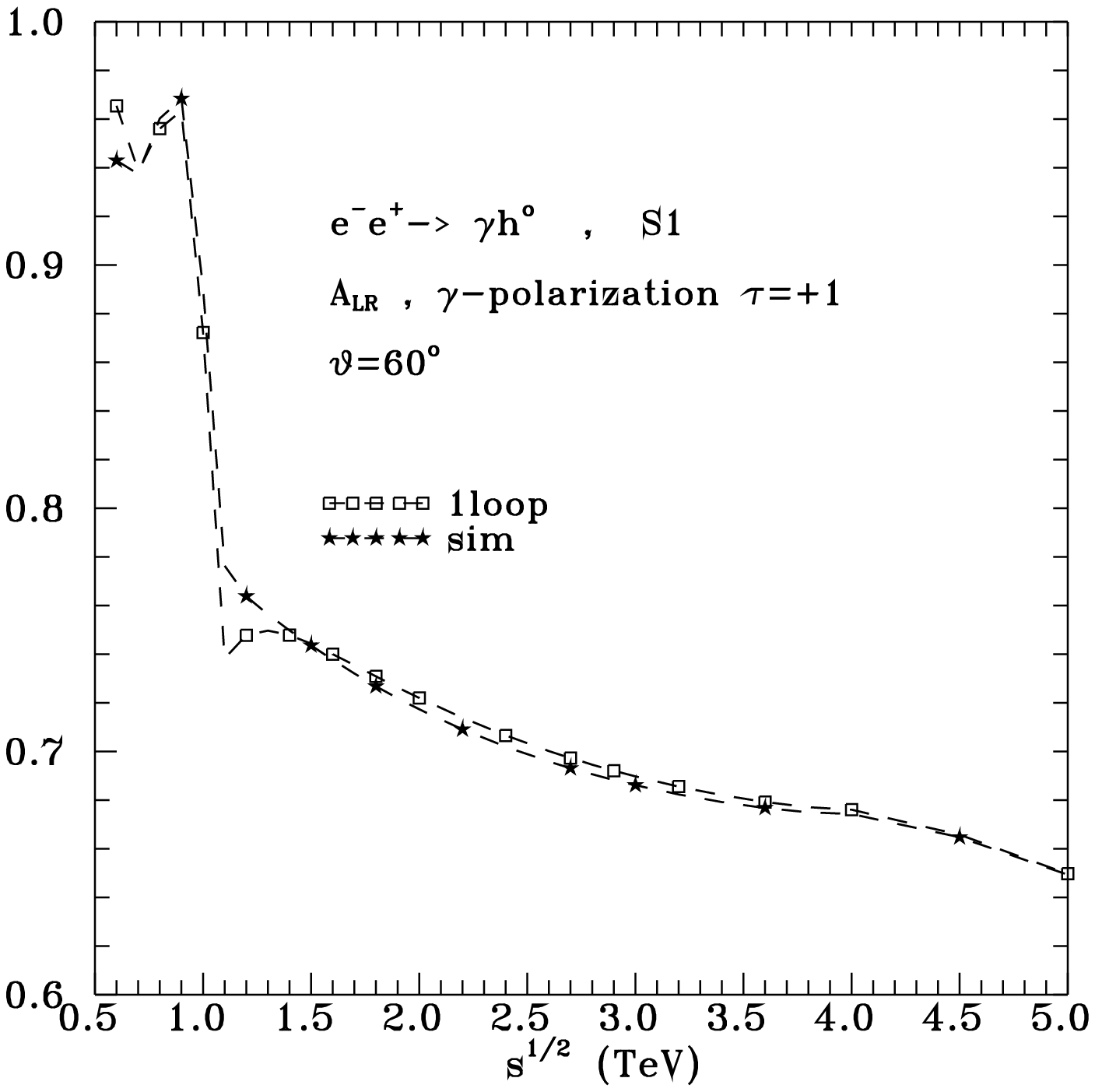,height=6.cm}
\]
\[
\epsfig{file=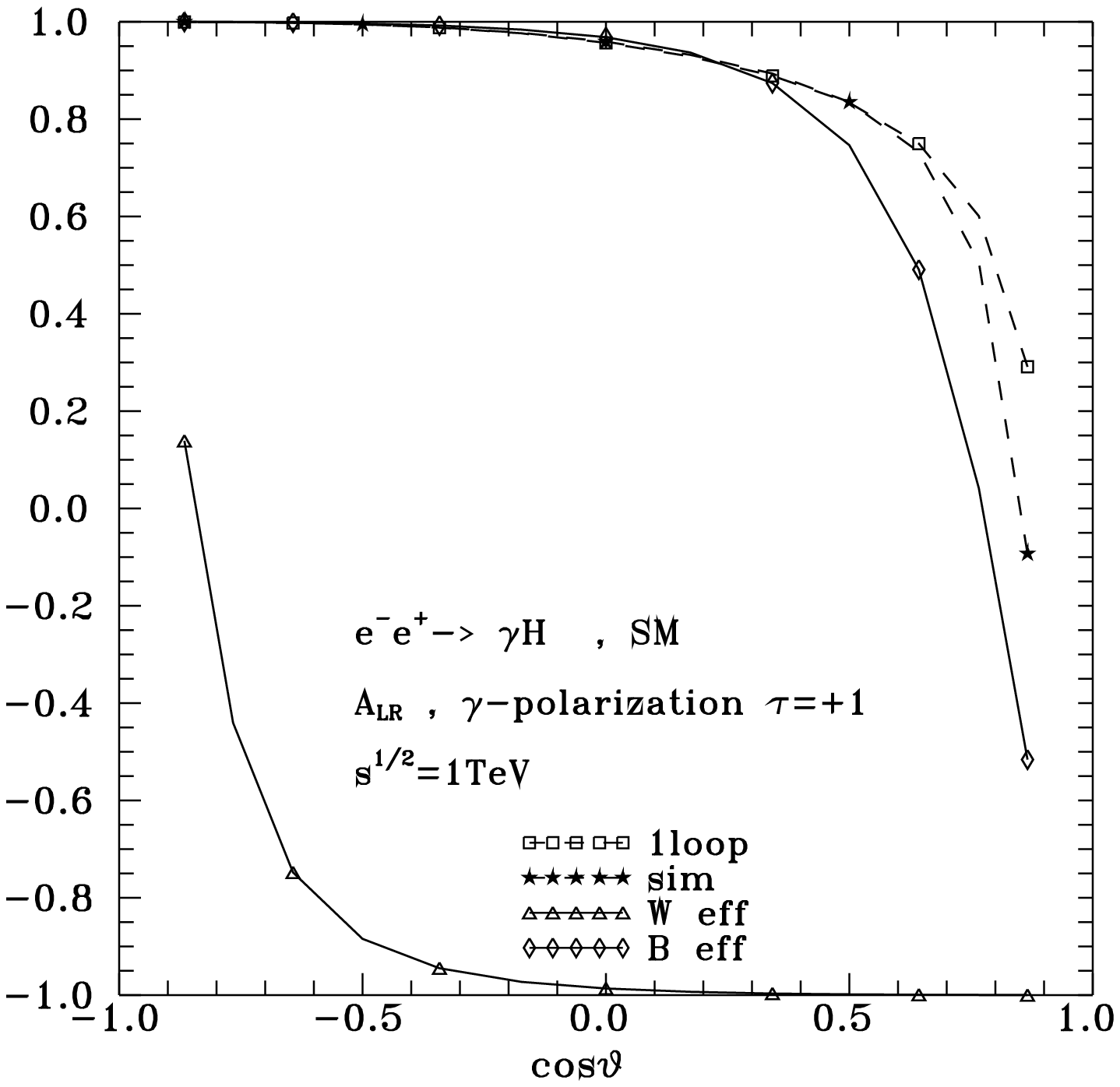, height=6.cm}\hspace{1.cm}
\epsfig{file=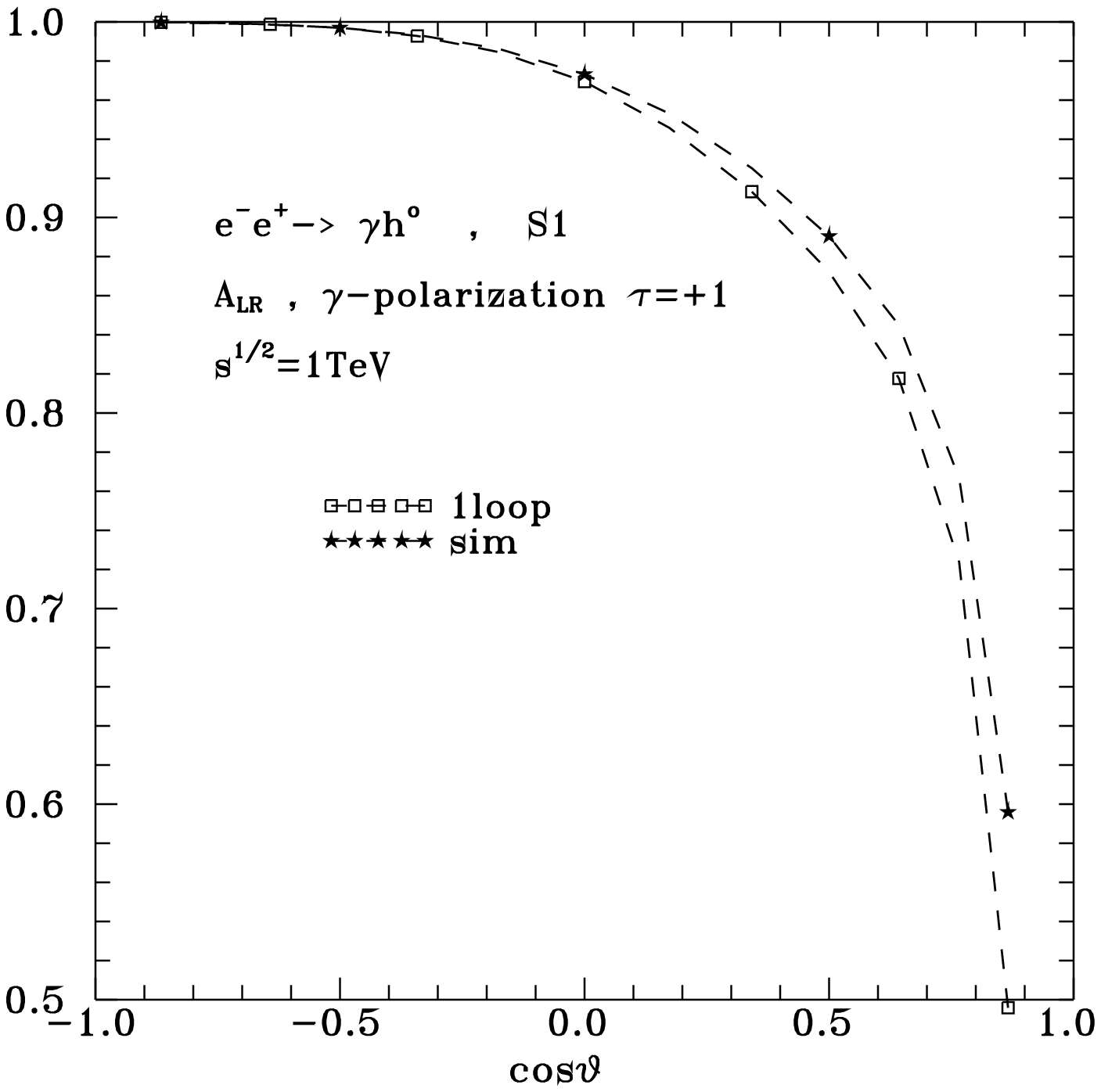,height=6.cm}
\]
\[
\epsfig{file=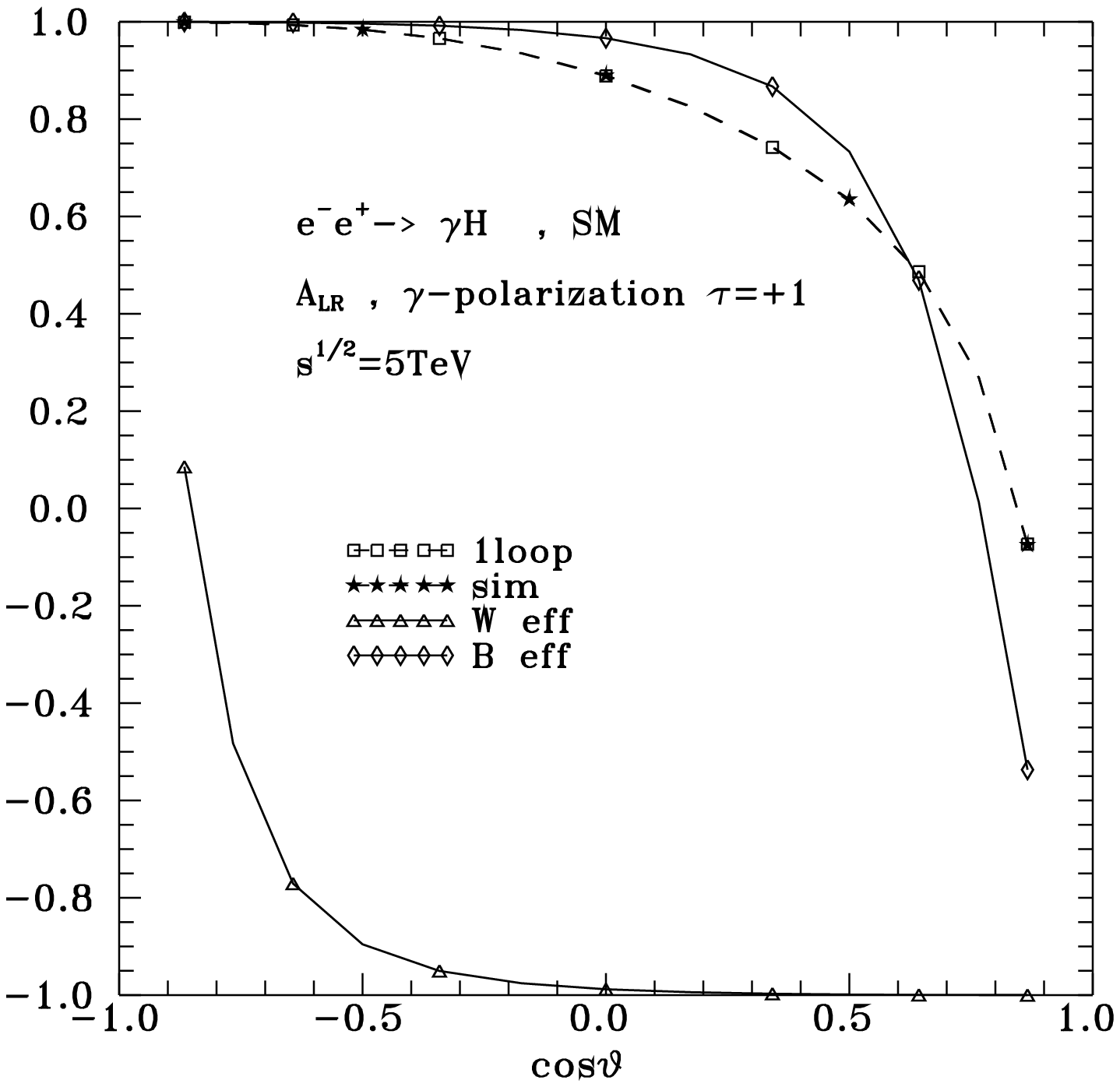, height=6.cm}\hspace{1.cm}
\epsfig{file=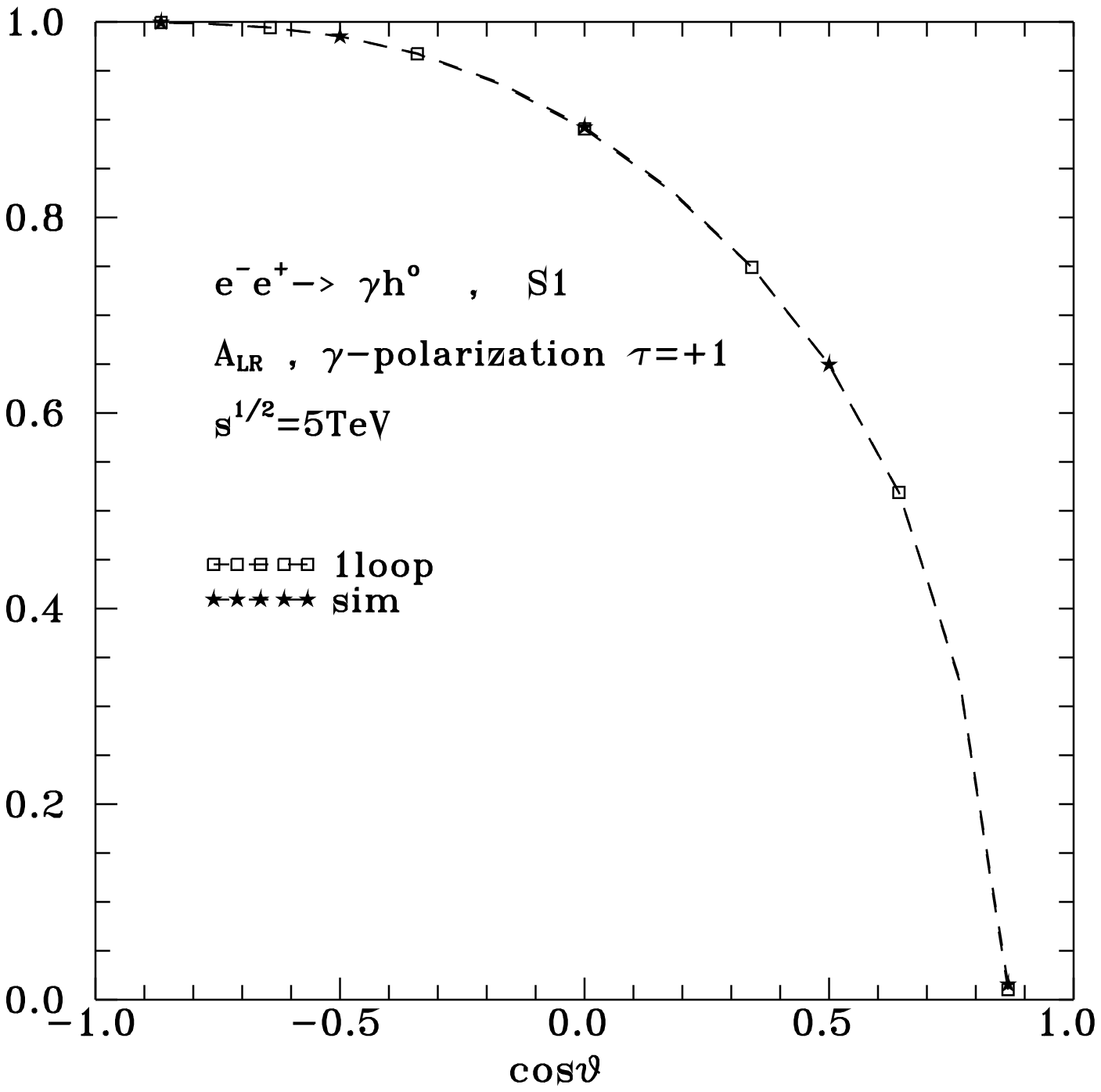,height=6.cm}
\]
\caption[1]{$A_{LR}$ for a photon with helicity $\tau=+1$,
for SM (left panels) and for $h^0$ in S1 MSSM (right panels).
BSM and MSSM parameters as in previous figures.}
\label{ALR-tauplus-SM-h0}
\end{figure}

\clearpage

\begin{figure}[h]
\vspace{-1cm}
\[
\epsfig{file=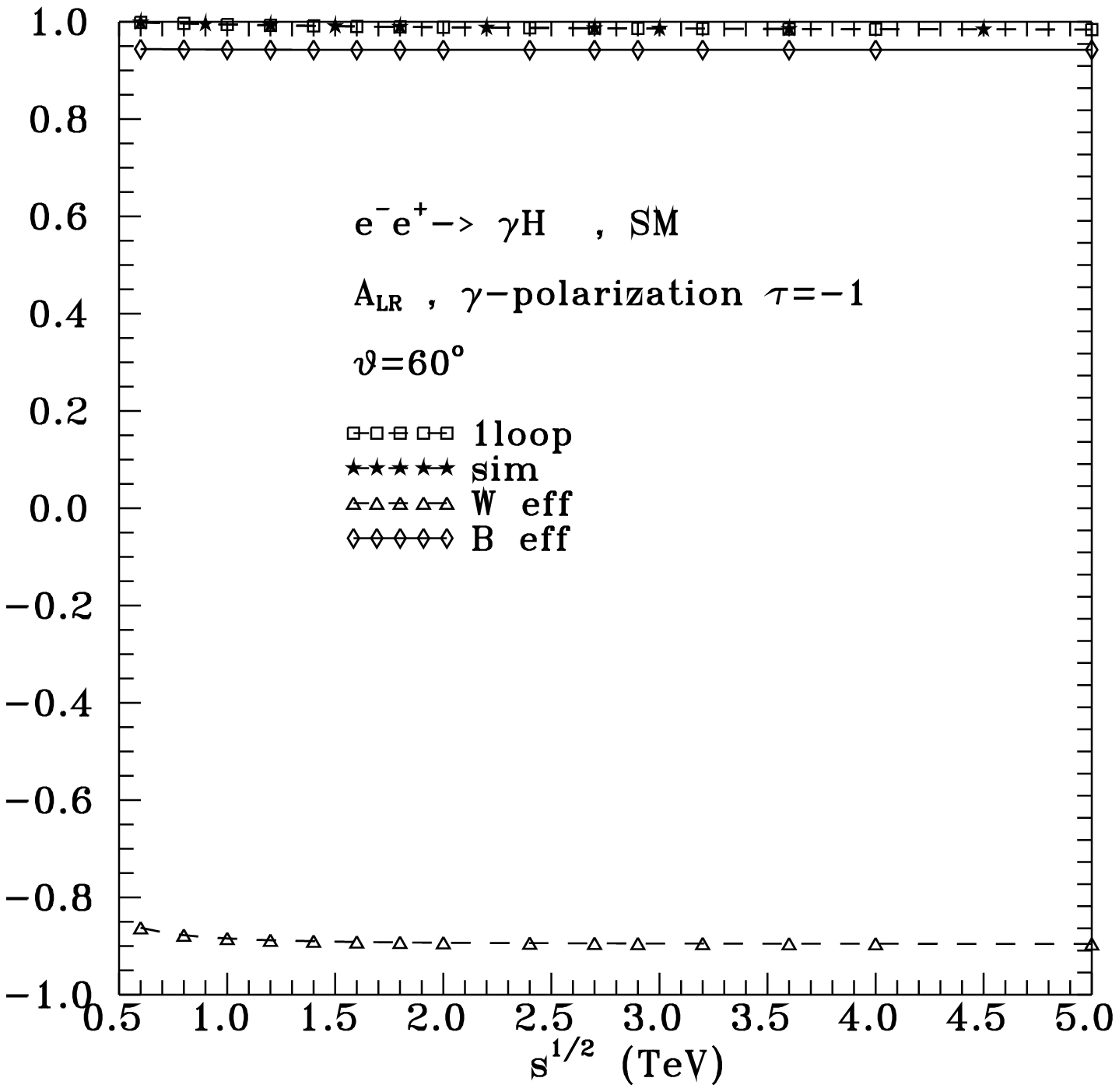, height=6.cm}\hspace{1.cm}
\epsfig{file=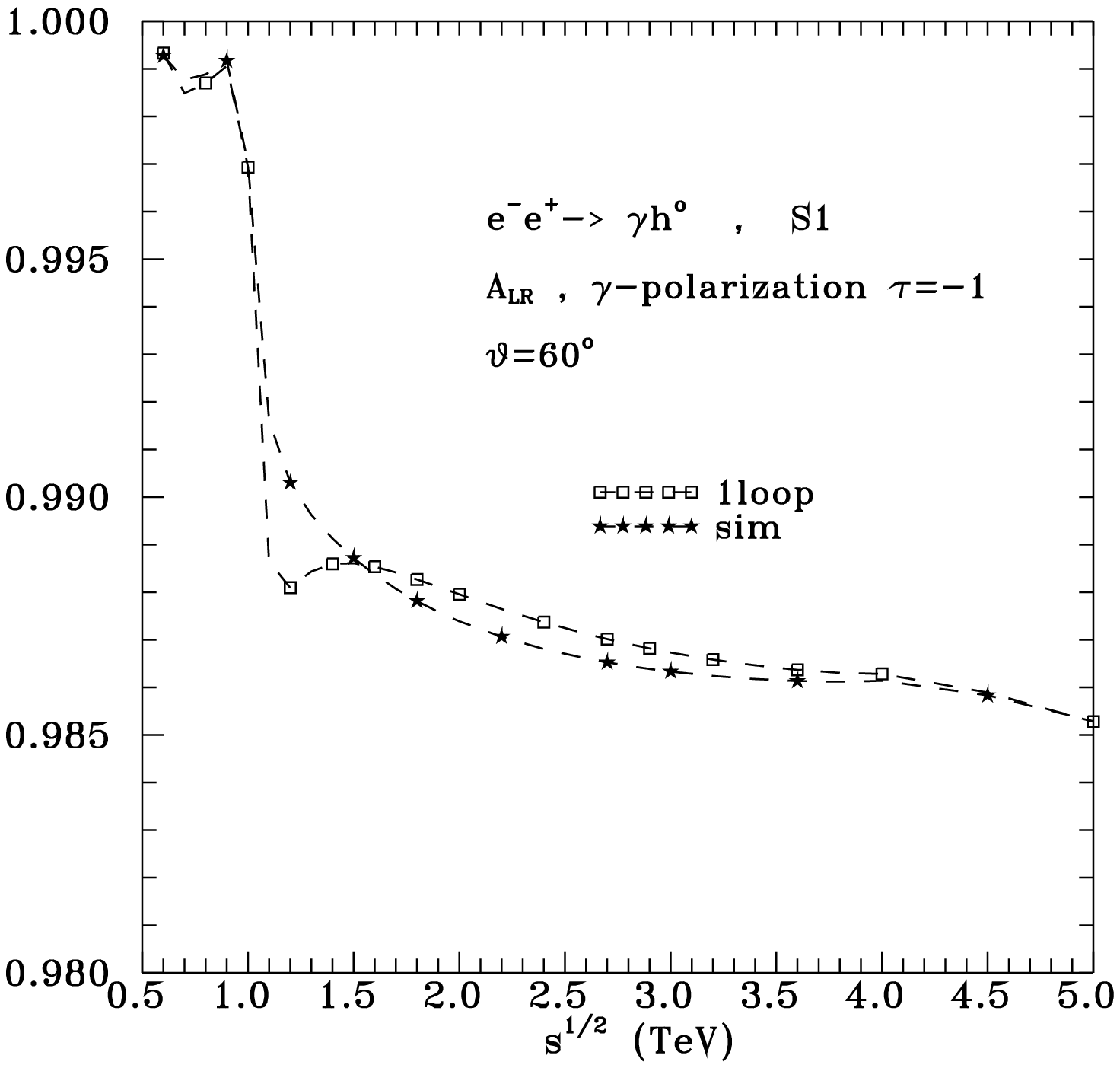,height=6.cm}
\]
\[
\epsfig{file=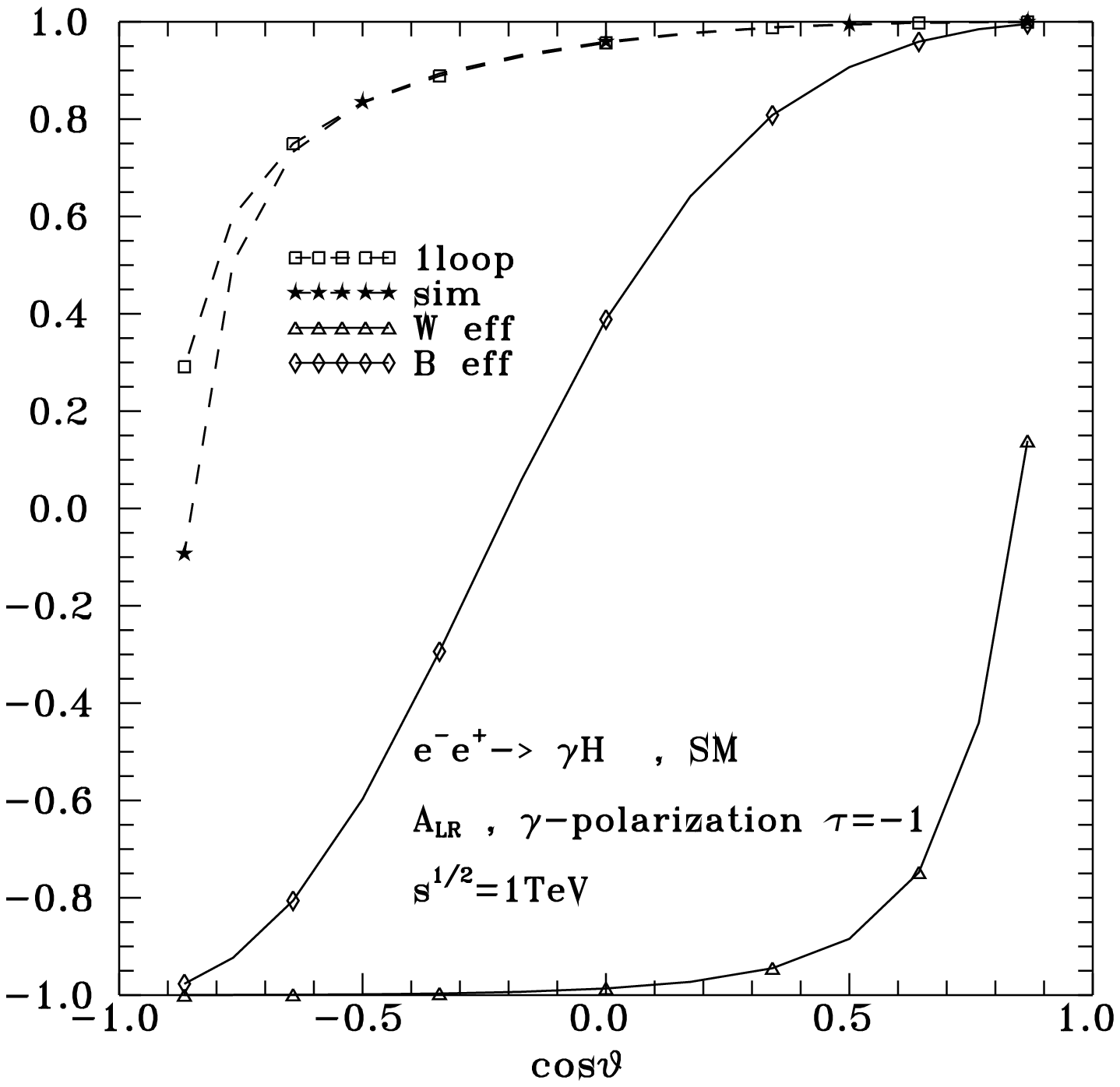, height=6.cm}\hspace{1.cm}
\epsfig{file=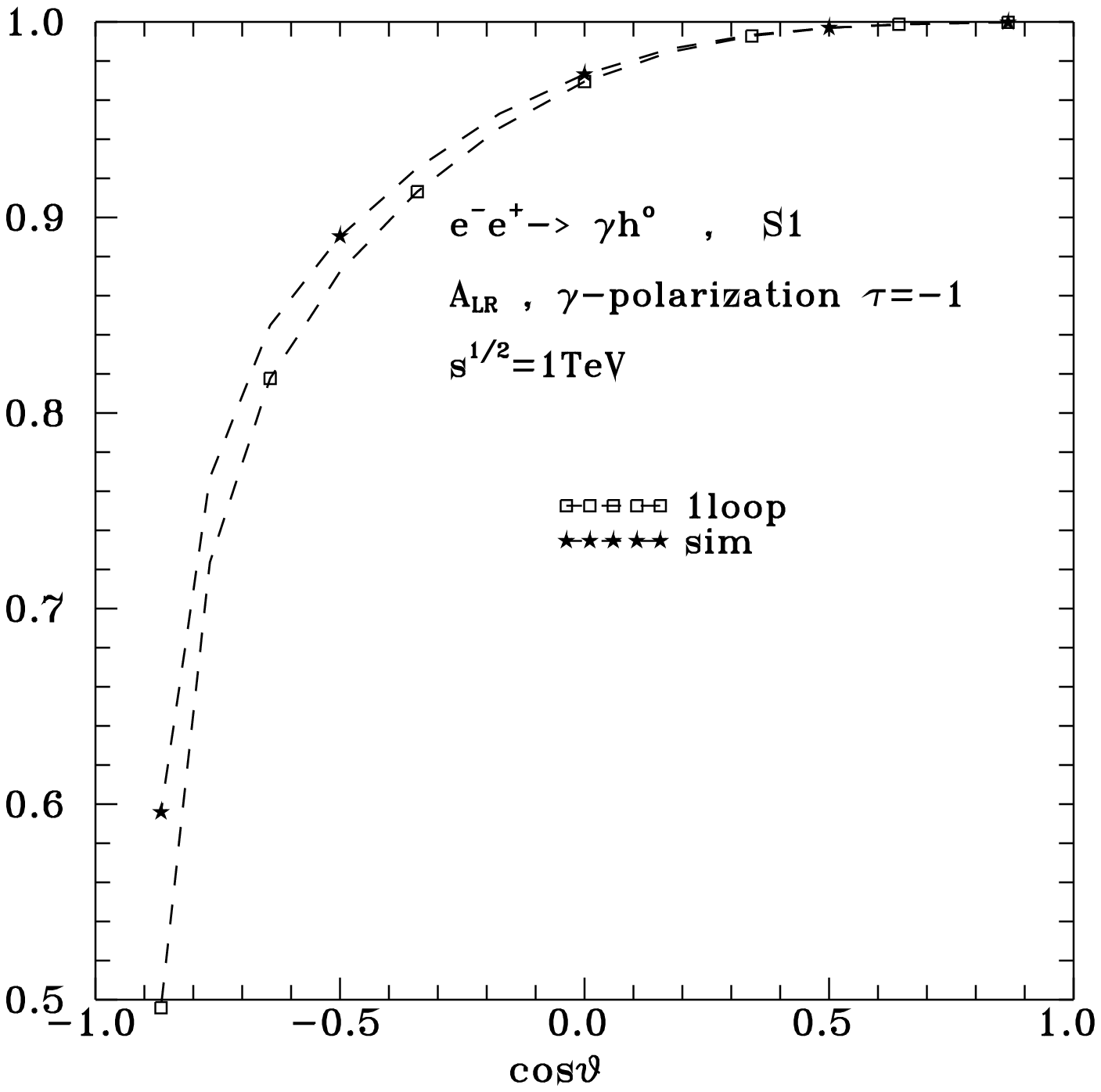,height=6.cm}
\]
\[
\epsfig{file=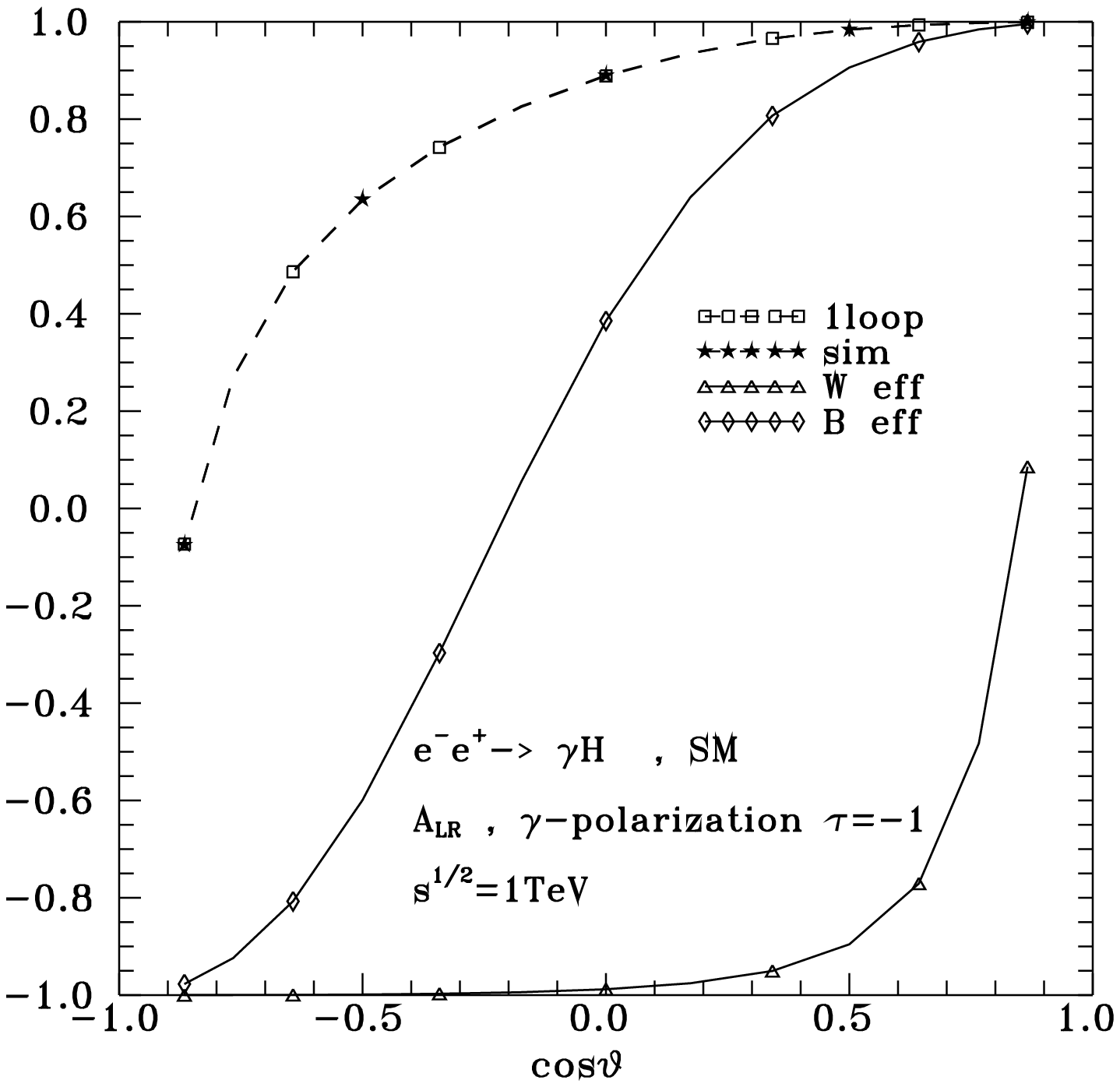, height=6.cm}\hspace{1.cm}
\epsfig{file=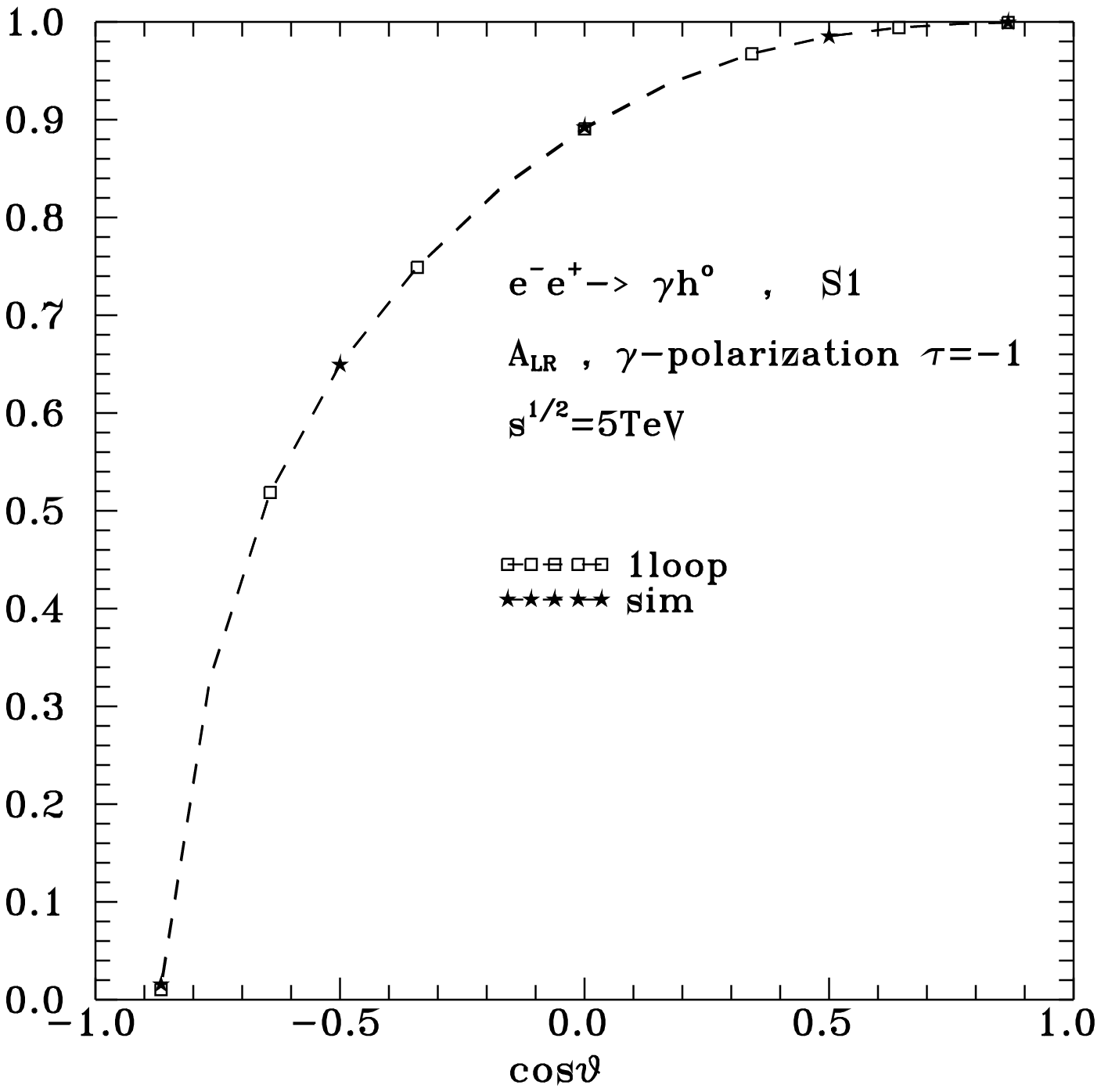,height=6.cm}
\]
\caption[1]{$A_{LR}$ for a photon with helicity $\tau=-1$,
for SM (left panels) and for $h^0$ in S1 MSSM (right panels).
BSM and MSSM parameters as in previous figures.}
\label{ALR-tauminus-SM-h0}
\end{figure}

\clearpage

\begin{figure}[h]
\vspace{-1cm}
\[
\epsfig{file=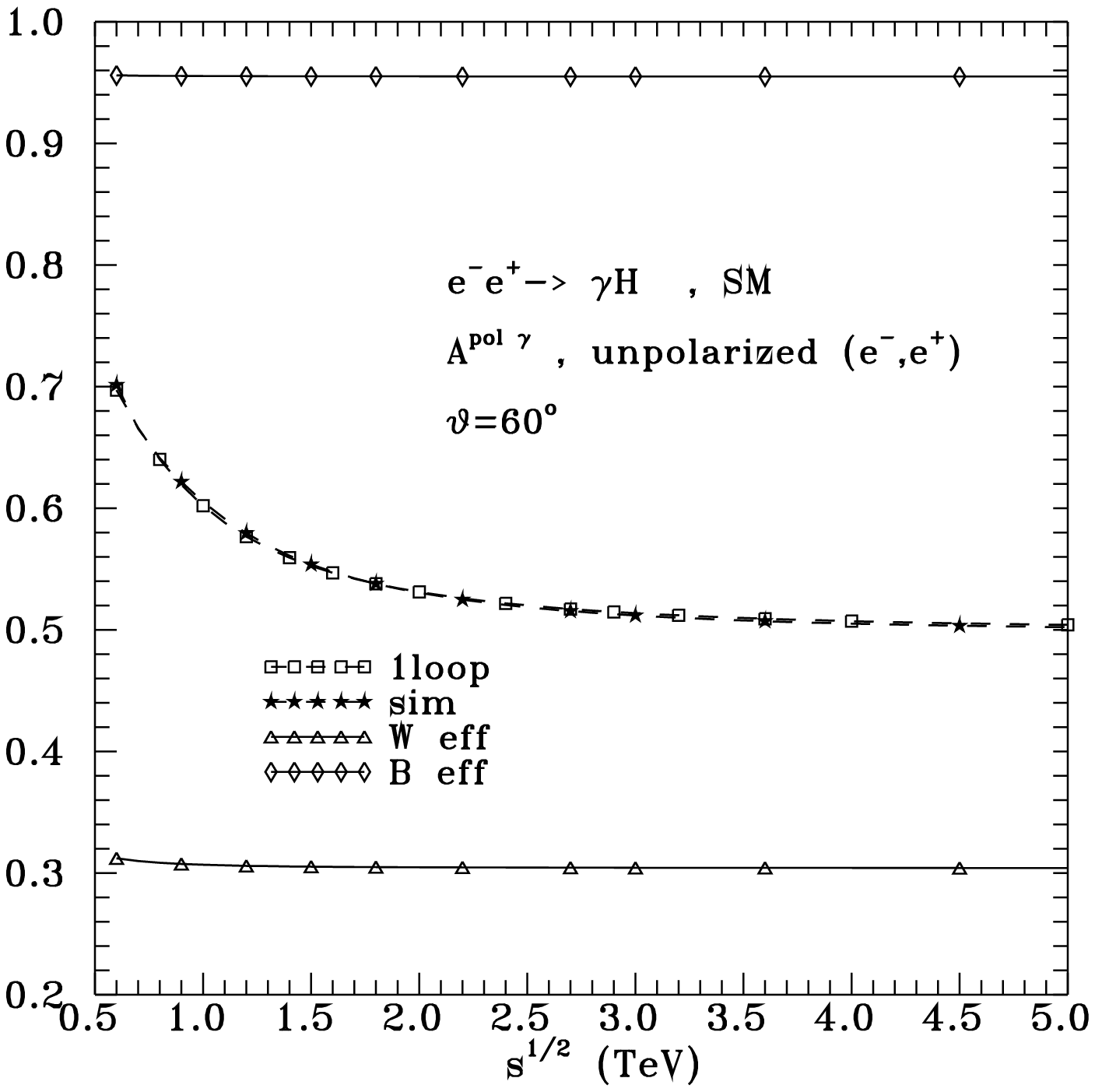, height=6.cm}\hspace{1.cm}
\epsfig{file=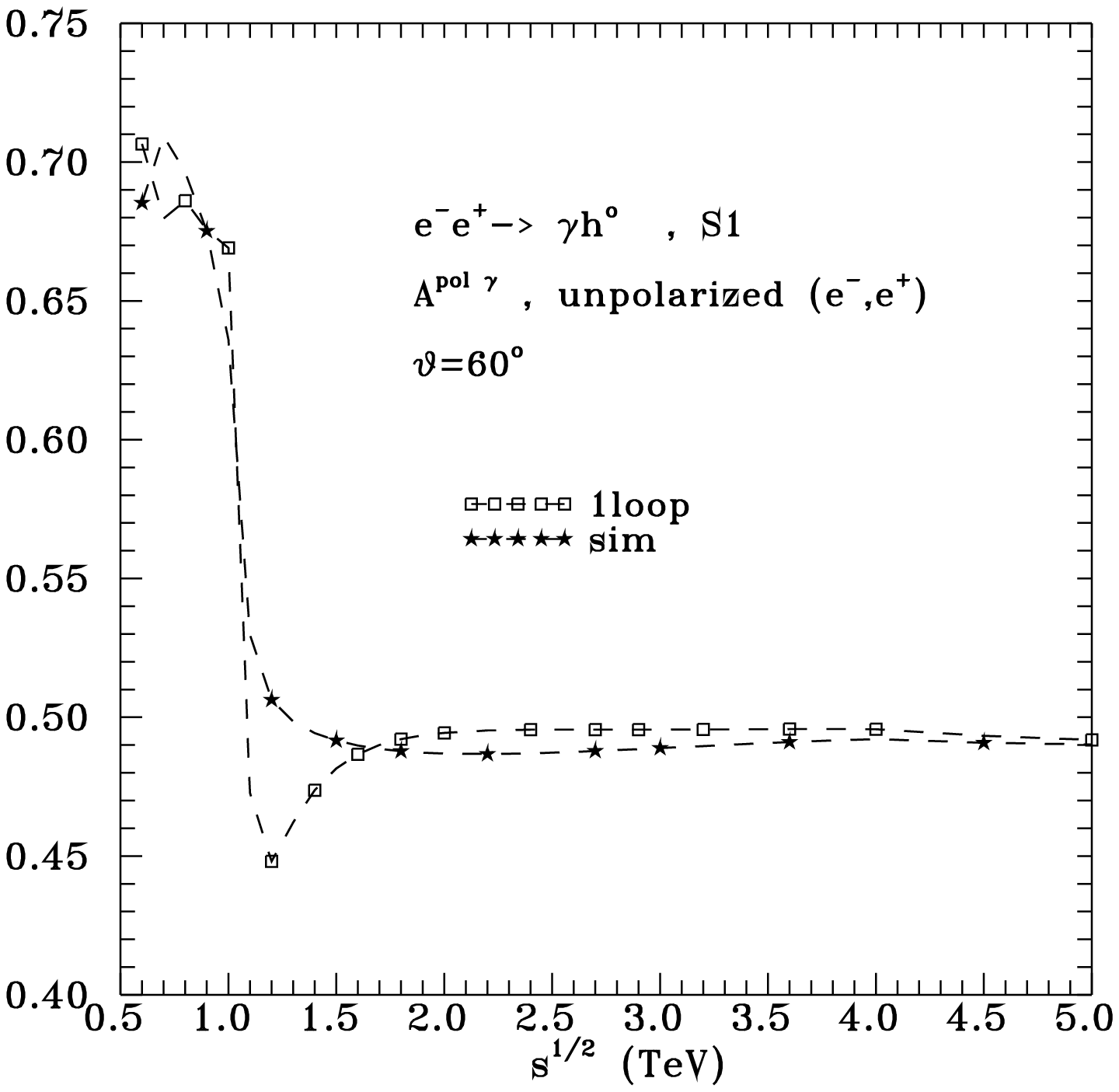,height=6.cm}
\]
\[
\epsfig{file=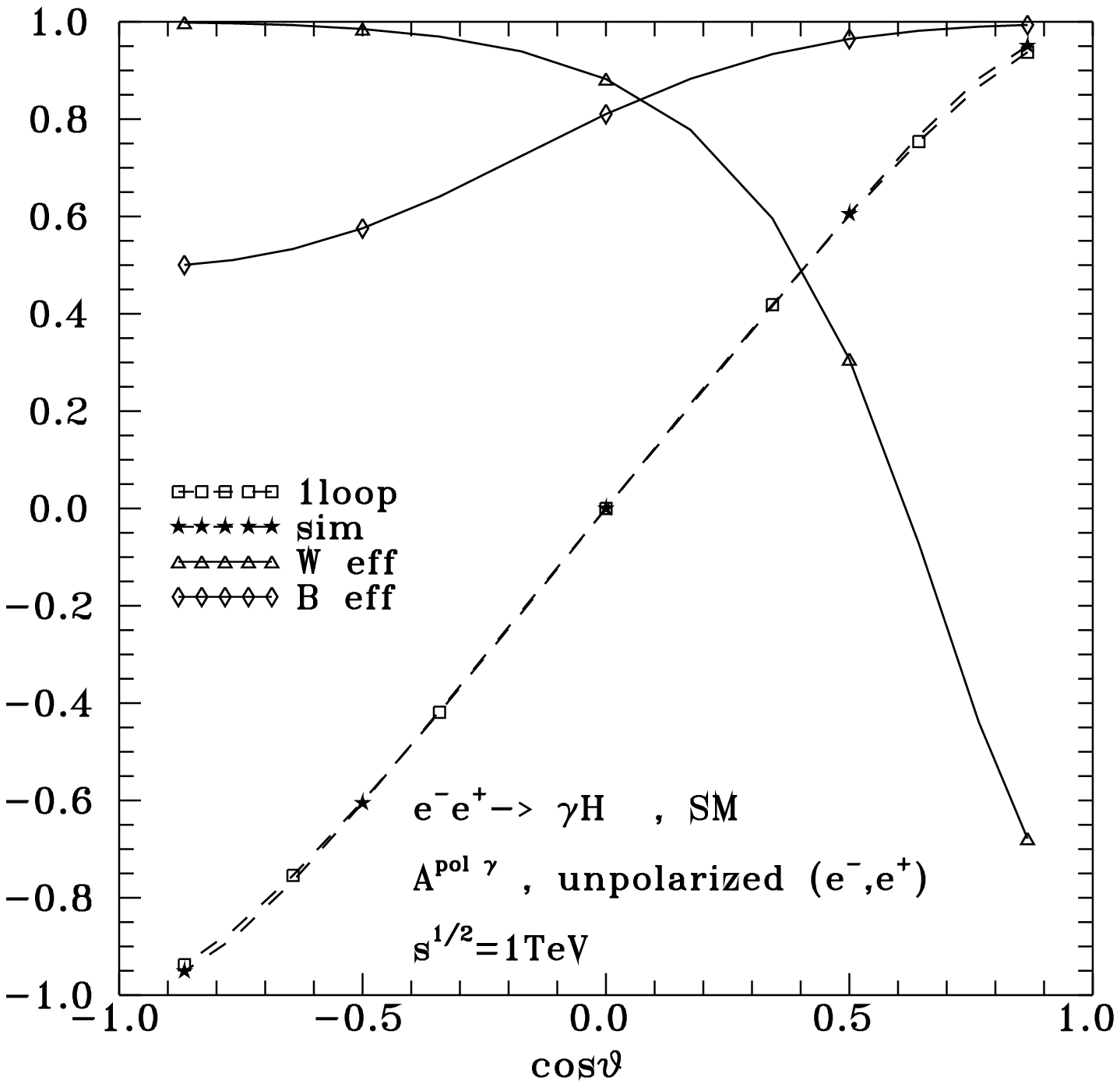, height=6.cm}\hspace{1.cm}
\epsfig{file=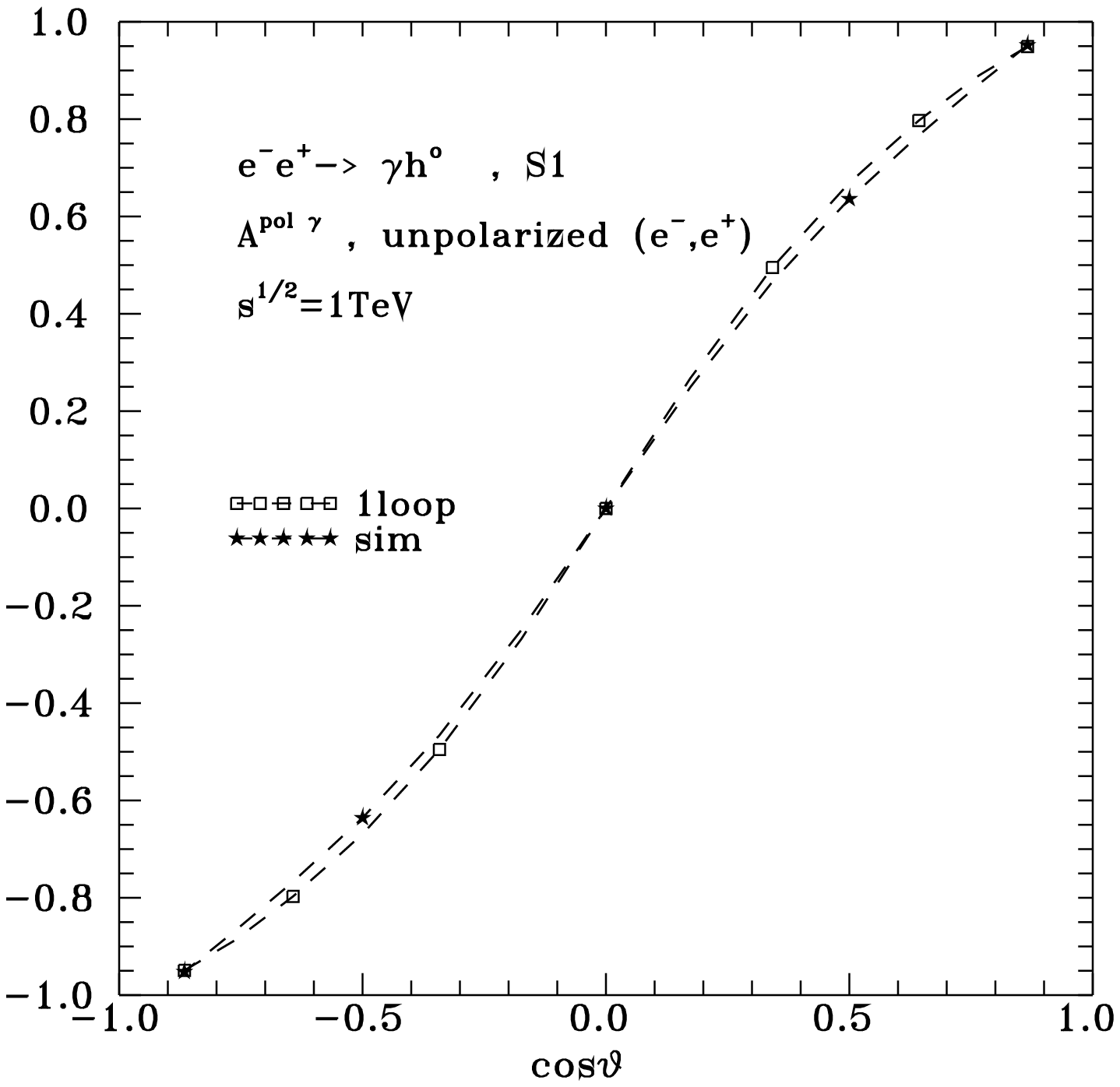,height=6.cm}
\]
\[
\epsfig{file=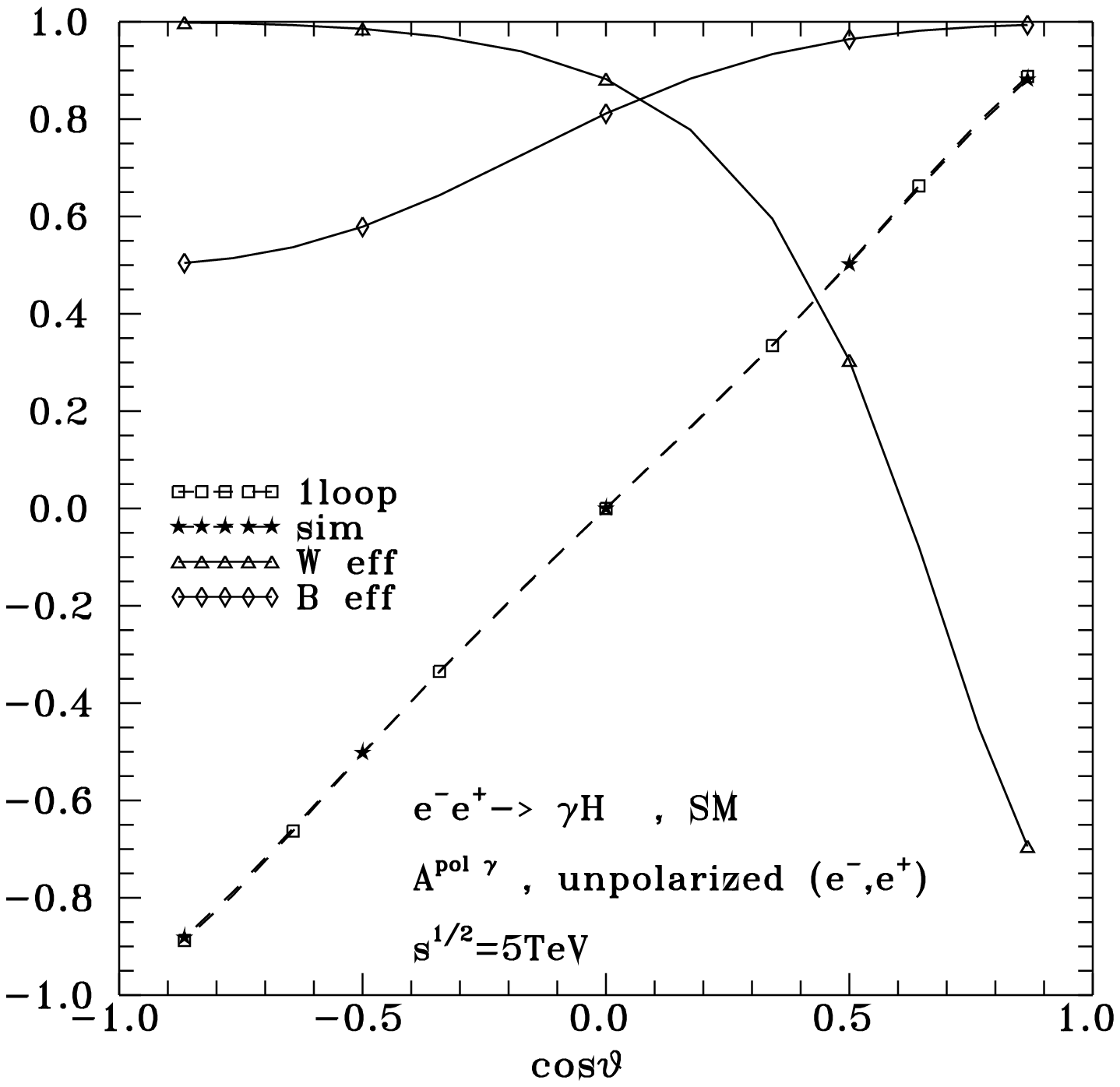, height=6.cm}\hspace{1.cm}
\epsfig{file=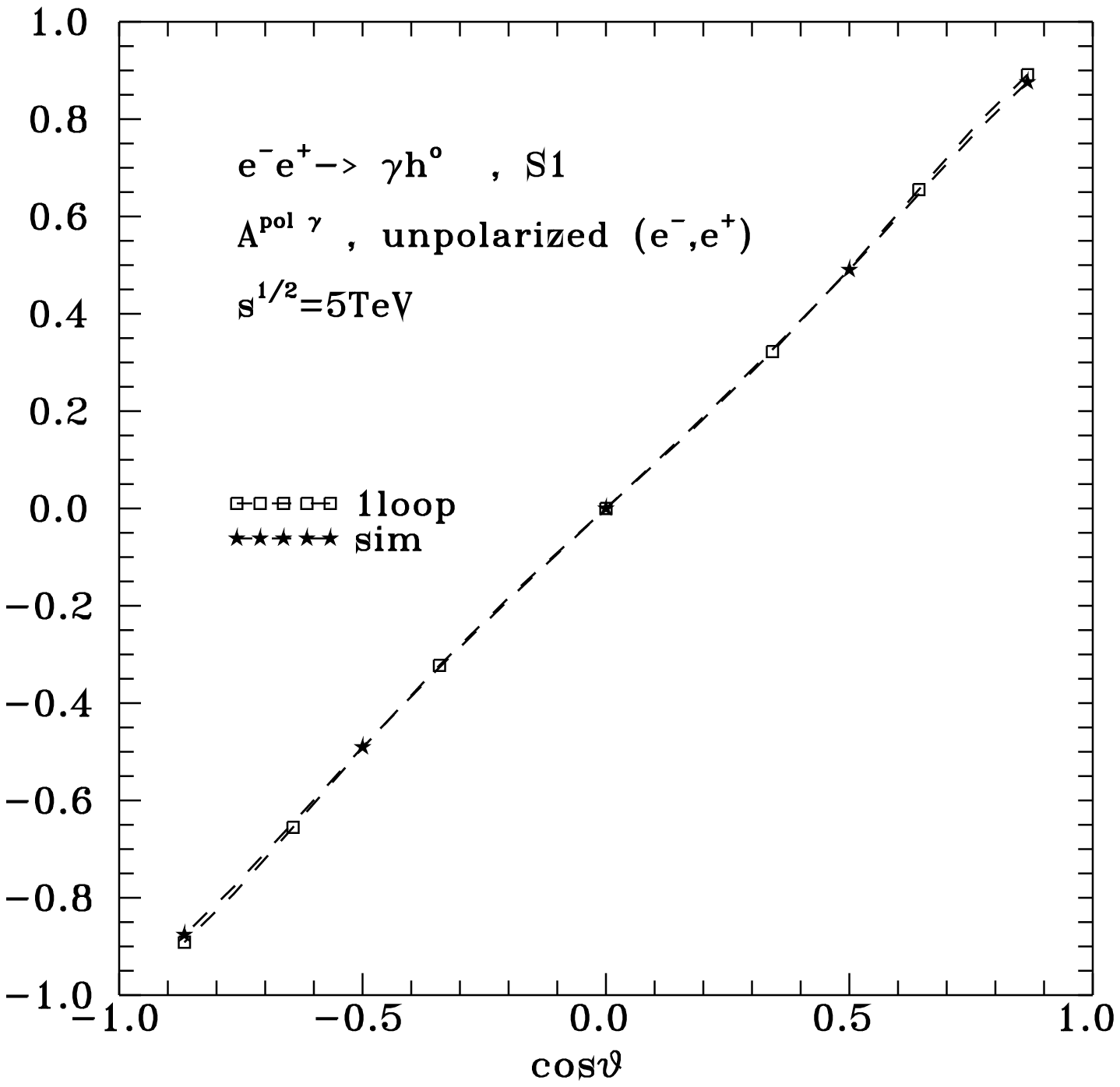,height=6.cm}
\]
\caption[1]{$A^{{\rm pol}~\gamma}$ defined in (\ref {A-polgamma})
 for unpolarized $e^\mp$, for SM (left panels)
and for $h^0$ in S1 MSSM (right panels).
BSM and MSSM parameters as in previous figures.}
\label{Apol-gamma-unpol-e-SM-h0}
\end{figure}

\clearpage

\begin{figure}[h]
\vspace{-1cm}
\[
\epsfig{file=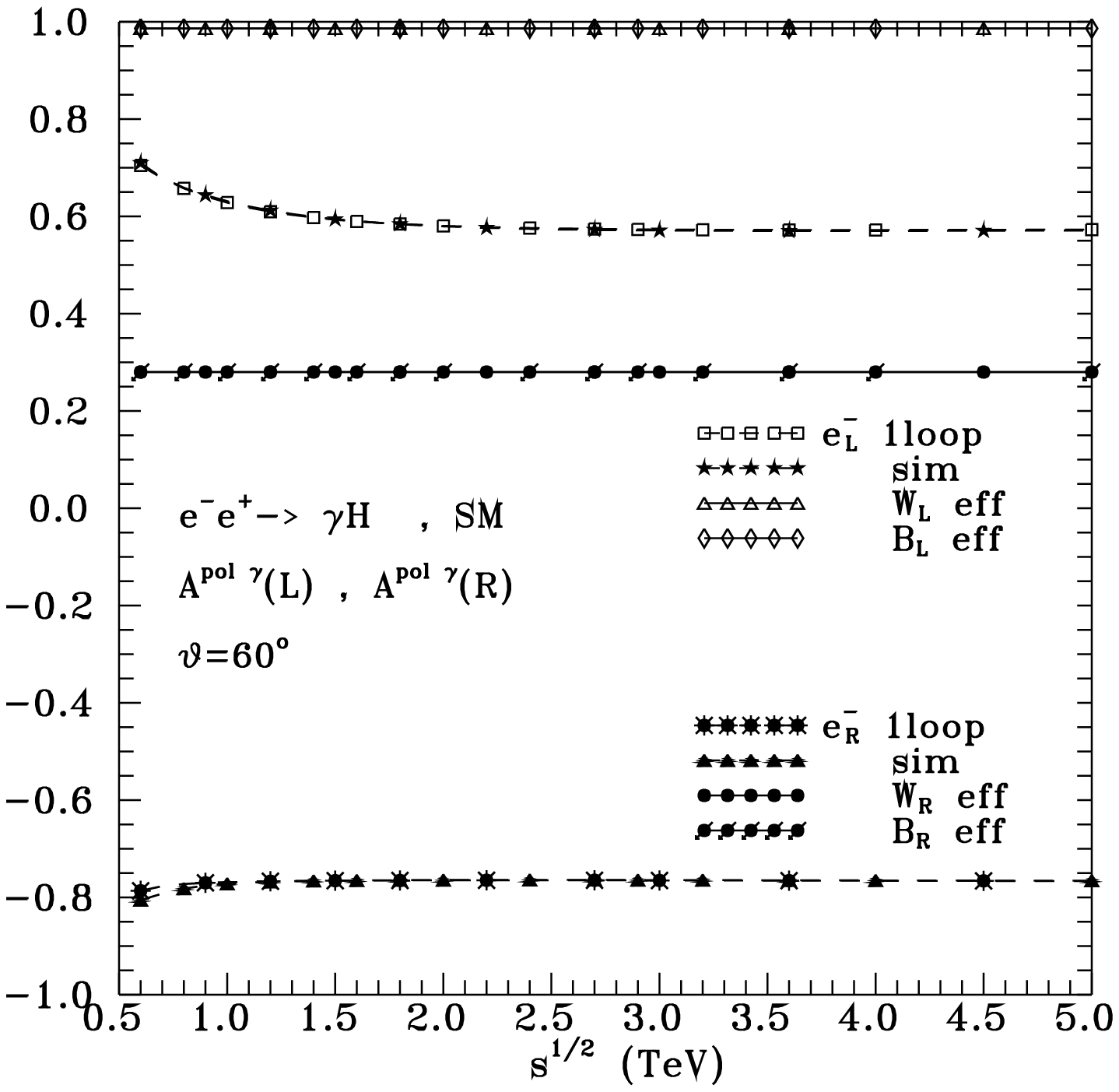, height=6.cm}\hspace{1.cm}
\epsfig{file=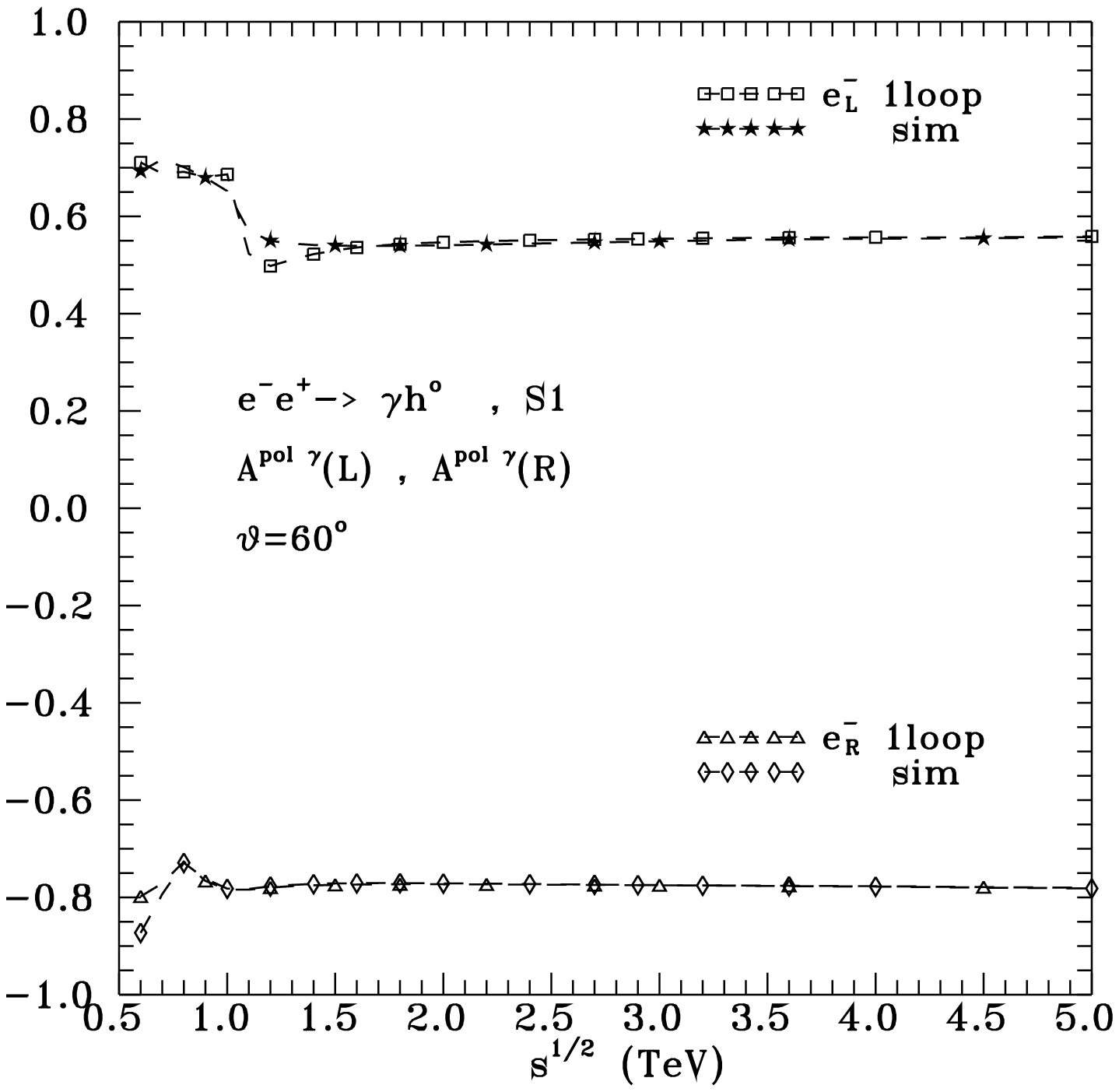,height=6.cm}
\]
\[
\epsfig{file=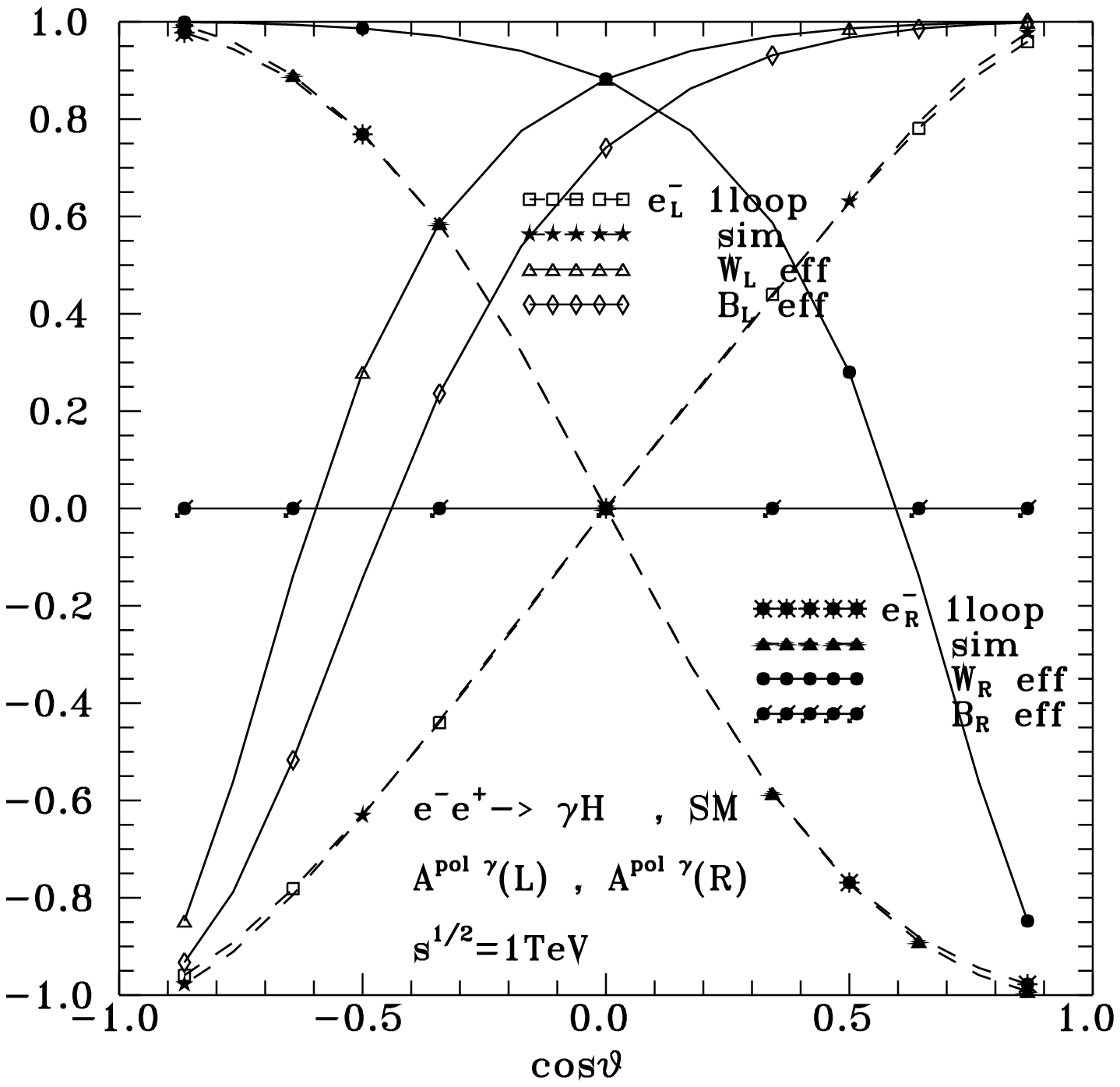, height=6.cm}\hspace{1.cm}
\epsfig{file=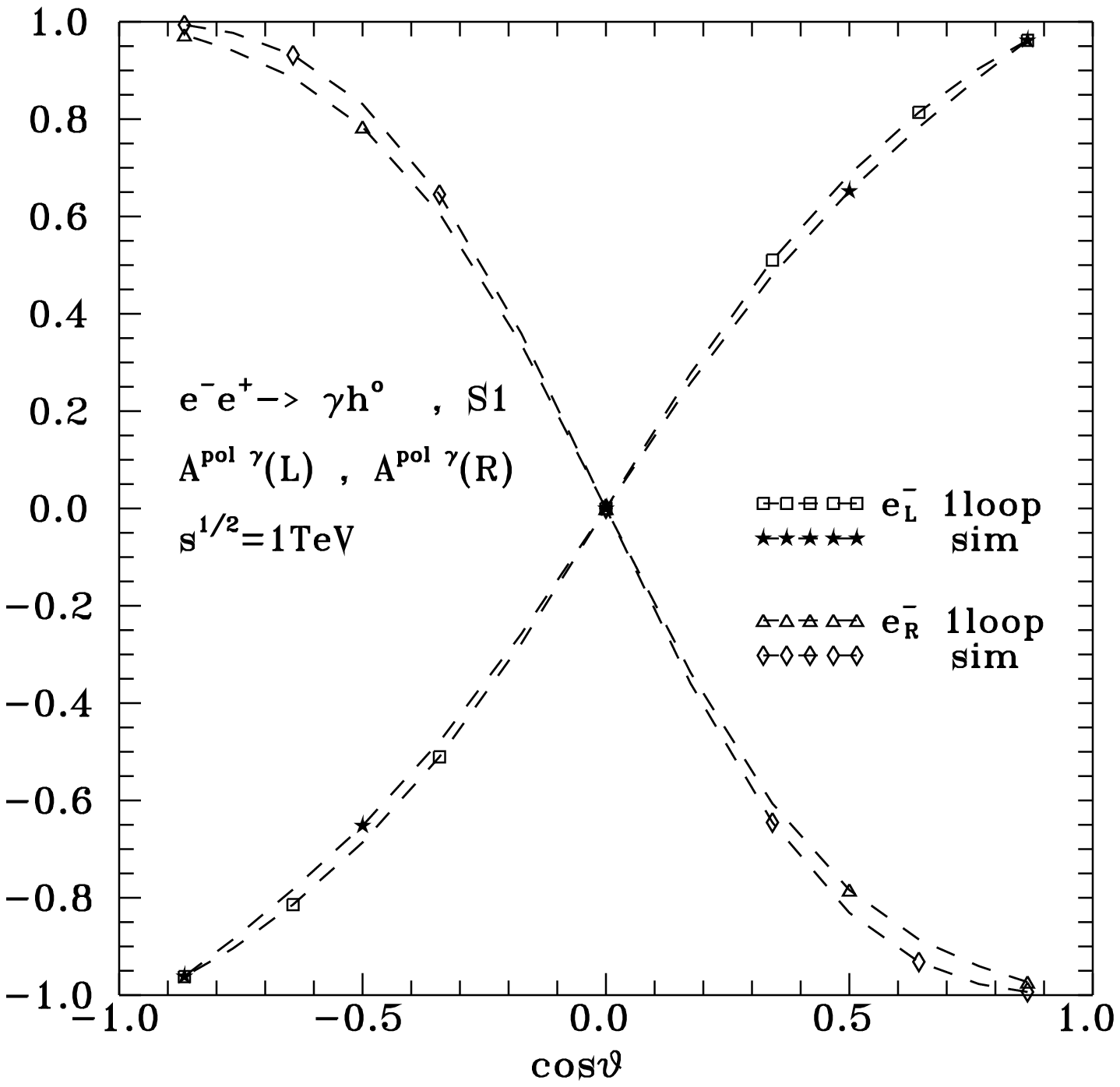,height=6.cm}
\]
\[
\epsfig{file=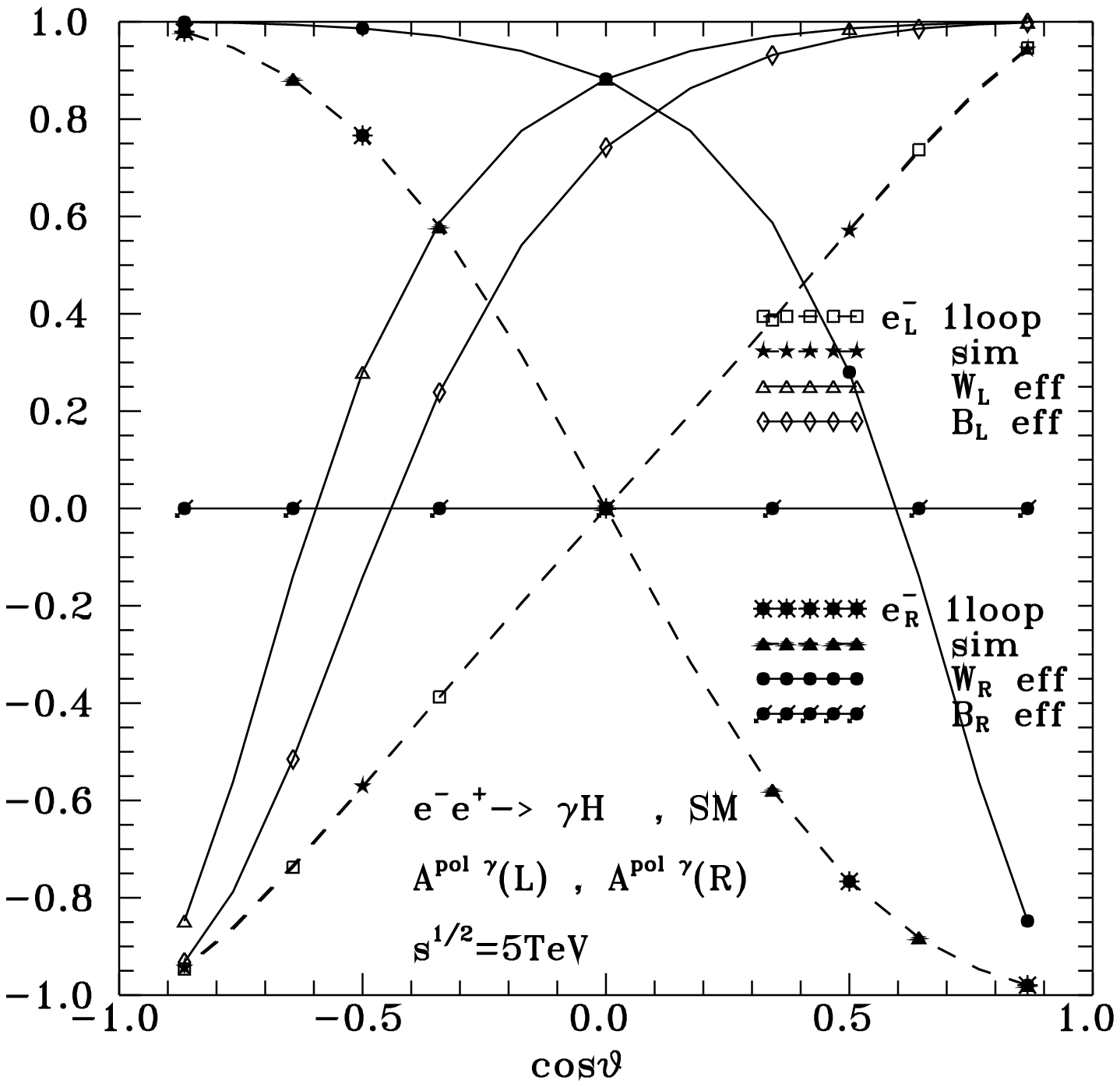, height=6.cm}\hspace{1.cm}
\epsfig{file=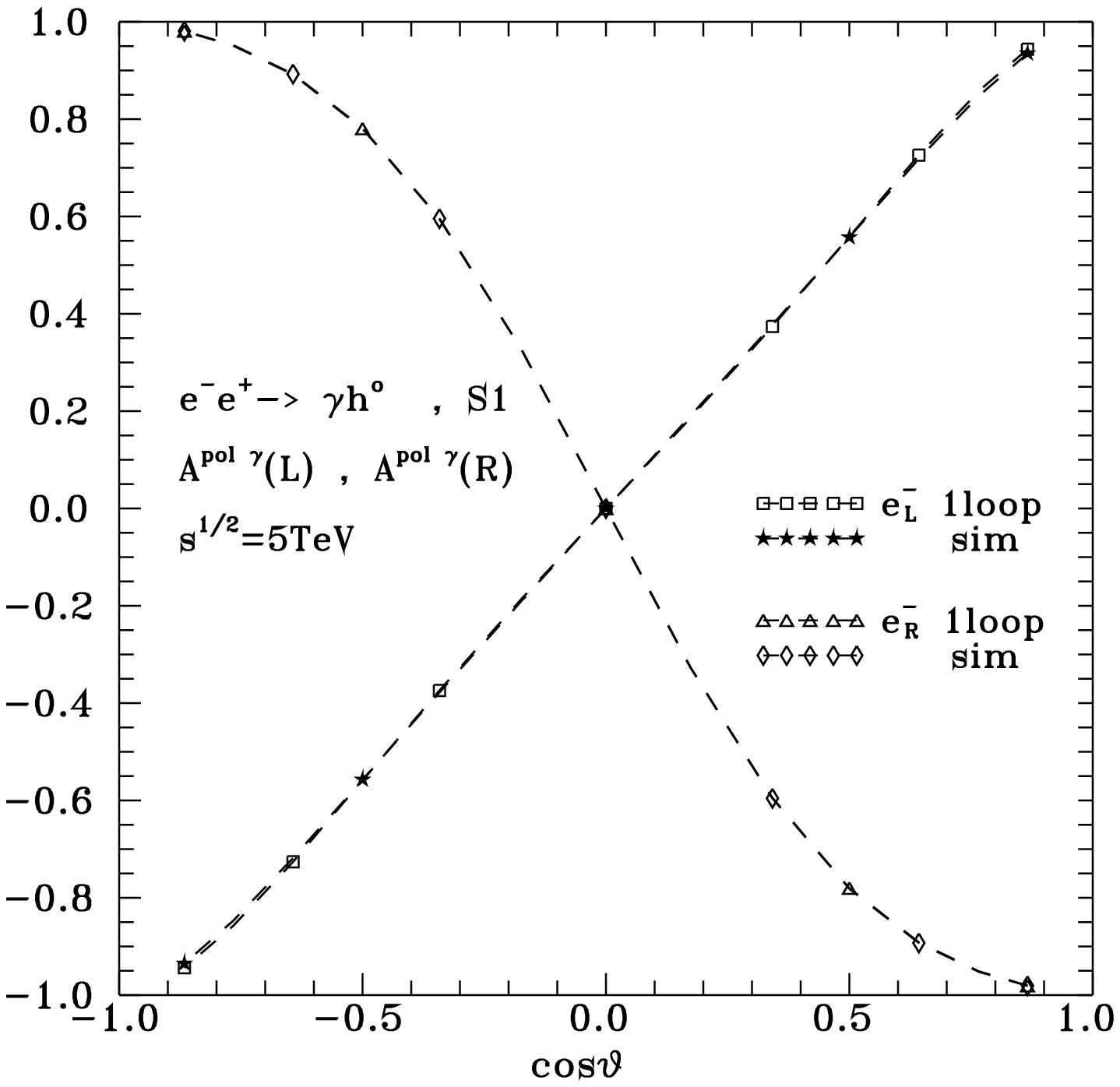,height=6.cm}
\]
\caption[1]{$A^{{\rm pol}~\gamma}$  defined in  (\ref {Alambda-polgamma})
for $e_L$ and $e_R$  beams,
in SM (left panels) and  in S1 MSSM for $h^0$ (right panels).
BSM and MSSM parameters as in previous figures.}
\label{Apol-gamma-eL-eR-SM-h0}
\end{figure}

\clearpage

\begin{figure}[h]
\vspace{-1cm}
\[
\epsfig{file=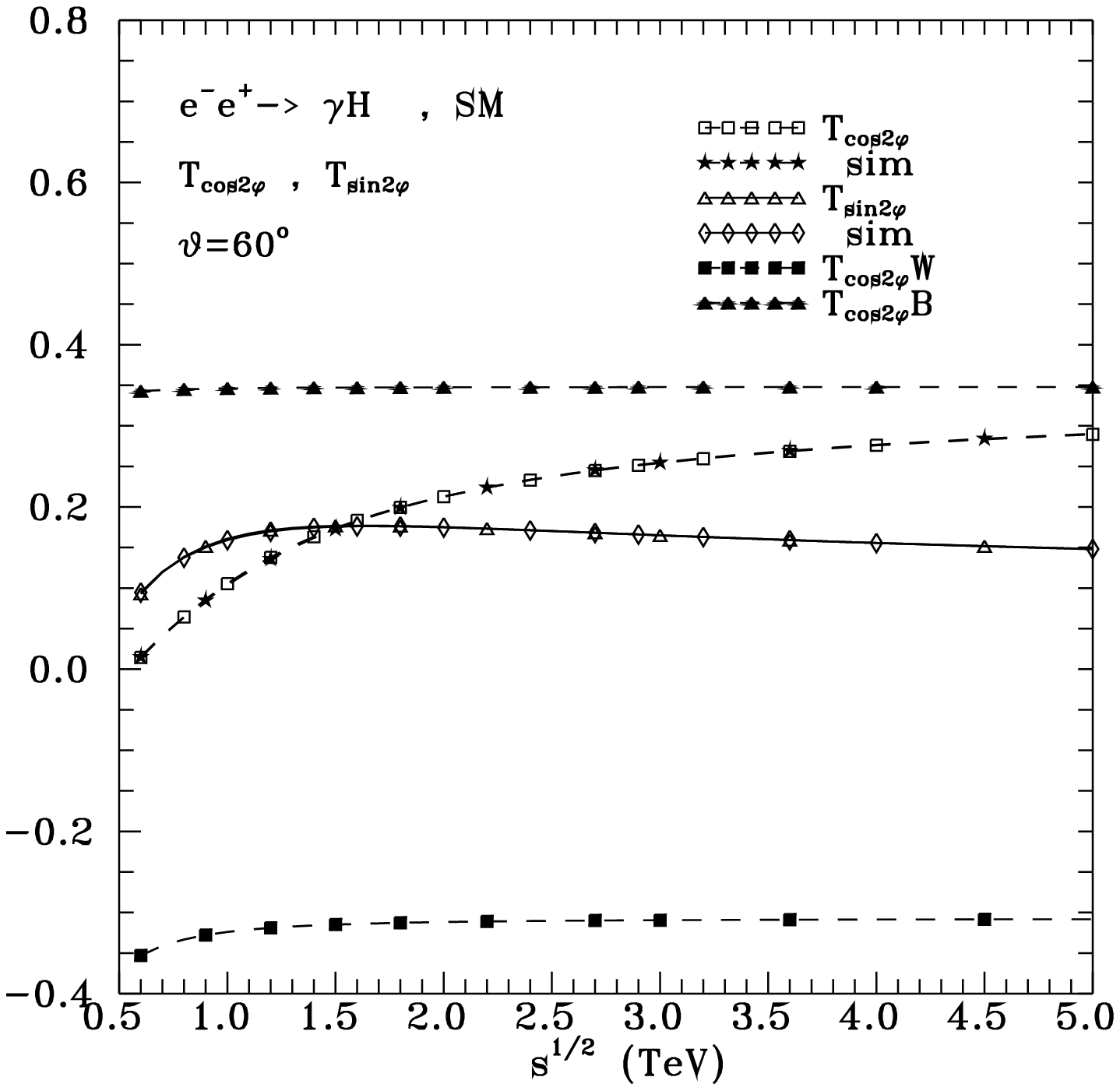, height=6.cm}\hspace{1.cm}
\epsfig{file=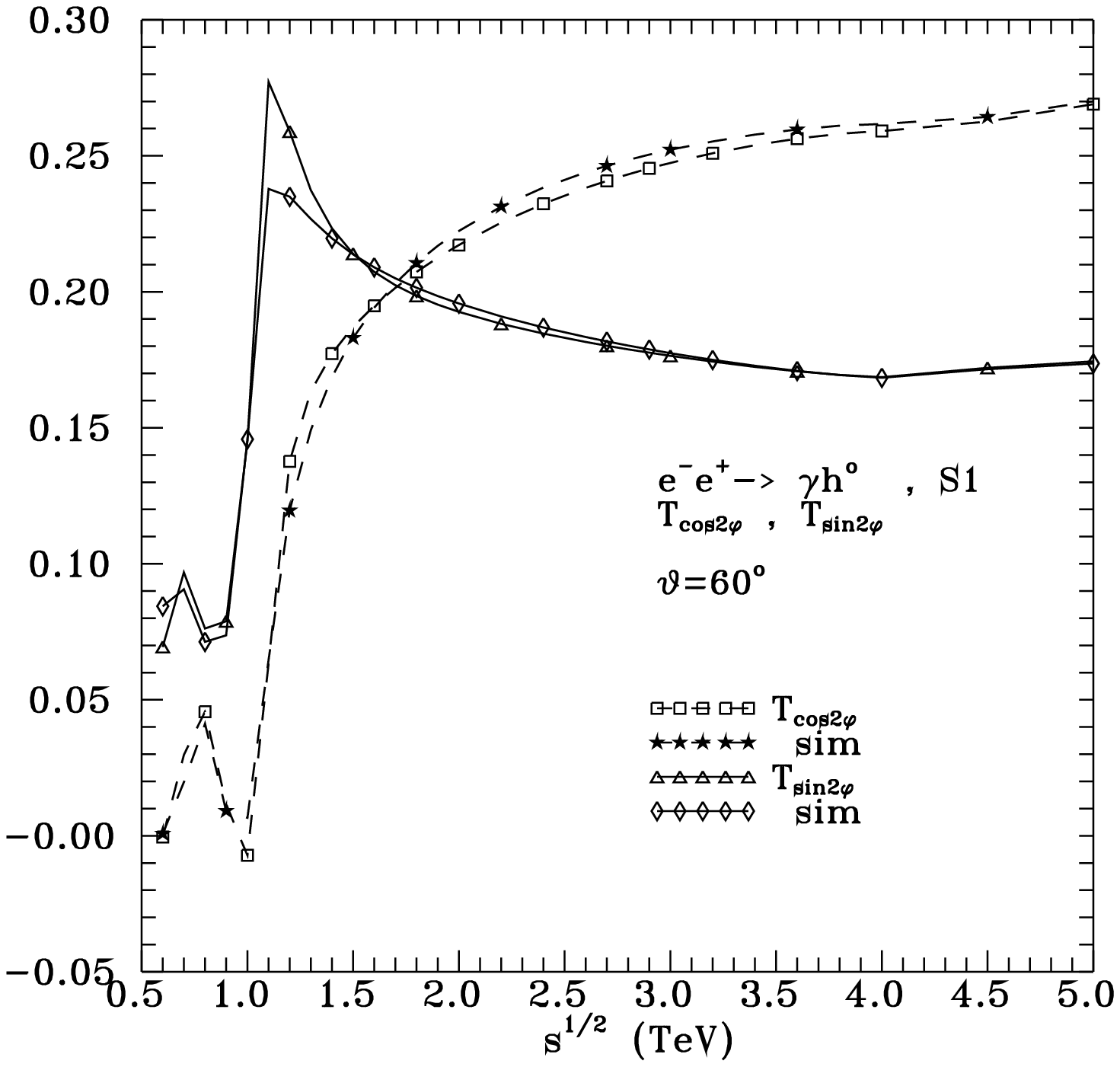,height=6.cm}
\]
\[
\epsfig{file=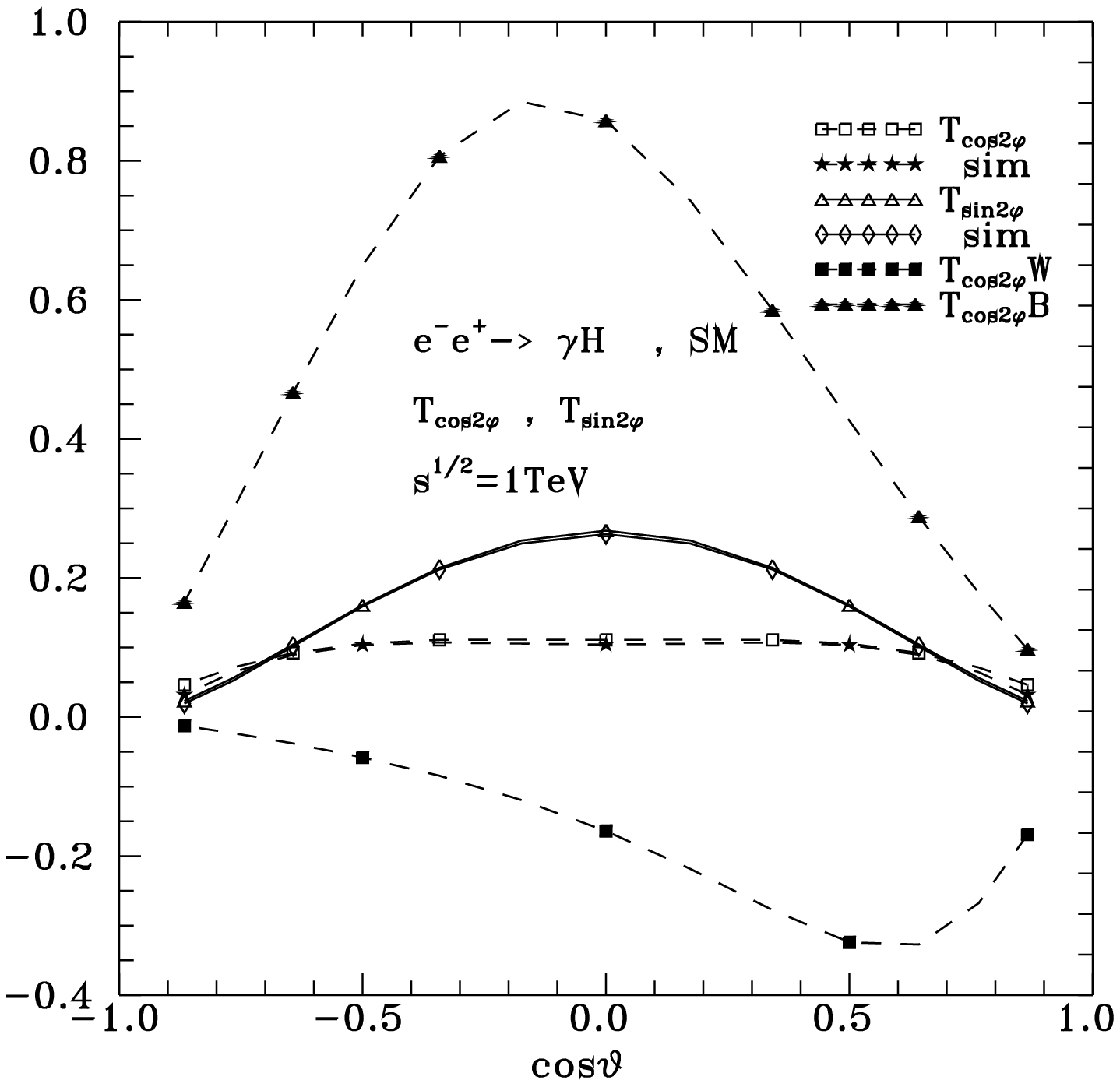, height=6.cm}\hspace{1.cm}
\epsfig{file=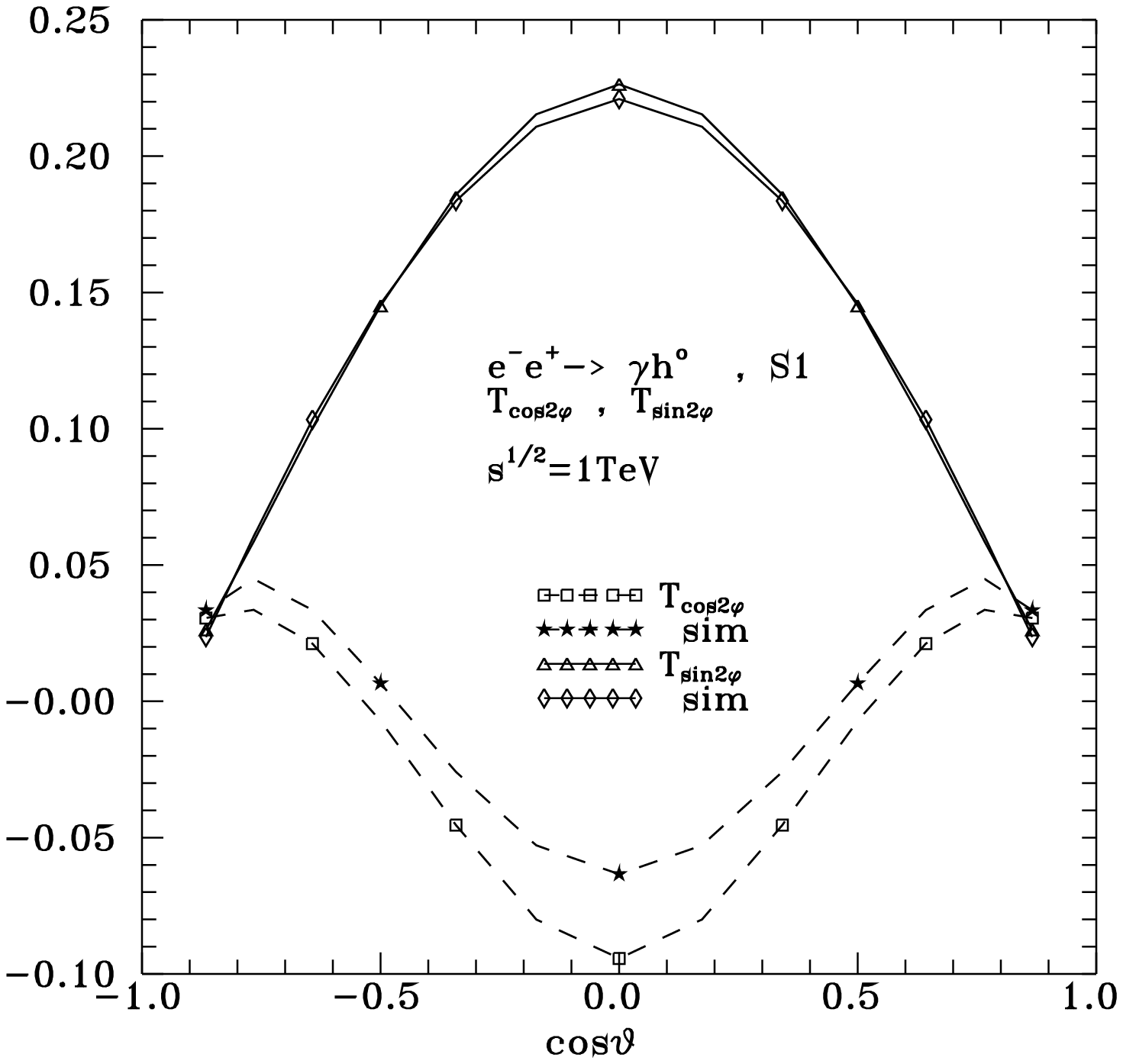,height=6.cm}
\]
\[
\epsfig{file=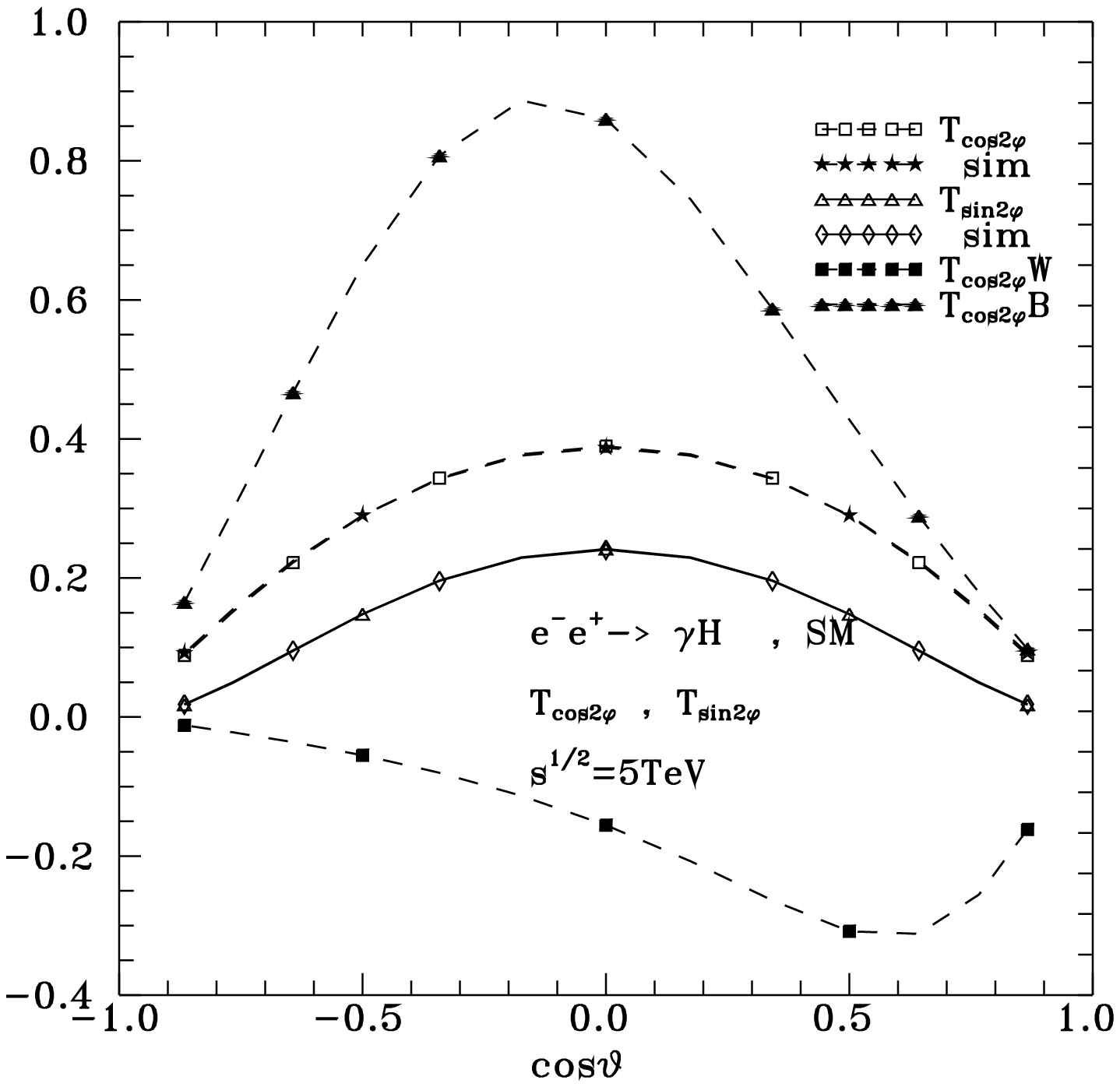, height=6.cm}\hspace{1.cm}
\epsfig{file=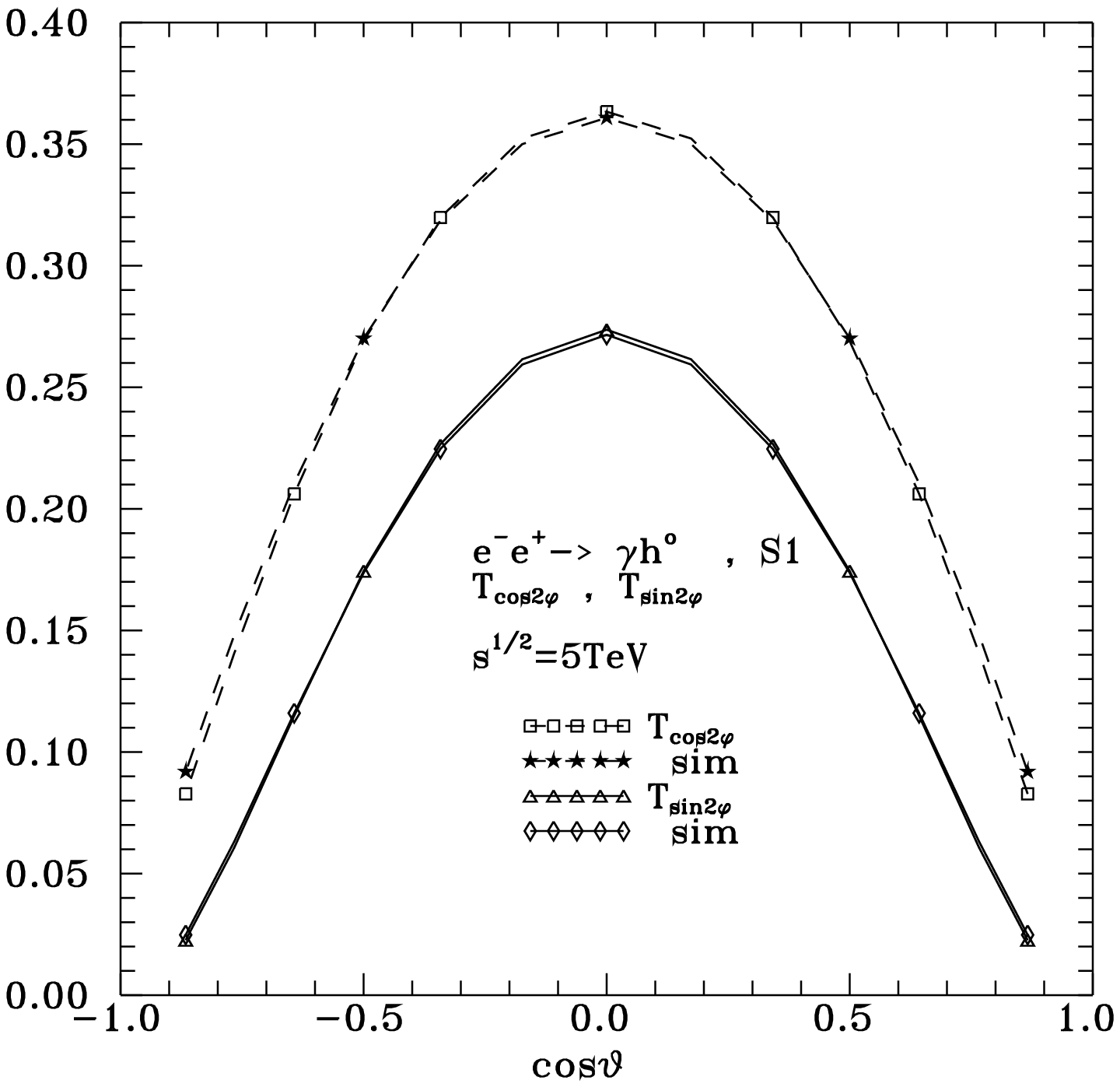,height=6.cm}
\]
\caption[1]{The coefficients $T_{\cos2\phi}$, $T_{\sin2\phi}$ defined in (\ref{T-terms}) for
SM (left panels) and for $h^0$ in S1 MSSM (right panels).
BSM and MSSM parameters as in previous figures.}
\label{Trans-SM-h0}
\end{figure}


\clearpage

\begin{thebibliography}{99}


\bibitem{Higgsdiscov} G. Aad et al. [ATLAS Collaboration], \pl{B716}{1}{2012}, arXiv:1207.7214 [hep-ex].
S. Chatrchyan et al. [CMS Collaboration], \pl{B716}{30}{2012}, arXiv:1207.7235
[hep-ex]. Gavin J. Davies for the  CDF and D0 Collaborations,
published in Front. Phys. China {\bf 8}, 270 (2013), arXiv:1207.0449 [hep-ex].
ATLAS Collaboration:
https://twiki.cern.ch/twiki/bin/view/AtlasPublic/HiggsPublicResults.
CMS Collaboration:
 https://twiki.cern.ch/twiki/bin/view/CMSPublic/PhysicsResultsHIG.
%
\bibitem{Higgs} P. Higgs, \pl{12}{132}{1964}; \prl{13}{508}{1964}; \pr{145}{1156}{1966)};
F. Englert and R. Brout, \prl{13}{321}{1964}; G. Guralnik, C. Hagen and
T. Kibble, \prl{13}{585}{1964}.
%
\bibitem{Higgsearch}John Ellis, [arXiv:1312.5672]. S. Dawson et al
(Higgs working group) [arXiv:1310.8361],
M. Klute, R. Lafaye, T. Plehn, M. Rauch and D. Zerwas, arXiv:1301.1322.
A. Djouadi, \prep{459}{1}{2008} 1 , arXiv:hep-ph/0503173.
J. Gunion, H. Haber, G. Kane and S. Dawson, The Higgs Hunter's Guide, Addison-Wesley,
1990.
S. Heinemeyer, \ijmp{A 21}{2659}{2006},   arXiv:0407244 [hep-ph].

%
\bibitem{ILCH} A. Ajaib et al, (Higgs Working Group),
arXiv:1310.8361 [hep-ex]; T.Behnke et al, arXiv:1306.6327.
%
\bibitem{simZH} G.J. Gounaris and F.M. Renard,  \pr{D90}{073007}{2014},
arXiv:1409.2596, [hep-ph].
%
\bibitem{ILC} H.Baer et al, ILC-Report-20136040.
%
\bibitem{CLIC} M.Aicheler et al, CERN-2012-007.
%
\bibitem{FCC-ee} see http://cern.ch/fcc
%
\bibitem{Barroso} A. Barroso, J. Pulido, J.C. Romao, \np{B267}{509}{1986}.
%
\bibitem{Abbasabadi} A. Abbasabadi, D. Bower-Chao, D.A. Dicus, W.W. Repko,
\pr{D52}{3919}{1995}.
%
\bibitem{PV} G. Passarino and M. Veltman \np{B160}{151}{1979}.
%
\bibitem{asPV} M. Beccaria, G.J. Gounaris, J. Layssac and F.M. Renard,
\ijmp{A23}{1839}{2008}.
%
\bibitem{super} G.J. Gounaris and F.M. Renard,
 \polon{42}{2107}{2011}, arXiv:1106.2707[hep-ph].
%
\bibitem{ttbar} G.J. Gounaris and F.M. Renard, \pr{D86}{013003}{2012},
arXiv:1205.4547 [hep-ph].
%
\bibitem{WW} G.J. Gounaris and F.M. Renard, \pr{D88}{113003}{2013},
arXiv:1309.3177 [hep-ph].
%
\bibitem{heli1}  G.J. Gounaris and F.M. Renard,
\prl{94}{131601}{2005},  hep-ph/0501046.
%
\bibitem{heli2} G.J. Gounaris and F.M. Renard,  \pr{D73}{097301}{2006},
hep-ph/0604041, (an Addendum).
%
\bibitem{JW} M. Jacob and G.C. Wick, \aop{7}{404}{1959}, \aop{281}{774}{2000}.
%
\bibitem{Denner} A. Denner, J. Kubleck, R. Mertig, M. Bohm, \zp{C56}{261}{1992}.
B.A. Kniehl, \zp{C55}{605}{1992}. A. Denner , B.A. Kniehl, \npps{29A}{263}{1992}.
R. Hempfling, B.A. Kniehl, \zp{C59}{263}{1993}.
%
\bibitem{anom} G.J. Gounaris, F.M. Renard and N.D. Vlachos,
\np{B459}{51}{1996}; G.J. Gounaris, D.T. Papadamou and F.M. Renard,
\zp{C76}{333}{1997}.
%
\bibitem{Chiappetta} P.Chiappetta, J. Soffer, P. Taxil, F.M. Renard,
\pr{D31}{1739}{1985}.
%
\bibitem{Renard} F.M. Renard, C85-06-10.1, Proc. Trieste Conf. 1985.
%
\bibitem{Rindani} S.D. Rindani, hep-ph/0409014; K. Rao, S.D. Rindani, \pl{B642}{85}{2006}
%
\bibitem{bench} M Arana-Catania, S. Heinemeyer, M.J. Herrero,
\pr{D88}{015026}{2013} arXiv:1304.2783 [hep-ph]. See also arXiv:1405.6960 [hep-ph].
%
\bibitem{Rosiek} J. Rosiek, \pr{D41}{3464}{1990}.
%
\end{thebibliography}
\end{document}